%% file: ms.tex
\documentclass[iop]{emulateapj}
\usepackage{rotating}
\usepackage{color} 
\usepackage{amsmath} 
\newcommand{\kms}       {\mbox{km s$^{-1}$}}%
\newcommand{\msun}      {\mbox{$M_\odot$}}%
\newcommand{\magarc}    {\ensuremath{\mbox{mag arcsec}^{-2}}}%
\newcommand{\cur}      {\mbox{$(u-r)$}}%
\newcommand{\cuj}      {\mbox{$(u-J)$}}%
\newcommand{\cuy}      {\mbox{$(u-Y)$}}%
\newcommand{\cuk}      {\mbox{$(u-K)$}}%
\newcommand{\cgr}      {\mbox{$(g-r)$}}%
\newcommand{\cgi}      {\mbox{$(g-i)$}}%

\shortauthors{Eckert et al.}
\shorttitle{RESOLVE and ECO Galaxy Mass Functions}

\begin{document}

\title{RESOLVE and ECO: The Halo Mass-Dependent Shape of Galaxy Stellar and Baryonic Mass Functions}

\author{Kathleen D. Eckert\altaffilmark{1},
        Sheila J. Kannappan\altaffilmark{1},
        David V. Stark\altaffilmark{1,2},
        Amanda J. Moffett\altaffilmark{1,3},
        Andreas A. Berlind\altaffilmark{3}, and
        Mark A. Norris\altaffilmark{1,5}}

\altaffiltext{1}{Department of Physics and Astronomy, University of
  North Carolina, 141 Chapman Hall CB 3255, Chapel Hill, NC 27599,
  USA; keckert@physics.unc.edu} 

\altaffiltext{2}{Kavli Institute for the Physics and Mathematics of the Universe (IPMU), The University of Tokyo, 5-1-5 Kashiwanoha, Kashiwa, 277-8583, Japan} 

\altaffiltext{3}{International Centre
  for Radio Astronomy Research (ICRAR), The University of Western
  Australia, 35 Stirling Highway, Crawley, WA 6009, Australia}

\altaffiltext{4}{Department of Physics and Astronomy, Vanderbilt
  University, PMB 401807, 2401 Vanderbilt Place, Nashville, TN
  37240-1807, USA}


\altaffiltext{5}{Jeremiah Horrocks Institute, University of Central Lancashire, Preston, PR1 2HE, United Kingdom}

\begin{abstract}

In this work, we present galaxy stellar and baryonic (stars plus cold
gas) mass functions (SMF and BMF) and their halo mass dependence for
two volume-limited data sets.  The first, RESOLVE-B, coincides with
the Stripe 82 footprint and is extremely complete down to baryonic
mass \mbox{M$_{bary}$ $\sim$ 10$^{9.1}$ \msun}, probing the gas-rich
dwarf regime below \mbox{M$_{bary}$ $\sim$ $10^{10}$~\msun}.  The
second, ECO, covers a $\sim$40$\times$ larger volume (containing
RESOLVE-A) and is complete to \mbox{M$_{bary}$ $\sim$ $10^{9.4}$
  \msun}. To construct the SMF and BMF we implement a new ``cross-bin
sampling'' technique with Monte Carlo sampling from the full
likelihood distributions of stellar or baryonic mass. Our SMFs exhibit
the ``plateau'' feature starting below \mbox{M$_{star}$ $\sim$
  $10^{10}$ \msun{}} that has been described in prior work. However,
the BMF fills in this feature and rises as a straight power law below
\mbox{$\sim$10$^{10}$ \msun}, as gas-dominated galaxies become the
majority of the population. Nonetheless, the low-mass slope of the BMF
is not as steep as that of the theoretical dark matter halo MF.
Moreover, we assign group halo masses by abundance matching, finding
that the SMF and BMF separated into four physically motivated halo
mass regimes reveal complex structure underlying the simple shape of
the overall MFs. In particular, the satellite MFs are depressed below
the central galaxy MF ``humps'' in groups with mass \mbox{$<$
  10$^{13.5}$ \msun{}} yet rise steeply in clusters. Our results
suggest that satellite destruction and/or stripping are active from
the point of nascent group formation. We show that the key role of
groups in shaping MFs enables reconstruction of a given survey's SMF
or BMF based on its group halo mass distribution.

\end{abstract}



\section{Introduction}
\label{sec:intro}

Galaxy luminosity and mass functions are key tools for understanding
the distribution of matter in the universe.  The shape of the
luminosity function (LF) reveals the mass assembly of galaxies through
hierarchical evolution, as the dwarf galaxies that dominate the galaxy
population eventually merge to form the rarer bright galaxy
population. As galaxies merge to form larger structures, their host
halos also merge and grow, implying a relationship between the LF and
halo mass function (HMF). Despite this link between galaxies and their
host halos, the faint-end slope for the LF has been measured to be
much shallower than the low-mass slope for the HMF.  While the
low-mass slope of the HMF is often reported to be
$\alpha_{HMF} = -2$ (e.g.,
\citealp{1974ApJ...187..425P} and \citealp{2005Natur.435..629S}), for
the Sloan Digital Sky Survey (SDSS, \citealp{2000AJ....120.1579Y})
\citet{2003ApJ...592..819B} measure the faint end of the LF to be
$\alpha_{LF} = -1.05$.


The faint-end slope of the LF, however, is dependent on environment
such that the faint-end slope in clusters is much steeper than in less
dense environments \citep{2002ApJ...569..573T}. In fact
\citet{2006A&A...445...29P} measured faint-end slopes
$\alpha_{LF} \sim -2$ in a large sample of
clusters. Other previous studies of the cluster LF, however, have
found varying faint-end slopes from $-1.0$ to $-1.4$
\citep{1978ApJ...223..765D,1997ApJ...479...90V,2002PASJ...54..515G}.
One proposed explanation for the discrepancy between low galaxy number
counts in the field vs.\ in clusters is that galaxies in more dense
environments formed earlier and faster, before the reionization of
hydrogen by the UV background, whereas galaxies in less dense
environments took longer to form. The latter galaxies were therefore
``squelched'' by reionization, which heated the gas so that the galaxy
formation was delayed until the reionized gas could re-cool and form
stars
\citep{1996ApJ...465..608T,1997ApJ...486..581G,2001AJ....122.2850B,2010MNRAS.408...57P}.

While the LF offers clues to the mass assembly of galaxies, it is also
sensitive to star formation histories. With the development of
multiband photometric surveys, we can now estimate stellar masses for
galaxies with population synthesis modeling and study the build up of
stellar mass through the stellar mass function (SMF). Working at
\mbox{z $< 0.05$} and \mbox{z $= 0.2-1$} respectively,
\citet{2008MNRAS.388..945B} and \citet{2009ApJ...707.1595D} find that
the traditional Schechter function does not adequately describe the
SMF, which exhibits a ``dip'' or ``plateau'' at stellar masses
\mbox{$\sim$10$^{10}$ \msun} before rising more steeply for lower mass
galaxies.  Instead the authors use a double Schechter function to fit
the SMF's more complex structure.

These discrepancies in mass function shape motivate the desire to
study galaxy mass functions less sensitive to galaxy star formation
properties. The baryonic mass function (BMF), or frequency
distribution of galaxies in stellar plus cold gas mass, is one step
closer to a total mass function than the SMF. In this work we define
baryonic mass as stars plus cold \textit{atomic} gas (neglecting the
cold molecular as well as warm and hot gas components). In general
cold atomic gas dominates the cold gas mass in galaxies, except in
large spirals, for which the total cold gas content is usually less
than the stellar mass (e.g.,
\citealp{1998A&A...331..451C,2013ApJ...777...42K,2014A&A...564A..66B}).

At high masses, we expect the BMF and SMF to be the same.  Above
stellar masses $\sim$10$^{10.5}$~\msun, the bimodality mass scale
\citep{2003MNRAS.341...54K}, galaxies are increasingly
spheroid-dominated with old stellar populations and little to no
recent star formation.  Since these galaxies have minimal cold gas
reservoirs, their baryonic masses are roughly equal to their stellar
masses. Below the bimodality mass scale, gas becomes increasingly
important, so we expect to see a divergence between the BMF and SMF
around or below this scale.  The divergence may be expected to become
more extreme below the gas-richness threshold mass identified by
\citet{2013ApJ...777...42K}, hereafter K13, as \mbox{M$_{star}$ $\sim$
  10$^{9.7}$ \msun} or \mbox{M$_{bary}$ $\sim$ 10$^{9.9}$~\msun},
below which gas-dominated galaxies become the norm (see also
\citealp{2004ApJ...611L..89K,2008AIPC.1035..163K,2009AJ....138..579K}).
The gas mass in such gas-dominated galaxies shifts them to more
massive bins by $\gtrsim$0.3 dex so we want to investigate whether
these galaxies fill in the dip feature seen in the SMF.

Early work by \citet{2003ApJ...585L.117B} investigated the SMF and
BMF, showing a divergence between the two at low masses. This work,
however, did not reveal or investigate any structure beyond a single
Schechter function form, perhaps due to systematics in stellar mass
estimation \citep{2007ApJ...657L...5K,2015MNRAS.452.3209R} and/or gas
mass estimation methodology.  Recent studies of the SMF and BMF find a
dip only in the SMF, which suggests the dip is purely due to neglect
of the cold gas mass
\citep{2008MNRAS.388..945B,2012ApJ...759..138P}. This result implies
that the BMF is a more fundamental way to characterize galaxy
populations than the SMF in that it better reflects the total galaxy
mass.

Thus to relate the mass assembly of galaxies and halos, we would like
to examine the BMF and in particular how the BMF depends on
environment. It has been predicted that the BMF might be invariant
across environments, as opposed to the faint-end slope of the LF,
which is observed to steepen in cluster versus field environments
\citep{2003ApJ...585L.117B}. This idea is based on the assumption of
gas re-cooling after reionization: galaxies in high density
environments form earlier and faster, using up their gas to form stars
before reionization hits, so the majority of cold baryons in the
cluster environment are associated with stars, while galaxies in
low-mass environments form later and are initially ``squelched'' by
reionization, eventually recovering their baryonic mass as the gas
re-cools over cosmic time
\citep{2000ApJ...539..517B,2002ApJ...572L..23S}. Thus the BMF (stars
plus cold gas) would now be constant across environment. While there
has been no measurement of an environment dependent low-mass slope for
the BMF, estimates for the low-mass slope of the overall BMF range
from $\alpha_{BMF} = -1.2$ \citep{2003ApJ...585L.117B,2012ApJ...759..138P}
to $\alpha_{BMF} = -1.8$ \citep{2008MNRAS.388..945B}, and the source of
these variations is unclear.

One possible reason for the differences between measurements of
low-mass slope could be that the methods used in each study have
varied widely. A steep low-mass slope is found in
\citet{2008MNRAS.388..945B}, where baryonic mass is inferred using the
stellar mass-metallicity relation combined with the relation between
metallicity and stellar mass fraction, i.e., fraction of baryons
locked up in stars, to obtain total baryonic masses that implicitly
include all gas, cold and warm.  Both \citet{2003ApJ...585L.117B} and
\citet{2005RSPTA.363.2693R} employ indirect methods of estimating the
cold atomic and molecular gas based on photometric properties,
respectively the $K$ band luminosity-$r_e$ plane and galaxy
morphology, both of which have large scatter.  The BMF from
\citet{2012ApJ...759..138P} is constructed from HI measurements from
the blind wide-area 21cm ALFALFA survey \citep{2011AJ....142..170H}
and does not include molecular gas (similar to this work).  Since
ALFALFA is flux-limited, however, gas rich but low-mass galaxies will
be detected only nearby, requiring large statistical corrections at
the faint-end of the mass function.

A second possible reason for the differences between studies could be
that environment actually does affect the BMF, via physics additional
to re-cooling since reionization. In high-mass halos, cold or warm gas
may be stripped as galaxies enter the group environment, effectively
moving the galaxies to lower baryonic mass, or causing them to lose
their future supply of cooling (sub)halo gas. In low-mass group halos,
cosmic accretion of gas onto the halo as well as halo gas re-cooling
may increase the gas content of galaxies and thus renew the pool of
gas as stars are formed. From these two examples, it is evident that
environment can play a multi-faceted role in shaping the galaxy BMF.
Any two given data sets may contain widely varying environment
distributions due to cosmic variance, so differences in the
environments sampled by previous BMF studies may contribute to
inconsistent low-mass slopes.



In this work we present the SMF and BMF for two volume-limited data
sets: the REsolved Spectroscopy of a Local VolumE (RESOLVE) survey
(Kannappan et al.\ in prep.) and the Environmental COntext (ECO)
catalog \citep{2015ApJ...812...89M}. Because these two data sets are
volume limited, we can examine the shape of the galaxy mass function
and its dependence on halo mass (the proxy for environment used in
this work) without the statistical completeness corrections required
for flux-limited surveys. Both data sets are more complete than the
SDSS main redshift survey, and one, the portion of the RESOLVE survey
overlapping Stripe 82 (RESOLVE-B), offers unprecedented completeness
that enables calculation of empirical completeness corrections for the
other, the ECO catalog containing RESOLVE-A. Our use of volume-limited
data sets enables robust group identification so that we can quantify
group halo mass and directly examine its effect on the mass function
shape. To obtain unbiased gas data for gas-rich yet low M$_{HI}$
galaxies, we require HI data that are fractional-mass limited, i.e.,
adaptively sensitive to a limiting atomic gas mass of 1.4M$_{HI} <
0.05 M_{star}$.  Thus for atomic gas measurements, we use a
combination of the highly complete fractional-mass limited RESOLVE HI
census from Stark et al.\ (submitted.), additional archival HI data
for ECO, and empirically estimated ``photometric gas fractions'' using
the probability density field approach of \citet{2015ApJ...810..166E},
hereafter E15.  Our stellar mass estimation applies SED fitting to
custom reprocessed NUV$ugrizYJHK$ photometry, optimized for recovery
of extended light (E15). Finally we take into account the full
likelihood distributions of both stellar and gas masses when computing
the SMF and BMF via a new cross-bin sampling approach.

As anticipated, we find that the BMF diverges significantly from the
SMF below the gas-richness threshold scale M$_{bary}$ $\sim$
$10^{9.9}$ \msun{} and rises as a straight power law, filling in where
the SMF dips. However, we find that the overall BMF hides significant
substructure as a function of group halo mass. We break down our mass
functions into four physically motivated halo mass regimes, finding
that although mass functions of central galaxies are discrete
``humps'' increasing in mass as halo mass increases, satellite galaxy
mass functions show much greater complexity. These shapes suggest a
connection between group formation and satellite destruction from the
point of first group formation. As evidence that the primary
environmental processes affecting the BMF occur on group scales, we
show that it is possible to combine the mass functions broken down by
group halo mass regime for ECO with the different frequency
distribution of group mass halos for RESOLVE-B to produce the observed
mass function of RESOLVE-B.

This work is laid out as follows.  In \S \ref{sec:datasets} we
describe the surveys used for this work.  In \S \ref{sec:data} we
describe the data, including photometric reprocessing, stellar mass
estimation, HI mass measurement, baryonic masses, and halo mass
determination.  We conclude \S \ref{sec:data} with a discussion of
completeness corrections and mass completeness limits. In \S
\ref{sec:statmassfunctions} we describe our new cross-bin sampling
technique to measure the SMF and BMF.  In \S \ref{sec:massfunctions}
we present the SMF and BMF and break them down by halo mass and
central/satellite designation.  In \S \ref{sec:discussion} we discuss
the role of the group halo mass environment in shaping the galaxy
population and the connection between the galaxy BMF and the
theoretical HMF. Finally in \S \ref{sec:conclusions} we summarize our
conclusions.

For distance measurements and other derived quantities in this work,
we assume a standard $\Lambda$CDM cosmology with $\Omega_m$ = 0.3,
$\Omega_\Lambda$ = 0.7, and $H_o$=70 \kms Mpc$^{-1}$.

\section{Data Sets}
\label{sec:datasets}

To measure the SMF and BMF, we use two volume-limited data sets, the
B-semester subvolume of the RESOLVE (REsolved Spectroscopy of a Local
VolumE) survey, RESOLVE-B (\citealp{2008AIPC.1035..163K}; Kannappan et
al.\ in prep.), and the ECO (Environmental COntext) catalog
(\citealp{2015ApJ...812...89M}, hereafter M15), which contains the
RESOLVE-A subvolume. RESOLVE is a volume and roughly baryonic mass
limited survey of $\sim$52,100 Mpc$^3$ of the z $\sim$ 0 universe. It
is smaller but more complete than ECO and is acquiring new 21cm and
optical spectroscopy to conduct a full mass census of stars, gas, and
dark matter. The RESOLVE-A 21cm census was used in E15 to calibrate
gas mass estimators based on photometric properties of galaxies. The
ECO catalog is $\sim$10 times larger than RESOLVE, providing better
statistics but reduced completeness, and is based on archival data
except in its overlap with RESOLVE-A.

\subsection{Common Features of RESOLVE and ECO}
\label{sec:common}

Both data sets are based on the SDSS main redshift survey, but include
additional redshifts from various archival sources: the Updated Zwicky
Catalog (UZC, \citealp{1999PASP..111..438F}), HyperLEDA
\citep{2003A&A...412...45P}, 6dF \citep{2009MNRAS.399..683J}, 2dF
\citep{2001MNRAS.328.1039C}, GAMA \citep{2011MNRAS.413..971D}, ALFALFA
\citep{2011AJ....142..170H}, and new RESOLVE observations (Kannappan
et al.\ in prep.). RESOLVE-B benefits from extra redshifts taken
during repeat SDSS observations of the Stripe 82 footprint. All of
these additional redshift sources help to recover both large and small
galaxies originally missed by the main SDSS survey for various
reasons, including fiber collisions, which affect galaxies of all
brightnesses, as well as ``shredding'' by the SDSS photometric
pipeline, which primarily affects low surface brightness galaxies
\citep{2005ApJ...631..208B}.

For both RESOLVE and ECO we have custom reprocessed near-UV, optical,
and near-IR photometry as described in E15 and M15. We have computed
stellar masses using the SED fitting routine described in K13. The
directly measured HI census is far more complete for RESOLVE than ECO,
but photometric gas fractions and upper limits have been computed in
the same way for both data sets. We have also performed group finding
using a modified version of the \citet{2006ApJS..167....1B}
Friends-of-Friends algorithm to assign groups and performed halo
abundance matching to assign halo masses. The halo masses in this work
are offset $\sim$0.15 dex toward higher masses than those reported in
M15 (see \S \ref{sec:halomassmeas}). When performing group finding, we
use expanded RESOLVE and ECO data sets that provide a buffer region in
cz space to recover galaxies in groups and clusters with large
peculiar velocities. For RESOLVE-B this buffer region is $\pm$250
\kms{} on either side of the volume. For ECO, which contains several
large groups and clusters (i.e., with large peculiar velocities) near
the survey redshift boundaries, the buffer region extends $\pm$470
\kms{} on either side of the volume. RESOLVE's largest cluster is
fortuitously centered in the survey redshift range. To determine final
membership within RESOLVE-B and ECO, we require that the group rather
than the galaxy redshift belongs within the defined volume (for N=1
groups, the group redshift is the galaxy redshift). In the following
sections we present a more detailed description of both data sets and
their relative strengths and weaknesses for the purposes of this
paper.

\subsection{RESOLVE-B}
\label{sec:stripe82}

\begin{figure*}
\plottwo{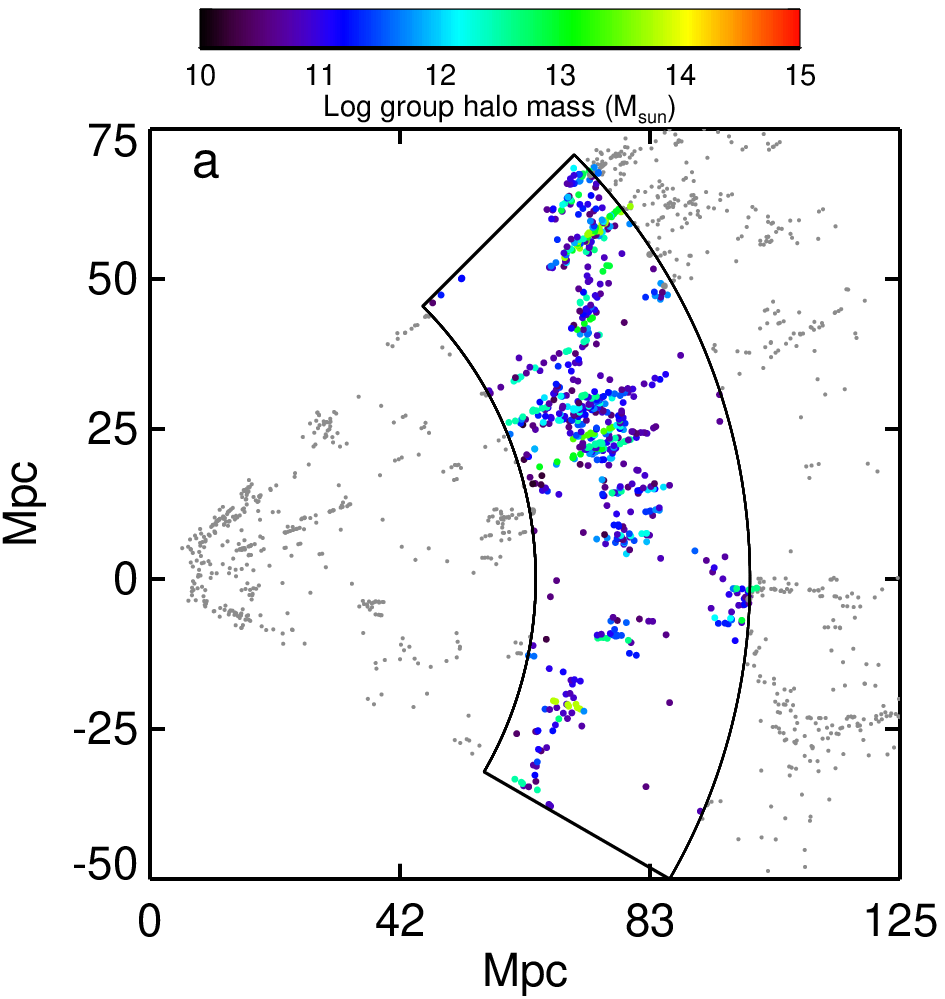}{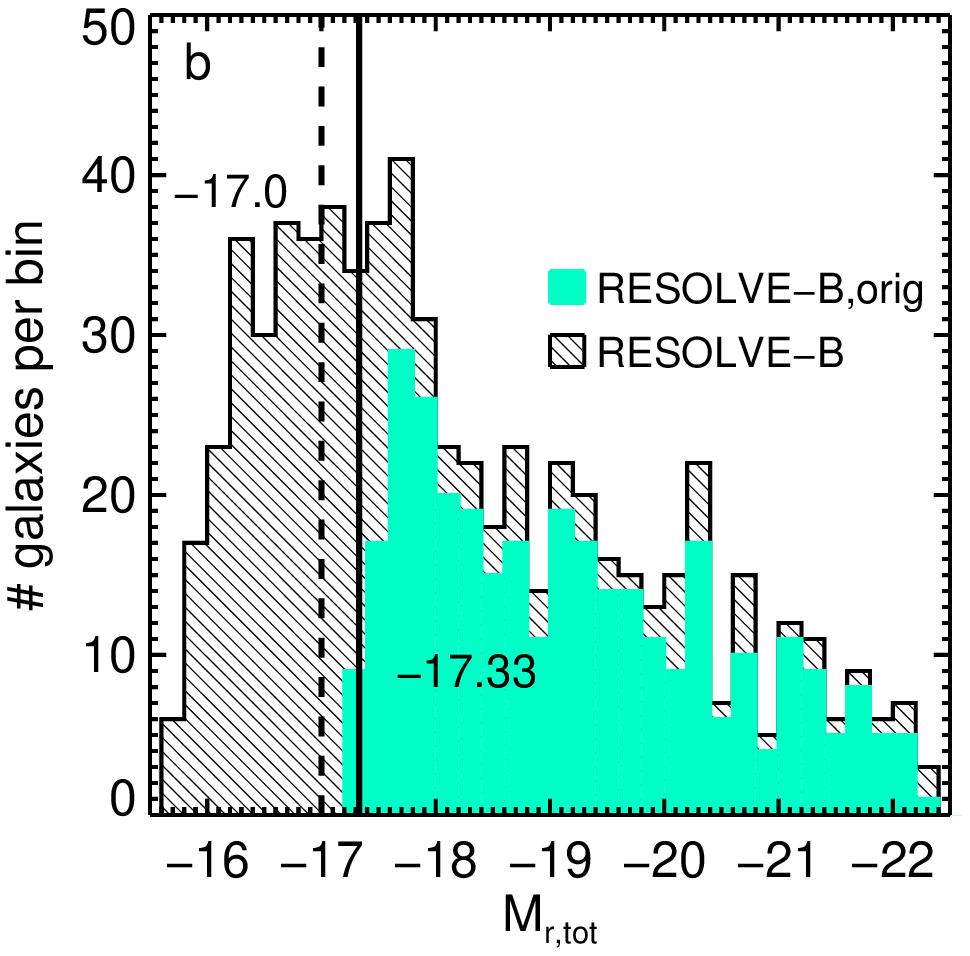}
\epsscale{1.0}
\caption{RA--cz and luminosity distributions for the Stripe 82
  subvolume of RESOLVE described in \S \ref{sec:stripe82} (RESOLVE-B).
  a) The black outline shows the edges of the $\sim$13,700 Mpc$^{3}$
  RESOLVE-B subvolume, which has been collapsed over the narrow
  Declination range from $-1.25$$^{\circ}$ to +1.25$^{\circ}$. Each
  point represents an individual galaxy.  Galaxies within the RESOLVE
  volume are color coded according to group halo mass (purple for
  galaxies in low-mass halos up to green for galaxies in RESOLVE-B's
  highest mass halos). Gray points show galaxies outside the RESOLVE
  volume.  b) Luminosity distributions for RESOLVE-B (black
  cross-hatch) and RESOLVE-B,orig (solid green) as defined in \S
  \ref{sec:stripe82}.  The black solid line shows the original SDSS
  luminosity completeness limit for RESOLVE-B,orig M$_{r,tot}$ =
  $-17.33$ using the reprocessed photometry described in \S
  \ref{sec:photdata}.  Additional redshifts (see \S
  \ref{sec:stripe82}) allow us to extend the luminosity completeness
  limit for RESOLVE-B to M$_{r,tot}$ = $-17.0$, as shown by the black
  dashed line, albeit with imperfect completeness below the original
  SDSS redshift survey limit.}

\label{fg:stripe82skydist}
\end{figure*}

RESOLVE-B is a subset of the RESOLVE survey located within the SDSS
Stripe 82 footprint and encompassing a volume of $\sim$13,700 Mpc$^3$,
with coordinate and redshift ranges of 22h $<$ RA $<$ 3h,
$-1.25^{\circ}<$ Dec $<+1.25^{\circ}$, and 4500 \kms $<$ cz $<$7000
\kms (see Figure \ref{fg:stripe82skydist}a).

RESOLVE-B is a powerful data set for measuring the galaxy SMF and BMF
due to its exceptional completeness above and beyond the main SDSS
redshift survey, allowing us to analyze mass functions well into the
gas-rich dwarf regime without statistical completeness corrections. In
Figure \ref{fg:stripe82skydist}b, the luminosity distribution of
galaxies from the main SDSS redshift survey with \mbox{M$_{r,petro}$
  $<$ $-17.23$} (corresponding to \mbox{m$_{r,petro}$ = 17.77} at the
far side of the RESOLVE volume) ``RESOLVE-B,orig'' is shown in
green. Note that the figure is plotted using our own M$_{r,tot}$
values (see \S \ref{sec:photdata} and E15), which are typically
$\sim$0.1 mag brighter than M$_{r,petro}$ (all magnitudes include
foreground extinction corrections), so the RESOLVE-B,orig luminosity
completeness limit occurs at \mbox{M$_{r,tot} = -17.33$}
mag. RESOLVE-B,orig consists of 329 galaxies brighter than this
limit. In contrast, the full RESOLVE-B data set is shown in black
crosshatch and consists of 426 galaxies brighter than $-17.33$ or 28\%
more galaxies than in the SDSS main redshift survey (see \S
\ref{sec:lumsbcompleteness} for more discussion on the reasons for
incompleteness). Due to the large number of extra redshifts, we adopt
a RESOLVE-B sample definition of M$_{r,tot}$ = $-17$ for a total of
487 galaxies brighter than this fainter luminosity completeness limit
and meeting the group cz criterion. Allowing all galaxies with known
redshift inside the volume and not cutting by absolute $r$-band
magnitude, there are 679 galaxies inside the RESOLVE-B subvolume.

In RESOLVE-B, we take advantage of the wealth of imaging data
available in the Stripe 82 legacy footprint, including deep SDSS
$ugriz$ coadds, MIS depth \textit{GALEX} NUV and/or \textit{Swift}
UVOT imaging for 98\% of RESOLVE-B galaxies, 2MASS $JHK$, and deeper
UKIDSS $YHK$. We have obtained 21cm coverage from both the ALFALFA
survey (covering the 0$^{\circ}$ to 1.25$^{\circ}$ northern strip) and
pointed observations with the Arecibo and Green Bank telescopes,
obtaining reliable detections or strong upper limits (1.4M$_{HI}$ $<$
0.05M$_{star}$) for 78\% of the RESOLVE-B data set with M$_{r,tot}$
$\leq$ $-17$, with the remaining 22\% of RESOLVE-B galaxies having
weak upper limits, confused detections, or no observations (Stark et
al.\ submitted; see also \S \ref{sec:hidata}). The HI data statistics
for all galaxies in the volume (no restriction on M$_{r,tot}$) yield
reliable detections or strong upper limits 60\% of the RESOLVE-B data
set, with the remaining 40\% of RESOLVE-B galaxies having weak upper
limits, confused detections, or no observations. In these cases we
fill in the HI census using the photometric gas fractions technique
(\S \ref{sec:hidata}).

\subsection{ECO}
\label{sec:eco}

\begin{figure*}
\epsscale{1.}
\plotone{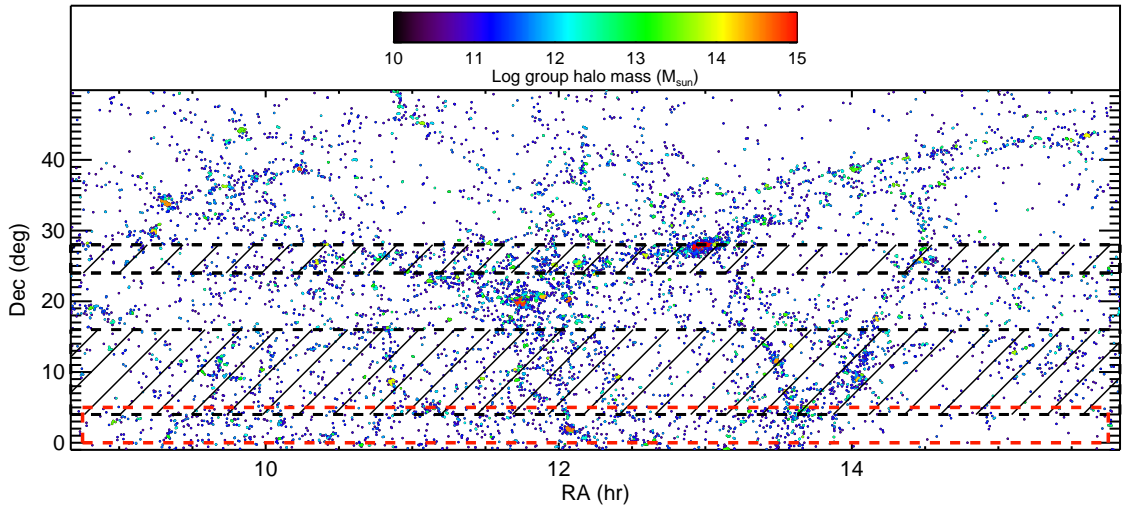}
\caption{RA--Dec distribution of the ECO catalog (M15), which includes
  galaxies within a $\sim$442,700 Mpc$^{3}$ volume from cz=3000--7000
  \kms\ as described in \S \ref{sec:eco}. Each point represents one
  galaxy and the galaxies are color coded by group halo mass according to
  the scale bar at the top (dark blue for low-mass groups and red for
  the highest mass groups). Overlap with the ALFALFA 21cm survey is
  shown as the black hash-marked region.  ALFALFA provides HI
  detections and upper limits for $\sim$1/3 of ECO. The RESOLVE-A
  subvolume footprint, where we have the most complete HI data for
  ECO, is outlined with the red dashed line.}
\label{fg:ecoskydist}
\end{figure*}

The full ECO (Environmental COntext) catalog, accounting for the
buffer region, is a volume-limited data set selected within coordinate
and redshift ranges from 8.7h $<$ RA $<$ 15.82h, $-1.^{\circ}<$ Dec
$<+49.85^{\circ}$, and 2530 \kms\ $<$ cz $<$ 7470 \kms\ (see Figure
\ref{fg:ecoskydist}), containing all known redshifts from the SDSS
survey, as well as from the other sources listed in \S
\ref{sec:common} (M15).  The full ECO catalog includes the RESOLVE-A
subvolume, which is marked by the red dashed line in Figure
\ref{fg:ecoskydist}, and encompasses a volume of $\sim$560,800
Mpc$^{3}$ ($\sim$41 times larger than RESOLVE-B) providing a much
larger data set for statistical analysis of the galaxy SMF and BMF,
but lacking RESOLVE-B's superior completeness. Attempts to fill in
completeness through inclusion of archival and RESOLVE redshift data
yield a survey $\sim$11\% more complete than the SDSS main redshift
survey alone. We take the stated luminosity limit of the SDSS main
redshift survey as the nominal luminosity completeness limit for
ECO. However, additional redshift incompleteness above this inherited
luminosity limit remains, which would compromise our mass function
analysis (see also \citealp{2005ApJ...631..208B}) so we will apply
empirical completeness corrections above the nominal luminosity
completeness limit in \S \ref{sec:compcorr}.


The volume of the ECO data set used in this work excludes the buffer
region and is $\sim$442,700 Mpc$^{3}$, containing 9456 galaxies with
M$_{r,tot}$ $<$ $-17.33$ mag. We limit ECO to galaxies belonging to
groups within a cz range of 3000-7000 \kms, allowing the remaining cz
range to serve as a buffer for identifying group members with peculiar
velocities outside this range. (A larger buffer from cz=1500-12,000
\kms\ is used to recover galaxies belonging to the largest groups and
clusters with very extended fingers-of-God; however, galaxies outside
2530-7470 \kms\ lack reprocessed photometry and are treated
statistically as described in more detail in \S \ref{sec:halomass}).

The ECO data set is uniformly covered by single-depth SDSS $ugriz$ and
2MASS $JHK$ imaging and also has 45\% MIS depth \textit{GALEX} NUV
coverage. Where the ECO catalog overlaps the A-semester of the RESOLVE
survey, we use the superior photometric data for RESOLVE (including
full MIS-depth \textit{GALEX} coverage and UKIDSS). We have obtained
21cm data from the public ALFALFA40 catalog
\citep{2011AJ....142..170H} that covers $\sim$1/3 of the ECO area and
provides detections (including confused detections) and enables
calculation of upper limits for 3572 galaxies. Since ALFALFA is a
flux-limited survey $\sim$90\% of the upper limits are weak, i.e.,
yield gas fractions greater than 5\% of the stellar mass.  In the area
of overlap with RESOLVE-A, we replace photometry, stellar masses, and
HI data with values from RESOLVE-A, including strong upper limits for
the HI. Thus the final HI statistics yield non-confused detections or
strong upper limits for 26\% of ECO, with the remaining 74\% of ECO
having weak upper limits (14\%), confused detections (6\%), or no
observations (54\%). For these ECO galaxies without HI observations,
with weak upper limits, or confused detections, we rely on the
photometric gas fractions technique presented in E15 to provide gas
mass estimates (\S \ref{sec:hidata}).

\subsection{Densities of RESOLVE-B and ECO}
\label{sec:cvofds}


The data sets used in this work have relatively small volumes, which
we have computed using a numerical method and defining the inner and
outer radii of the volume using the comoving distance. Because their
volumes are small, RESOLVE-B and ECO may be affected by cosmic
variance. The primary effect of cosmic variance is on the overall
galaxy density in the volume. To gauge this effect for RESOLVE-B and
ECO, we compute the number density of galaxies in each volume
belonging to the main SDSS redshift survey (to eliminate any
difference due to overall completeness) within both M$_{r,tot}$ $<$
$-20$ and M$_{r,tot}$ $<$ $-21$ and we compare with the corresponding
densities measured by \citet{2006MNRAS.373..469B} for the entire SDSS
DR4 volume.\footnote{The volume calculation in
  \citet{2006MNRAS.373..469B} has been done with the same cosmological
  parameters as used in this work ($\Omega_m$ = 0.3, $\Omega_\Lambda$
  = 0.7, and $H_o$=70 \kms Mpc$^{-1}$). Using our numerical volume
  calculation method, which determines the volume of the spherical
  shell defined by the inner and outer radii and then multiplying the
  shell volume by the ratio of the solid angle subtended by the survey
  to the solid angle of a sphere, we reproduce their survey volume of
  2.3 $\times$ 10$^{7}$ Mpc$^3$.}  We find that log($\rho_{<-21}$) and
log($\rho_{<-20}$) for RESOLVE-B are $-2.46$ and $-2.15$
log(Mpc$^{-3}$), and for ECO they are $-2.76$ log(Mpc$^{-3}$ and
$-2.33$ log(Mpc$^{-3}$). For SDSS DR4, \citet{2006MNRAS.373..469B}
measure log($\rho_{-21}$) and log($\rho_{-20}$) to be $-2.8$ and
$-2.35$ log(Mpc$^{-3}$) respectively at z $\sim$ 0.02, which is
roughly in the middle of the redshift coverage of the RESOLVE and ECO
volumes. Thus while ECO is similar in density to SDSS DR4, RESOLVE-B
is overdense compared to the larger SDSS region, which results in an
overall higher normalization of its mass functions.


To gauge whether the relative difference in overall density is within
the expected amount of cosmic variance for the RESOLVE-B and ECO
volumes, we have considered the cosmic variance recipes of
\citet{2008ApJ...676..767T}, \citet{2010MNRAS.407.2131D}, and
\citet{2011ApJ...731..113M}, which are designed for deep high-redshift
pencil beam surveys much different from the RESOLVE-B and ECO
volumes. These works yield a wide range in cosmic variance estimates,
respectively 33\%, 49\%, and 45\% for RESOLVE-B and 16\%, 25\%, and
19\% for ECO. To better assess the cosmic variance of RESOLVE-B and
ECO, we use mock galaxy catalogs customized for the RESOLVE-B and ECO
volumes (V. Calderon, private communication). We estimate cosmic
variance by measuring the total number of galaxies meeting our survey
limits in each mock and dividing the standard deviation of those
values by the mean value. We obtain cosmic variance estimates of 58\%
and 12.5\% for RESOLVE-B and ECO respectively. These values (along
with the results of \citealp{2010MNRAS.407.2131D} and
\citealp{2011ApJ...731..113M}) yield consistent overall number
densities in ECO and RESOLVE-B after cosmic variance is taken into
account. We defer a more detailed discussion of the cosmic variance
within RESOLVE-B and ECO to ongoing work using these custom mock
catalogs (Cisewski et al.\ in prep.).

A secondary effect of cosmic variance is related to the fact that more
dense regions tend to have larger halos and structures.  We
investigate the group halo mass distributions of our two data sets in
\S \ref{sec:halomassmeas} and find possible evidence for this effect
in RESOLVE-B.



\section{Data And Methods}
\label{sec:data}

RESOLVE-B and ECO have largely homogeneous data products with key
differences in quality of HI and degree of completeness.  In \S
\ref{sec:photdata} we summarize the newly reprocessed photometry from
SDSS, 2MASS, UKIDSS, and \textit{GALEX}, presented in E15 and M15. We
then describe the stellar population modeling used to estimate stellar
masses from the improved photometry in \S \ref{sec:mstar}. In \S
\ref{sec:hidata} we review the HI data used in this work as well as
the photometric gas fractions technique from E15, used to predict HI
masses for galaxies without HI data or with inadequate HI data. We
describe the computation of baryonic masses in \S \ref{sec:barymass}.
In \S \ref{sec:halomass} we discuss the group finding and halo
abundance matching method used to determine halo masses so that we may
investigate the mass functions in different group halo mass regimes.
Lastly, in \S \ref{sec:completeness} we compare the completeness of
RESOLVE-B with estimates of SDSS completeness from the literature,
describe our calculation of empirical completeness corrections for
ECO, and determine the stellar and baryonic mass completeness limits
for RESOLVE-B and ECO.

\subsection{Photometric Data}
\label{sec:photdata}

For both the RESOLVE and ECO surveys, we use reprocessed photometry
from the UV to the near-IR to obtain consistent and well-determined
total magnitudes as described in E15 and M15. We use SDSS optical $ugriz$
data \citep{2011ApJS..193...29A}, NIR $JHK$ and $YHK$ from 2MASS
\citep{2006AJ....131.1163S} and UKIDSS \citep{2008MNRAS.384..637H}
respectively, and NUV data from the \textit{GALEX} mission
\citep{2007ApJS..173..682M}, as well as uvm2 data from the
\textit{Swift} UVOT telescope for 19 RESOLVE-B galaxies lacking MIS
depth NUV data.

We make several key improvements in the photometric reprocessing to
obtain consistent photometry with well characterized errors.  First we
use the improved sky subtraction for SDSS data from
\citet{2011AJ....142...31B} and our own additional sky subtraction for
2MASS and UKIDSS data. Second, we apply the same ellipse fits to each
band using the sum of the high S/N $gri$ images to define the
ellipses.  This approach allows us to determine the PA and axial ratio
of the galaxy's outer disk and measure magnitudes even in bands in
which catalog photometry has no detection. Third, we compute
\textit{total} magnitudes extrapolated to infinity (not aperture
magnitudes) via three non-parametric methods for each band
independently, allowing us to obtain systematic uncertainties on the
magnitude measurements, key for robust stellar mass estimation through
SED fitting (described in \S \ref{sec:mstar}).  These total magnitude
measurements also allow for color gradients, which are explicitly not
allowed in the SDSS model magnitude system (see \S 3.1 in E15 for more
information).

A key difference between the two surveys is the quality of the data
available. For RESOLVE-B deep optical data from SDSS are available for
all galaxies, since the Stripe 82 footprint was repeatedly imaged with
$\sim$20 coadds over each area of the sky. Additionally, 98\% of
RESOLVE-B galaxies are covered by either MIS depth ($\sim$1500s)
\textit{GALEX} NUV or \textit{Swift} uvm2. Finally, 97\% of RESOLVE-B
galaxies have deep near-IR photometry in at least one UKIDSS band. For
ECO, we are limited to shallow data from SDSS and 2MASS over the
entire data set and MIS depth \textit{GALEX} for only $\sim$45\% of the
data set (M15).

We provide a more in-depth comparison of the reprocessed photometry
with the catalog SDSS DR7 photometry in E15.  The new photometry
yields brighter magnitudes and larger $r_e$ values than the SDSS DR7
catalog model and Petrosian calculations. These differences increase
for galaxies with larger radii and are in line with the expectations
from \citet{2011AJ....142...31B} due to improved sky subtraction.  We
also find generally bluer colors and a less tight red sequence, which
we attribute to allowing color gradients (as mentioned, disallowed in
the standard SDSS catalog pipeline).

We have demonstrated consistency between RESOLVE-B and ECO photometry
in E15, finding rms $\sim$ 0.04 mag between the two in $r$-band
magnitude. We attribute small differences to different masking
procedures (for RESOLVE we check all masks by hand, while for ECO mask
checking is only done for auto-flagged galaxies) and to the algorithm
for merging final magnitudes from our three methods.


\subsection{Stellar Masses}
\label{sec:mstar}

To calculate stellar masses and colors, we use the spectral energy
distribution (SED) modeling code described in
\citet{2007ApJ...657L...5K}, \citet{2009AJ....138..579K}, and K13,
which fits our newly reprocessed total NUV$ugrizYJHK$ magnitudes to a
grid of stellar population models. With up to 9 or 10 bands of
photometric data in ECO and RESOLVE-B respectively, we can estimate
robust stellar masses. In RESOLVE-B we exclude the UKIDSS data if they
are flagged due to sky background or image artifact issues (E15). We
do not use UKIDSS $J$ and we exclude $H$ and $K$ band data if they are
fainter than 18 and 17.5 mag respectively. We exclude 2MASS $JHK$ data
if they are fainter than 16, 15, and 14.5 mag respectively. We exclude
any NUV data fainter than 24 mag. Galaxy magnitudes fainter than these
cuts are unreliable as determined by examination of the SED fits. The
IR magnitude cuts are also similar to the 10$\sigma$ point source
detection limits of 2MASS and the UKIDSS Large Area Survey.

In this work we consider two model sets to determine how robust the
mass function shape is to changes in IMF and star formation
history. Both model grids consist of an old and young population,
yielding a composite stellar population (CSP). The first model grid
(model grid \textit{a}) from K13, models the old population with a
simple stellar population (SSP) that can range in age from 2-12
Gyr. Model grid \textit{a}'s young population may be described by
continuous star formation that started 1015 Myr ago and turned off
between 0 and 195 Myr ago or as a single quenching burst (SSP) with
age 360, 509, 641, 806, or 1015 Myr. The young population can
contribute 0-94.1\% of the stellar mass. The second model grid (model
grid \textit{b}) also uses for its old stellar population an SSP that
can range in age from 1.4-13.5 Gyr. Model grid \textit{b}'s young
population is an SSP that can range in age from 5-1000 Myr and
contribute 0-64\% of the stellar mass. The model grids are built using
the \citet{2003MNRAS.344.1000B} stellar population models. Model grid
\textit{a} uses a Chabrier IMF \citep{2003PASP..115..763C} with four
possible metalliticies (Z = 0.004, 0.008, 0.02, or 0.05), while model
grid \textit{b} uses a diet Salpeter IMF \citep{2001ApJ...550..212B}
with three possible metallicities (Z = 0.008, 0.02, or 0.05). Both
model sets allow for eleven reddening values ranging from 0-1.2 that
are applied to the young stellar population using the dust law from
\citet{2001PASP..113.1449C}. While the two model sets have a few
smaller differences, the largest physical difference between the two
model sets is the inclusion of a continuous star formation mode for
the young stellar population in model grid
\textit{a}.

For each CSP model in the grid, a stellar mass and likelihood based on
the model fit to the data are computed. We combine the likelihoods
over all models, yielding the galaxy's stellar mass likelihood
distribution for each galaxy. The nominal value of stellar mass of a
galaxy is the median of the stellar mass likelihood
distribution. However, for the SMF and BMF calculated later in this
work, we use the entire stellar mass likelihood distributions to take
into account the large uncertainties on galaxy stellar mass estimates.

From the SED modeling code, we also obtain likelihood weighted model
colors for each galaxy, which are effectively smoothed and k-corrected
by the model fits.  To denote these modeled colors we use a
superscript m (following the notation from K13, E15, and M15). We have
shown in E15 that the RESOLVE-B and ECO color-stellar mass plots are
very similar, even though ECO lacks deep optical and IR data and
complete NUV coverage.

Comparing the two model sets used in this work, we find an overall
offset of $\sim$0.08 dex (such that model set \textit{a} masses are
smaller) with rms scatter of $\sigma$ $\sim$ 0.1 dex, consistent with
K13 and well within the typical uncertainties on stellar mass of 0.15
dex.  Model set \textit{a} is designed to mimic the essential features
of more complex multi-burst star formation histories (e.g.,
\citealp{2007ApJS..173..267S}) with reduced computational demand. We
note that the less physically motivated model set \textit{b} from
\citet{2009AJ....138..579K} yields extremely similar results. As a
gauge of consistency with literature mass estimates, we note that the
slightly altered \textit{b} model set described in
\citet{2009AJ....138..579K} produces stellar masses consistent with
those from \citet{2003MNRAS.341...33K}, which have also been
demonstrated to be consistent with the stellar masses from
\citet{2005ApJ...619L..39S}. All of these comparisons yield rms
scatter $\sim$0.1 dex.  In contrast, comparison of masses from an
earlier version of model set \textit{b} described in
\citet{2007ApJ...657L...5K} with masses estimated with the $g-r$
vs.\ M$_{*}$/L$_{K}$ relation of \citet{2003ApJS..149..289B} found
that the \citealt{2003ApJS..149..289B} masses are offset toward higher
mass by $\sim$0.2 dex for high mass galaxies and $\sim$0.3 dex for
low-mass galaxies. \citet{2015MNRAS.452.3209R} also find that the
\citet{2003ApJS..149..289B} mass scale and mass-to-light ratios are
different from other stellar mass systems.

\subsection{HI Masses}
\label{sec:hidata}

To measure baryonic masses we must include the cold neutral gas, the
dominant component of which is typically found in atomic hydrogen
(although for large spiral galaxies molecular gas may dominate, see
Figure 8a of K13). The HI masses and limits for RESOLVE and ECO come
from both the blind 21cm ALFALFA survey \citep{2011AJ....142..170H}
and from new pointed observations with the GBT and Arecibo telescopes
(Stark et al.\ submitted).  All galaxies have optical diameters much
smaller than the smallest beam size of these telescopes (3.5\arcmin{}
for ALFALFA). Additionally, due to missing data, we estimate masses
for a significant portion of the ECO catalog (and a small percentage
of RESOLVE-B) using the ``photometric gas fractions'' technique
described in E15. To account for the contribution from helium to the
cold neutral gas mass, we multiply the HI mass by 1.4, i.e.,
M$_{gas}$=1.4M$_{HI}$.



\subsubsection{RESOLVE-B Gas Inventory}
\label{sec:resgmass}

RESOLVE-B has 21cm data from the ALFALFA survey covering the northern
0$^{\circ}$ to 1.25$^{\circ}$ strip. Data reduction and source
extraction have been performed according to
\citet{2011AJ....142..170H}. Since the ALFALFA survey is flux-limited
with a fixed HI mass sensitivity of $\sim$10$^{9}$ \msun{} at RESOLVE
redshifts, many of these northern targets have upper limits weaker
than our desired goal (M$_{gas}$ $<$ 0.05M$_{star}$). To fill in the
southern Stripe 82 strip and obtain deeper data for these weak upper
limits, we have obtained pointed observations with the GBT and Arecibo
telescopes for 385 galaxies in RESOLVE-B, aiming for detections with
S/N $>$ 10 or strong upper limits (Stark et al.\ submitted, see also
E15). RESOLVE's HI survey only targets galaxies brighter than
M$_{r,tot}$ = $-17.0$ or which have predicted M$_{bary}$ $>$
10$^{9.0}$ \msun, so many lower mass galaxies (below our completeness
limits) in RESOLVE-B have no HI data.

To measure HI masses and upper limits, we use the algorithms described
in K13 and Stark et al.\ (submitted).  Confused sources are determined
based on the telescope used for their measurement; we use a search
radius of 4' for the ALFALFA smoothed resolution element, 9' for
the GBT, and 3.5' for Arecibo pointed observations.  We also perform
deconfusion of the HI profiles when possible as described in Stark et
al.\ (submitted), which builds off of techniques described in K13.

For this work (as in E15) we consider an HI detection okay if the HI
detection has S/N $>$ 5. If the HI detection is confused, we use the
deconfused HI data if the systematic uncertainty is $<$25\% of the
deconfused HI mass, as this value is not significantly worse than the
error on our weakest S/N $\sim$ 5 detections. HI upper limits are
considered strong if M$_{gas}$ $<$ 0.05M$_{star}$. For galaxies not
meeting these requirements, we use gas mass estimates as described in
\S \ref{sec:gmassest}.


In RESOLVE-B, limiting the data set at \mbox{M$_{r,tot}$ $<$ $-17$
  mag}, there are currently 274 galaxies with unconfused detections
and an additional 32 detections that have been successfully
deconfused. Along with the 74 strong upper limits that yield
\mbox{M$_{gas}$ $<$ 0.05M$_{star}$}, 78\% of RESOLVE-B galaxies have
HI data meeting our requirements. Of the remaining 22\%, 34 galaxies
cannot be deconfused, 3 have not been observed in HI, 4 have
unreliable detections with S/N $<$ 5, and 66 have weak upper limits.
For the entire RESOLVE-B data set, not restricted to M$_{r,tot}$ $<$
$-17$, there are 334 galaxies with unconfused or deconfused detections
and 74 galaxies with strong upper limits, yielding 60\% of the full
RESOLVE-B data set with HI data meeting our requirements.  For the
remaining 40\% of galaxies, 44 are impossible to deconfuse, 95 have
not been observed in HI, 8 have low S/N detections, and 124 have weak
upper limits.

\subsubsection{ECO Gas Inventory}
\label{sec:ecogmass}

Within the ECO catalog, there is a region that overlaps with the
ALFALFA40 public catalog (see Figure \ref{fg:ecoskydist}), allowing us
to obtain HI detections for galaxies with HI masses $>$10$^{9}$
\msun. We calculate HI upper limits as in Stark et al.\ (submitted) and
K13 using the typical declination dependent rms from ALFALFA. We
obtain upper limits for $\sim$40\% of the ECO-ALFALFA crossmatched
data set, however only $\sim$11\% of these limits are strong limits
(M$_{gas}$ $<$ 0.05M$_{star}$).  Confused sources are flagged if there
are neighboring galaxies within the smoothed resolution element of 4'.
For ECO, we cannot attempt to extract better fluxes in these cases,
which are treated as if they do not have HI data.  Confused sources
account for 15\% of the ECO-ALFALFA crossmatched catalog.

Within the region of ECO overlapping with RESOLVE-A, we substitute
RESOLVE HI data from Stark et al.\ (submitted).  The RESOLVE-A
footprint has been completely covered by ALFALFA and we have followed
up with GBT and Arecibo pointed observations as in RESOLVE-B,
obtaining reliable HI detections or strong upper limits for 76\% of
galaxies with \mbox{M$_{r,tot}$ $<$ $-17.33$} (Stark et al.\ submitted,
E15).

Including the RESOLVE-A data with the ECO-ALFALFA regions produces an
inhomogeneous HI data set.  For this work, however, we wish to include
as much real HI data as possible to measure the BMF. Out of the full
ECO data set, 5126 galaxies ($\sim$54\%) have not been observed in
HI. Of the 4330 galaxies with HI observations, 2148 have reliable
detections (including 34 successfully deconfused HI profiles), 247
galaxies are strong upper limits, 1331 galaxies are weak upper limits,
16 have low S/N detections, and 588 galaxies cannot be
deconfused. Thus $\sim$75\% of ECO requires photometric gas fractions.

\subsubsection{Estimating Gas Masses}
\label{sec:gmassest}

We require complete HI data for RESOLVE-B and ECO to
measure the BMF. Thus we turn to the photometric gas fractions (PGF) technique
to empirically estimate gas masses for galaxies in RESOLVE-B and
ECO that have not been observed, or for which we have not obtained a
reliable detection or strong upper limit. We use the probability
density field approach to the photometric gas fractions technique
presented in E15, which uses RESOLVE-A as a calibration data set.

The probability density field method given in E15 fits a 2D model to
the density field of log gas-to-stellar mass ratio, log(G/S), vs.\ a
linear combination of color and axial ratio ($b/a$),
which we call ``modified color'' or $mc$.  From this 2D model it is
possible to construct a probability distribution of log(G/S) for each
galaxy given its color and axial ratio.\footnote{See E15 for more
  detail on the PGF distributions. IDL and Python codes to generate
  log(G/S) distributions for a given color or modified color are
  provided at https://github.com/keckert7/codes/.}

The calibration from E15 is ideal for this work for three reasons.
First the PGF calibration from E15 provides a probability distribution
in log(G/S) for each galaxy making it easy to integrate into the
statistical analysis of the SMF and BMF discussed in \S
\ref{sec:statmassfunctions}. Second, because the 2D model is defined
to include the population of HI upper limits, we can use this
calibration even for red quenched galaxies for which color-limited
versions of the PGF calibration break down (see \S 4 of E15). Lastly
the calibration data set, RESOLVE-A, is also a volume and absolute
$r$-band magnitude limited survey with a similar selection as
RESOLVE-B and ECO.

This last point about using a similarly selected data set for the
calibration is key because the prediction values will change if we use
a differently selected calibration sample. Since RESOLVE-B and ECO are
volume limited, they are dominated by gas-rich low-mass galaxies.
Other calibrations in the literature are not based on volume-limited
data sets and thus do not predict gas masses well for the low-mass
galaxies in RESOLVE-B and ECO.  In E15, we show that other PGF
calibrations underestimate (\citealp{2013MNRAS.436...34C}; K13) or
overestimate \citep{2012ApJ...756..113H} the actual log(G/S) of
galaxies at a given color.

A limitation of the 2D model from E15 is that it is calibrated to work
on data with all values of log(G/S) below $-1.3$ set equal to $-1.3$
($\sim$log(0.05)). Thus we can only predict a galaxy's gas content
down to 5\% of its stellar mass, even if we know the upper limit to be
lower. The minimum baryonic mass for galaxies with PGF gas mass
estimates is therefore 1.05$\times$M$_{star}$, which in log space adds
only 0.02 dex to the stellar mass. This shift is much smaller than
the typical bin size used in this work (0.2 dex), meaning that we do
not have to worry about any significant effect on the mass functions.


\subsection{Baryonic Masses}
\label{sec:barymass}

To compute baryonic masses, we perform a ``pseudo-convolution'' of the
full stellar mass distributions that are provided by the SED fitting
code (see \S \ref{sec:mstar}) and the HI likelihood distributions
built from either the HI data or the PGF calibration chosen for each
galaxy, yielding a full baryonic mass likelihood distribution for each
galaxy. We then take the median of the baryonic mass likelihood
distribution to be the nominal baryonic mass of the galaxy but use the
full distribution in constructing the mass functions in this work
(just as we do for stellar mass). Below we describe the decision tree
determining which PGF calibration is chosen for each galaxy that
requires gas mass estimation.  Then we detail the pseudo-convolution
algorithm.

\subsubsection{Choice of PGF Estimator}
\label{sec:calchosen}

In choosing the PGF estimator, we prefer to use those calibrations
including colors with the longest baselines, since typical magnitude
errors $\sim$0.05 mag can change the log(G/S) value significantly in
shorter baseline calibrations, e.g., \cgr$^m$ has a prediction
baseline of $\sim$0.5 mag while \cuj$^m$ has a prediction baseline of
1.6 mag.  However, the near-IR colors that provide the longest
baselines may be suspect or non existent if the galaxy is too faint to
be detected and/or the UKIDSS photometry is flagged.  To decide
whether the \cuj$^m$ color is reliable enough for use in the PGF
relation, we compare the SED modeled \cuy$^m$ and \cuj$^m$ colors with
the \cuy$_{90}$ and \cuj$_{90}$ aperture colors measured within the
90\% light radius in the $r$-band.  If either the \cuy{}$^m$ or
\cuj{}$^m$ SED modeled and aperture colors agree within 2$\sigma$ of
the general relation, we proceed with using the modified \cuj$^m$
color (including $b/a$) PGF calibration.  If not, we move on to using
the PGF calibration based on modified \cuk$^m$ color, performing the
same analysis (using both 2MASS and UKIDSS $K$) of comparing \cuk$^m$
and \cuk$_{90}$ to determine whether the photometry is acceptable.  In
the case that both \cuj$^m$ and \cuk$^m$ are unsuitable, we revert to
the PGF calibration based on modified \cur$^m$ color, performing the
same comparison between \cur$^m$ and \cur$_{90}$.  Lastly, for the
handful of galaxies with both unreliable $u$ and unreliable
2MASS/UKIDSS we use the modified \cgr$^m$ color PGF relation, which
provides the shortest baseline. Of RESOLVE-B (ECO) galaxies requiring
gas mass estimates, 95\% (91\%) use the modified \cuj$^m$ calibration,
3\% (4\%) use the \cuk$^m$ calibration, 1\% (2\%) use the modified
\cur$^m$ calibration, and 1\% (3\%) use the \cgr$^m$ calibration.

For galaxies that have been identified as confused (without
possibility of deconfusion), have low S/N detections, or have not been
observed in HI, we use the PGF estimator log(G/S) distribution to
perform the baryonic mass likelihood distribution calculation
described in \S \ref{sec:computembary}. For galaxies with weak upper
limits, we also use the selected log(G/S) distribution, but we cut off
the distribution at the weak upper limit value and renormalize the
distribution (see Figure \ref{fg:mbarycalc}b inset).

\subsubsection{Calculation of Baryonic Masses}
\label{sec:computembary}

\begin{figure*}
\plotone{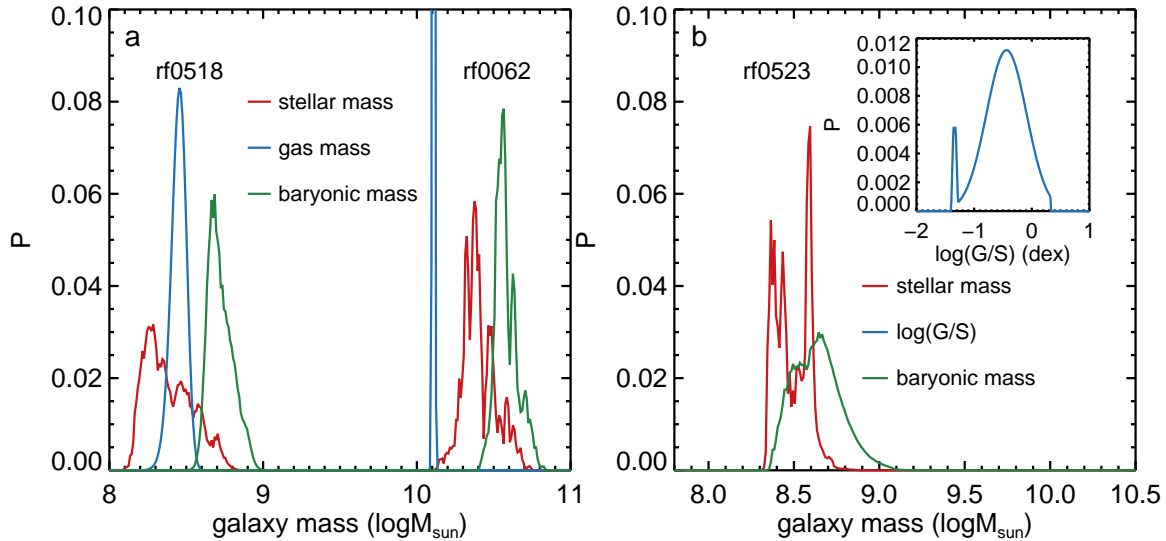}

\epsscale{1.0}
\caption{Likelihood distributions for stellar (red), gas (blue), and
  baryonic (green) mass for three RESOLVE-B galaxies. a) Likelihood
  distributions for two galaxies with HI detections. The gas mass
  distribution of rf0518 is located at similar values as the stellar
  mass distribution and dominates the shape of the baryonic mass
  distribution. In opposition, the gas mass distribution is at lower
  values than the stellar mass distribution for rf0062, and so the
  stellar mass distribution dominates the shape of the final baryonic
  mass distribution. b) Likelihood distribution for a galaxy with a
  weak upper limit. The inset plot shows the log(G/S) likelihood
  distribution from the PGF calibration, which cuts off the
  distribution at the upper limit value (log(G/S) = $0.3$) and for
  values $<-1.3$. The green shows the resulting baryonic mass
  likelihood distribution.}

\label{fg:mbarycalc}
\end{figure*}

To compute baryonic masses, we start by constructing the stellar mass
likelihood distribution for each galaxy. The SED fitting code outputs
a mass and likelihood for each model in the grid (model sets
\textit{a} and \textit{b} contain 26,932 and 9,855 models
respectively). The final stellar mass likelihood distribution for each
galaxy is binned in 0.01 dex intervals, much smaller than the typical
1$\sigma$ width of the distribution of $\sim$0.25 dex. Examples of
stellar mass likelihood distributions are shown in red in Figure
\ref{fg:mbarycalc}.

For galaxies with clean or successfully deconfused HI detections, we
model the gas mass likelihood distribution as a Gaussian with $\mu$
equal to the gas mass and $\sigma$ equal to the systematic uncertainty
on the gas mass measurement.  Two examples of the gas mass likelihood
distributions for HI detections are shown in blue in Figure
\ref{fg:mbarycalc}a. We resample the stellar and gas mass likelihood
distributions in linear spacing with $\Delta$M$_{star}$ =
$\Delta$M$_{gas}$, set equal to a fraction between $0.01$--$0.5$ of
the minimum stellar or gas mass with likelihood $>$1e-4 (keeping the
spacing as small as possible without causing the calculation to be too
time consuming).  We then determine the combined likelihood for each
stellar and gas mass combination in linear mass units from their
linearly spaced likelihood distributions. Summing up the likelihoods
for all resulting baryonic masses, we obtain a baryonic mass
likelihood distribution.

For galaxies with weak upper limits, low S/N detections, no HI
observations or severely confused detections, we use the log(G/S)
distributions from the PGF calibration described in \S
\ref{sec:calchosen}. An example of a log(G/S) distribution for a weak
upper limit is shown in the inset of Figure \ref{fg:mbarycalc}b. To
find the baryonic mass likelihood distribution for galaxies, we
resample the stellar mass likelihood distribution into linear spacing
with $\Delta$M$_{star}$ equal to a fraction between $0.01$--$0.5$ of
the minimum stellar mass value having likelihood $>$1e-4. For each
possible stellar mass, we compute a new spacing for the log(G/S)
distribution such that $\Delta$M$_{gas}$ = $\Delta$M$_{star}$. The
likelihood and baryonic mass are then computed at each possible
stellar mass and G/S value. We then sum the likelihoods at each
baryonic mass to produce the baryonic mass distribution.

Finally for galaxies with strong upper limits, we set the likelihood
in log(G/S) to be 1 at the upper limit value of log(G/S). Since for
strong upper limits this value can be at most 0.05, the baryonic mass
will be at most 0.02 dex larger than the stellar mass.

The final baryonic mass likelihood distribution for each galaxy comes
from the array of likelihoods and baryonic masses output by these
procedures and is binned up finely into 0.01 dex bins. Three examples
are shown in green in Figure \ref{fg:mbarycalc} for the different
cases. To compute single value baryonic masses, we determine the
median value of the resulting baryonic mass likelihood distribution
for each galaxy, just as for stellar mass.

\subsection{Group Identification and Halo Masses}
\label{sec:halomass}

To investigate the SMF and BMF in different group halo mass regimes,
we associate the galaxies in the RESOLVE-B and ECO data sets to groups
using the Friends-of-Friends (FOF) group finding algorithm from
\citet{2006ApJS..167....1B} following the algorithm described in M15.
We are able to use this algorithm because our data sets are limited on
absolute magnitude and volume-limited. In the following sections we
describe the choice of linking lengths and the group finding and halo
mass assignment. We also compare the halo mass distributions for
RESOLVE-B and ECO and examine the relationship between galaxy stellar
and baryonic mass and halo mass.

\subsubsection{Choice of Linking Lengths}
\label{sec:llchoice}

The FOF algorithm links together galaxies that lie within a cylinder
defined by a tangential linking length (in projected physical distance)
and a line-of-sight linking length (in cz space), which are determined
by the mean spacing between objects in the volume.  In
\citet{2006ApJS..167....1B}, the best tangential and line-of-sight
linking lengths are determined to be 0.14 and 0.75 times the mean
spacing between galaxies. Using mock catalogs, the linking lengths are
optimized to reproduce the multiplicity function and projected sizes
of groups with $>$10 galaxies.

However, we prefer a larger line-of-sight linking length and smaller
tangential linking length for this work. Larger line-of-sight linking
lengths are better for recovering the full finger-of-God of
groups. For this reason, analysis of mock catalog data finds that a
line-of-sight linking length of 1.3 best optimizes recovery of group
velocity dispersions (A.\ Baker 2014, B.S.\ honors
thesis\footnote{http://resolve.astro.unc.edu/pages/pdf/ashbake\_thesis3.0.pdf}). We
also find that using a tangential linking length of 0.14 over-links
low-N groups, i.e., singleton galaxies are linked into false pairs and
triplets.  We have used the mock galaxy catalog described in M15 to
compare the distribution in distances between galaxies in pairs and
triplets with the distribution of distances between single galaxies
and their nearest neighbors. Based on this analysis a tangential
linking length of 0.07 minimizes breaking up truly paired galaxies
($<$5\%), while preventing the overgrouping of truly single
galaxies. The independent analysis of linking length parameter space
provided by \citet{2014MNRAS.440.1763D} suggests optimal tangential
and line-of-sight linking lengths of 0.07 and 1.1 for studies of
galaxy properties in the context of environment. Their study showed
that these linking lengths result in low group merging fractions
across all group halo masses, although they do result in high
fragmentation for the largest groups (of which there are relatively
few in RESOLVE and ECO). Overall galaxy completeness and reliability
are found to be high using these linking lengths to find groups. These
linking lengths are also similar to those found in the analysis of
\citet{2011MNRAS.416.2640R}, which determined that while low-N (N $<$
5) groups do suffer from contamination, their integrated group
luminosities are not strongly affected. Analysis of mock galaxy
catalogs customized for the RESOLVE and ECO surveys shows that these
linking lengths yield high purity and completeness ($>$0.75) for
centrals and satellites in low-mass, low-N groups. The purity and
completeness decrease for higher mass halos to $\sim$0.5 (V. Calderon,
private communication). Given these results, we adopt tangential and
line-of-sight linking lengths 0.07 and 1.1 to create our group
catalog. These linking lengths were also used in M15.

\subsubsection{Group Finding and Halo Mass Assignment}
\label{sec:halomassmeas}

\begin{figure*}
\plotone{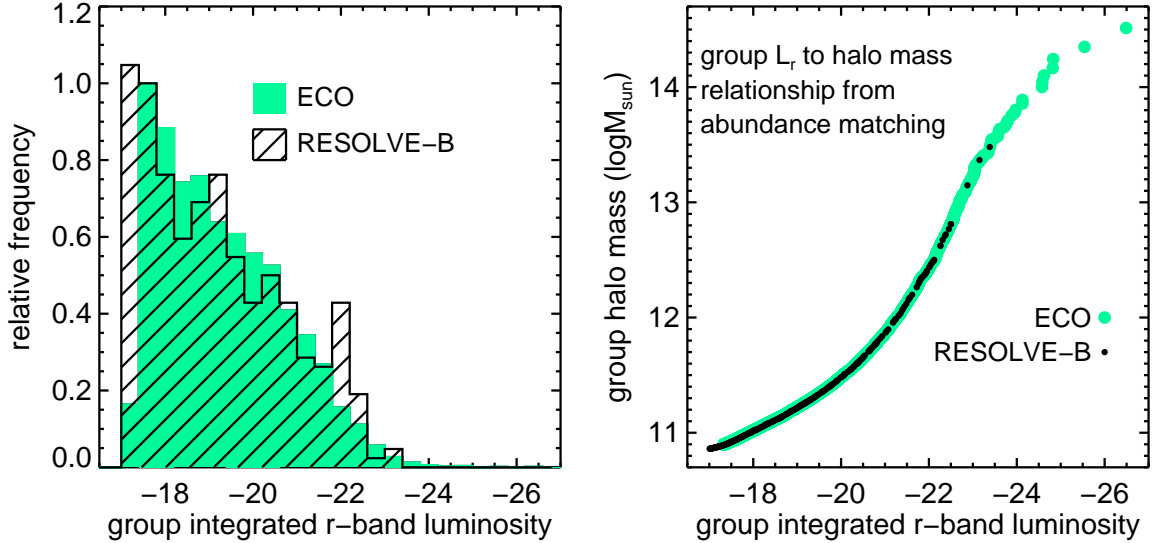}

\epsscale{1.0}
\caption{a) The frequency distributions of group integrated $r$-band
  luminosity in RESOLVE-B and ECO. The distributions have been
  normalized by the number of groups in the last complete bin for
  ECO. ECO's distribution is much less noisy than that of
  RESOLVE-B. b) The group halo mass to group integrated $r$-band
  luminosity relationship for ECO and RESOLVE-B determined through
  abundance matching to the theoretical group HMF from
  \citet{2006ApJ...646..881W}. The relationship for RESOLVE-B is
  determined by a spline fit to the abundance matching relation for
  the RESOLVE-B-analog version of ECO which extends down to
  M$_{r,tot}$ = $-17.0$ mag. Thus groups of similar luminosity in
  RESOLVE-B and ECO are matched to the same group halo mass.}
\label{fg:grplum}
\end{figure*}

To find groups in RESOLVE we use the same general algorithm described
in M15 and outlined below.  For ECO, we perform the group finding on
the full data set over the redshift range from 2530 \kms\ to 7470
\kms, limiting the data set to galaxies brighter than M$_{r,tot}$ =
$-17.33$ mag.  We also construct a RESOLVE-B-analog data set from ECO,
including galaxies down to $-17.0$ mag.  We create this analog data
set to determine the physical linking lengths for the RESOLVE-B data
set, for which the volume is too small and subject to cosmic variance
to determine the linking lengths dynamically.  For RESOLVE-B, we
perform the group finding with these fixed physical linking lengths
for galaxies brighter than $M_{r,tot}$ = $-17.0$ mag and over the
range 4250 \kms\ to 7250 \kms, allowing for a 250\kms\ buffer on
either side of the subvolume. In Figure \ref{fg:grplum}a, we show the
resulting group luminosity distributions for ECO and RESOLVE-B.

Halo masses are inferred from the total group $r$-band luminosity
using halo abundance matching between the identified groups and the
theoretical group HMF from \citet{2006ApJ...646..881W}, assuming
cosmological parameters consistent with WMAP5
\citep{2009ApJ...701.1804D}. The algorithm assumes zero scatter
between group luminosity and group halo mass (this scatter has been
estimated to range from $\sigma_{L}$ = 0.13--0.17 at fixed halo mass
for central galaxies, see
\citealp{2006MNRAS.365..842C,2008ApJ...676..248Y,2009MNRAS.392..801M}).
Because the RESOLVE-B volume is small and subject to cosmic variance,
we fit a spline to the abundance matching result (group L$_{r}$
vs.\ halo mass relation) from the RESOLVE-B-analog version of the ECO
catalog and use the fit to assign halo masses to the groups in
RESOLVE-B. Therefore, the halo masses assigned to ECO and RESOLVE-B
groups are consistent as shown in Figure \ref{fg:grplum}b.

Note that in this work, we have not performed any correction to the
group halo masses, unlike M15. In M15, group finding and halo
abundance matching were performed for both ECO and a large mock
catalog. The mock catalog was used to assess whether there was an
offset between the assigned group halo masses and the true halo masses
in the simulation. An offset of $-0.15$ dex was found and applied to
the ECO group halo masses. Further investigation has revealed,
however, that the overall simulation used for the mock catalog is
under-dense compared to ECO (shown to be similar to the overall SDSS
in \S \ref{sec:cvofds}), and this under-density leads to groups being
assigned larger masses than their true masses in the
simulation. Performing the comparison of assigned to true group halo
masses within a subvolume of the mock catalog that has similar density
to ECO results in no offset between the true and assigned group halo
masses. Thus we do not apply any offset and our group halo masses are
$\sim$0.15 dex larger than those reported in the M15 ECO catalog. We
refer to these masses as halo masses throughout this paper; however,
we emphasize that errors in group finding can cause significant
scatter (\mbox{$\sigma$ $\sim$ 0.1}) between our estimated masses and
the true masses of the underlying halos in addition to the neglected
intrinsic scatter between L$_{r,tot}$ and group halo mass.

To test the robustness of our galaxy mass functions computed in
different halo mass bins (see \S \ref{sec:condmfs}), we have also
performed halo abundance matching based on group stellar mass rather
than group $r$-band luminosity. The results shown in \S
\ref{sec:condmfs} are not affected by whether we use group luminosity
or group stellar mass for halo abundance matching.

\begin{figure*}
\plottwo{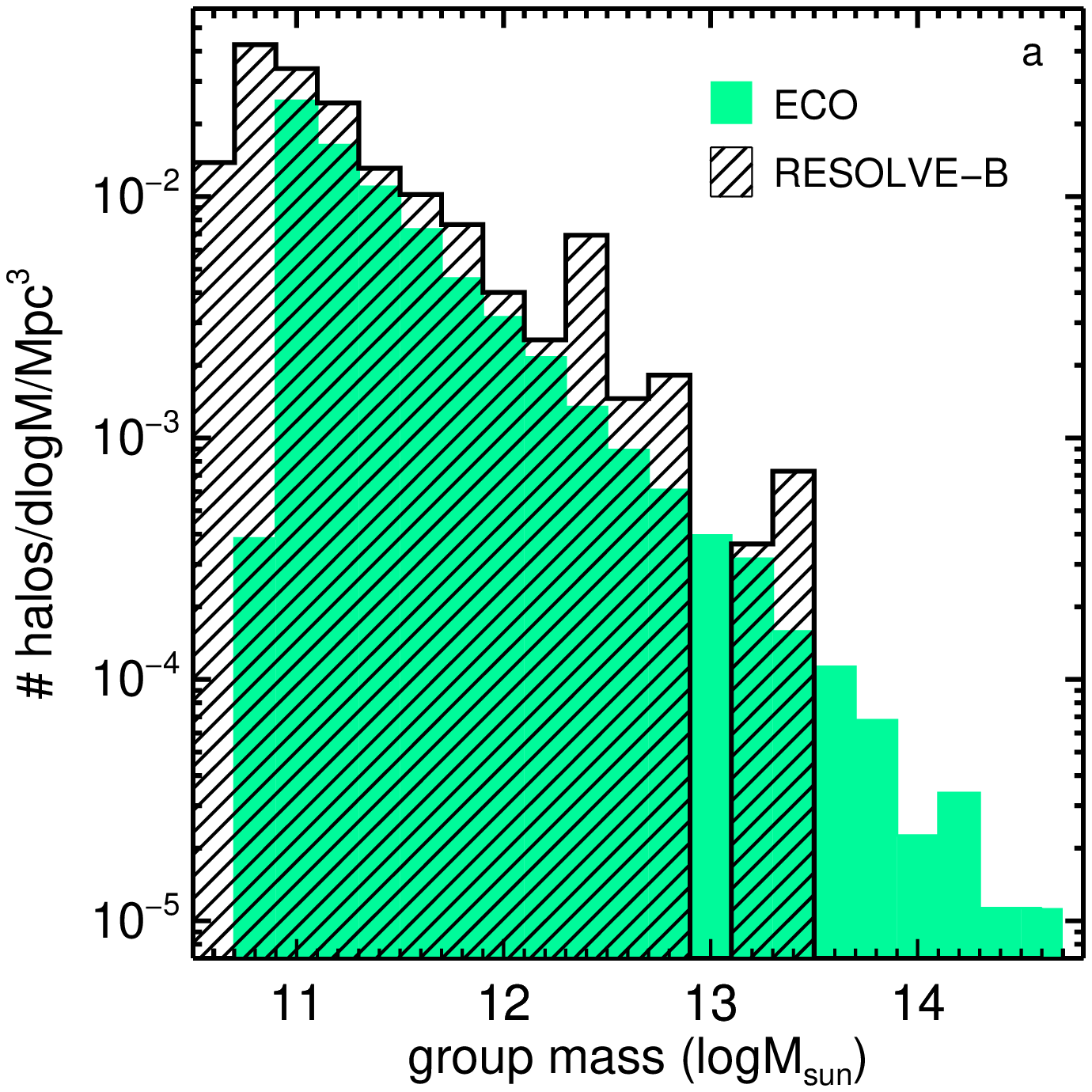}{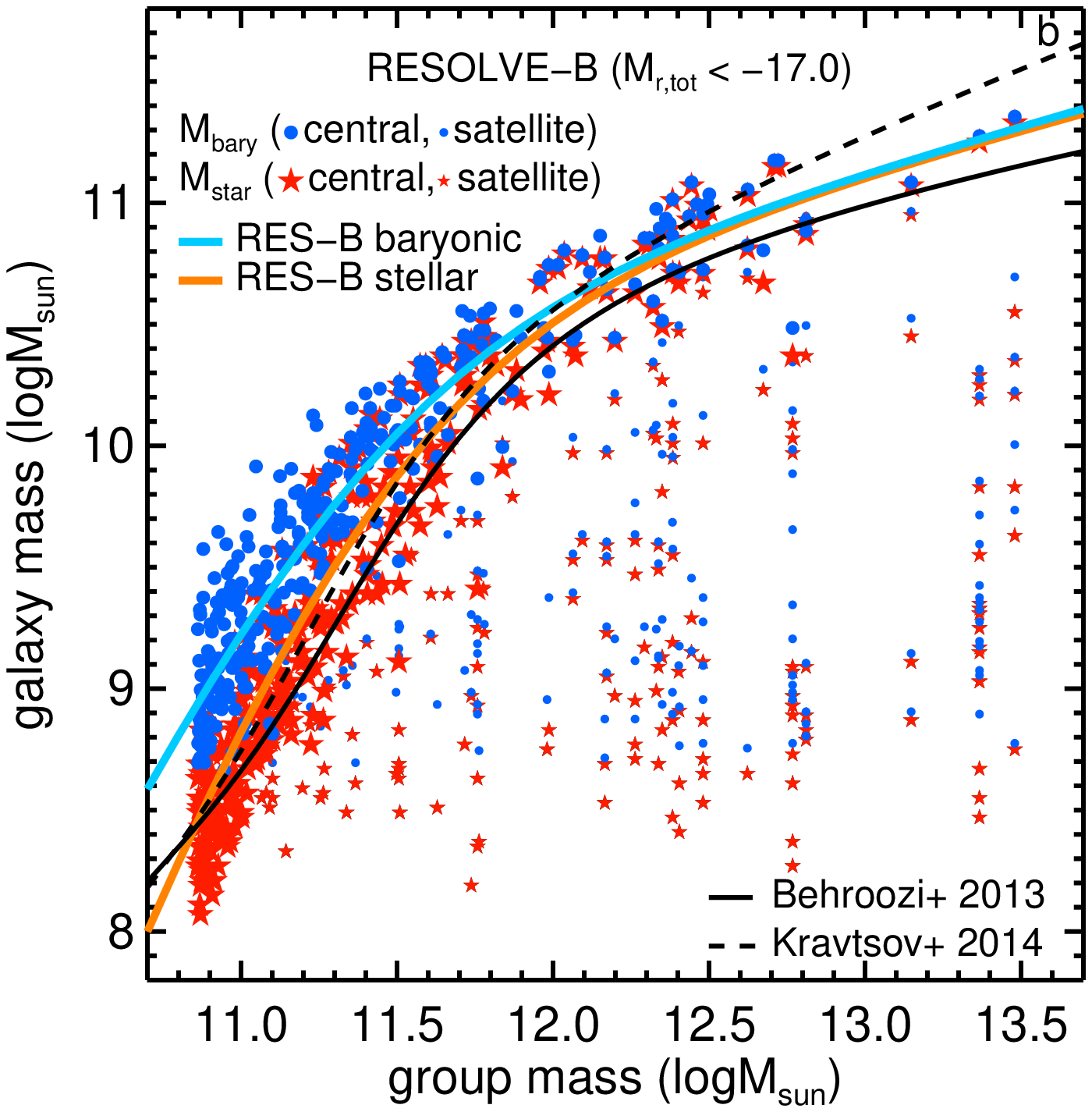}

\epsscale{1.0}
\caption{The group halo mass distribution and relationship between
  galaxy stellar or baryonic mass and group halo mass. a) RESOLVE-B
  (black cross-hatch) and ECO (M$_{r,tot}$ $<$ $-17.33$; solid green)
  HMFs using the FOF algorithm of \citet{2006ApJS..167....1B} to find
  groups and halo abundance matching to assign masses based on a total
  group luminosity-to-halo mass conversion factor (\S
  \ref{sec:halomass}).  RESOLVE-B has an overabundance of group halos
  of mass $\sim$10$^{12.5}$ and $\sim$10$^{13.5}$ \msun{} and no
  groups more massive than 10$^{13.5}$ \msun{}, unlike the smoother
  ECO halo mass distribution. b) Galaxy stellar or baryonic mass to
  group halo mass relation for RESOLVE-B (limited to galaxies brighter
  than $-17$).  The stellar and baryonic masses are computed from the
  median of the likelihood weighted mass distributions described in \S
  \ref{sec:computembary}. The plot shows both the central galaxy
  stellar (large red stars) and baryonic (large blue dots) masses and
  the satellite galaxy masses (smaller symbols). The orange and blue
  solid lines show our fits to the RESOLVE-B central stellar and
  baryonic mass to halo mass relationships according to equation \ref
  {eq:cent} (parameters are given in Table \ref{tb:cent} along with
  the similar results for ECO). We find that the central
  stellar-to-halo mass relation is in agreement with central
  stellar-to-halo mass relationships from \citet{2013ApJ...770...57B}
  and \citet{2014arXiv1401.7329K}, splitting the difference at large
  group halo masses (\S \ref{sec:halomassmeas}).}

\label{fg:halodist}
\end{figure*}

In Figure \ref{fg:halodist}a we compare the HMFs for RESOLVE-B (black
striped histogram) and ECO (green solid histogram) normalized by each
data set's respective number of halos. The ECO group halo mass
function is smooth by definition as it is directly matched to the halo
mass function of \citet{2006ApJ...646..881W}. RESOLVE-B has a noisier
distribution, since we assign group halo masses using the group
luminosity to group halo mass relation for the RESOLVE-B-analog
version of ECO. While RESOLVE-B has no groups more massive than
\mbox{10$^{13.5}$ \msun}, it does have an overabundance of halos of
mass $\sim$10$^{12.5}$ \msun{} and 10$^{13.5}$ \msun.  Since RESOLVE-B
is overdense, as described in \S \ref{sec:cvofds}, its HMF is slightly
elevated over that of ECO. Although RESOLVE-B does not contain any
clusters, the fact that it is overdense may contribute to the large
number of intermediate and large group halos in RESOLVE-B, as more
dense areas tend to have larger structures.

We also look at the relationship between galaxy stellar or baryonic
mass and group halo mass in Figure \ref{fg:halodist}b. Central
galaxies are determined to be the brightest galaxy in a group and
denoted by larger symbols. Centrals show a monotonic relationship with
halo mass, which we model as a function of two power laws given in
equation \ref{eq:cent}.

\begin{equation}
M_{gal} = \phi_{0}\frac{(\frac{M_{grp}}{M_{0}})^{\alpha}}{(x_{0}+\frac{M_{grp}}{M_{0}})^{\beta}}
\label{eq:cent}
\end{equation}

Using MPFITFUN (an IDL code that implements a Levenberg-Marquardt
least squares fit, see \citealp{more} and
\citealp{2009ASPC..411..251M}), we fit this model to the central
stellar and baryonic mass to halo mass relationships for RESOLVE-B and
ECO. The parameters of the fits are given in Table \ref{tb:cent} and
the RESOLVE-B stellar and baryonic fits are shown in Figure
\ref{fg:halodist}b as orange and blue lines respectively. Our stellar
mass to halo mass relationship lies between those of
\citet{2013ApJ...770...57B} and \citet{2014arXiv1401.7329K}.  (The
steeper relationship of \citealp{2014arXiv1401.7329K} reflects
photometry from \citealp{2013MNRAS.436..697B}, which recovers extended
light around brightest cluster galaxies.) Below a halo mass of
$\sim$10$^{12}$ \msun{} the baryonic mass of centrals starts to become
significantly larger than the stellar mass.

\input{table1.tex}

Satellite galaxies are denoted by smaller symbols in Figure
\ref{fg:halodist}b and are seen cascading down from the central galaxy
mass-to-halo mass relationship. There is no clear relationship between
the stellar or baryonic masses of satellites and halo mass, although
it is evident that the number of satellites increases for larger
halos.

Since the RESOLVE-B data set extends to luminosities fainter than
$-17$ mag, we perform an extra step to determine whether any galaxies
with M$_{r,tot}$ $>$ $-17$ belong to previously identified
groups. First, we determine whether the faint galaxy is within the
virial radius of a group center. If so, we determine whether the faint
galaxy's recessional velocity is within the larger of the
line-of-sight linking length or 3 times the group velocity dispersion
from that group center. If the galaxy meets both the radius and
velocity criteria, it is matched to the group. If it does not match to
any group it is placed in a group by itself and given a halo mass
based on the extrapolation of the halo mass to group integrated
$r$-band luminosity relation used in abundance matching. Of the 192
galaxies in RESOLVE-B fainter than $-17$ mag, 47 are associated to
identified RESOLVE-B groups and 145 are in halos by themselves.

\subsection{Completeness of Data Sets}
\label{sec:completeness}

To ensure that we interpret the SMF and BMF correctly, it is important
that we understand the stellar and baryonic mass completeness limits
of the RESOLVE-B and ECO data sets.  In \S
\ref{sec:lumsbcompleteness}, we compare the surface brightness
completeness of RESOLVE-B with estimates of SDSS completeness from the
literature to show that RESOLVE-B is a highly complete data set. In \S
\ref{sec:compcorr}, we present empirical completeness corrections
derived for the ECO data set based on RESOLVE-B. In \S
\ref{sec:masscompleteness}, we determine the stellar and baryonic mass
completeness limits for RESOLVE-B and ECO.

\subsubsection{RESOLVE-B Completeness}
\label{sec:lumsbcompleteness}

Since RESOLVE-B has the benefit of additional redshift coverage from
several sources, we wish to compare its added completeness with the
estimated incompleteness of the SDSS main redshift survey determined
in \citet{2005ApJ...631..208B}. The SDSS main redshift survey is known
to have spectroscopic incompleteness $\sim$6-10\% due to a mechanical
issue limiting the minimum spacing between fibers to 55\arcsec{} from
each other \citep{2003AJ....125.2276B}. This incompleteness estimate,
however, is limited to galaxies that were targeted for spectroscopic
followup (m$_{r,petro} < 17.77$).

Other sources of incompleteness arise from known issues with the SDSS
photometric pipeline that cause galaxies to be omitted as targets in
the redshift survey. These problems are oversubtraction of sky around
the galaxy, causing the amount of flux to be underestimated
\citep{2002AJ....124.1810S,2011AJ....142...31B}, and ``shredding'' of
galaxies, which means that rather than identifying and measuring the
flux for one galaxy, the pipeline breaks up the galaxy into several
individual pieces measuring the flux for each piece
\citep{2002AJ....123..485S}.  This ``shredding'' means that no one
piece of the galaxy is bright enough to be included in the redshift
survey, even if the galaxy is truly bright enough. In addition low
surface brightness galaxies ($\mu_{50}$ $<$ 24.5 \magarc) were excluded from
spectroscopic follow-up deliberately despite meeting the magnitude cut
\citep{2002AJ....124.1810S}.

\begin{figure*}
\plotone{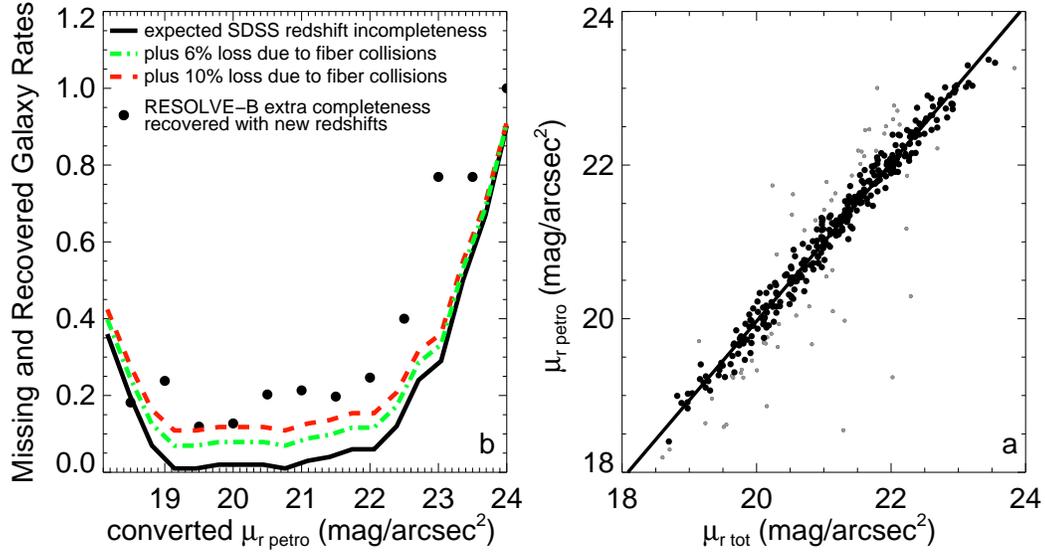}
\epsscale{1.0}
\caption{a) Evaluation of the extra completeness of RESOLVE-B due to
  redshift completion as a function of surface brightness (see panel
  b) and comparison to expectations from
  \citet{2005ApJ...631..208B}. Black dots show our missing galaxy
  recovery rates, i.e., one minus the ratio of the number of galaxies
  in RESOLVE-B,orig divided by the number of galaxies in the final
  RESOLVE-B for all galaxies with m$_{r,tot}$ $<$ 17.67 (which
  corresponds to the SDSS redshift survey limit m$_{r,petro}$ $<$
  $17.77$). The black solid line shows the estimated incompleteness of
  the SDSS spectroscopic survey due to photometry as described in
  \citet{2005ApJ...631..208B}; see \S \ref{sec:lumsbcompleteness}. The
  green dotted-dashed and red dashed lines include additional
  incompleteness due to the 6-10\% rate of fiber collisions. We
  recover at least the expected number of galaxies missed by the
  spectroscopic survey (more for galaxies fainter than 20 \magarc). b)
  Comparison of Petrosian surface brightness, $\mu_{r,petro}$, with
  surface brightnesses measured for this work, $\mu_{r,tot}$ for
  galaxies in RESOLVE-B with SDSS main redshift survey spectra. The
  gray points show points $>$2$\sigma$ away from the general trend.
  We fit a line to the two surface brightness measurements within
  2$\sigma$ of the general relation (black points) to use the measured
  values from this work to assess incompleteness, as not all galaxies
  in our data set have a reliable catalog $\mu_{r,petro}$ (since many
  galaxies are ``shredded'' by the SDSS pipeline).}

\label{fg:sbcomp}
\end{figure*}

To gauge the spectroscopic incompleteness caused by photometric
pipeline issues that underestimate the flux in galaxies that could
otherwise be bright enough to qualify for the redshift survey,
\citet{2005ApJ...631..208B} used simulated galaxies to test the
effectiveness of the SDSS photometric pipeline over a range of galaxy
surface brightness.  They took into account galaxies lost via sky
oversubtraction and ``shredding'' to
determine the survey completeness as a function of surface brightness.
Figure \ref{fg:sbcomp}a shows their spectroscopic incompleteness (or
1-completeness) as a function of surface brightness (black solid
line).  Green dash-dotted and red dashed lines show the higher survey
incompleteness taking into account a 6\% or 10\% loss due to fiber
collisions respectively.

To compare with the results of \citet{2005ApJ...631..208B}, we plot
one minus the ratio of the number of galaxies in RESOLVE-B,orig
divided by the number of galaxies in the final RESOLVE-B as a function
of surface brightness. Because the SDSS photometric measurements are
not adequate for many of these additional galaxies, we use our
reprocessed photometry and limit both data sets to an $r$-band
apparent magnitude of m$_{r,tot} < 17.67$ mag which corresponds to
m$_{r,petro} < 17.77$ mag.  We then perform a fit between
$\mu_{r,petro50}$ and $\mu_{r,50}$ to translate the surface brightness
measured in this work to that of the SDSS for comparison with
\citet{2005ApJ...631..208B} (Figure \ref{fg:sbcomp}b).

We find that for galaxies with converted Petrosian surface
brightnesses brighter than $\sim$20 mag/arcsec$^2$, the added
RESOLVE-B completeness is consistent with the results of
\citet{2005ApJ...631..208B}, i.e., we have recovered the expected
number of galaxies lost due to photometric errors and fiber
collisions. For galaxies fainter than $\sim$20 mag/arcsec$^{2}$, we
find a higher than expected recovery rate of missing galaxies in
RESOLVE-B. We treat this result as evidence that the RESOLVE-B data
set is as close to complete as possible, making RESOLVE-B a powerful
data set for galaxy population studies.


\subsubsection{ECO Completeness Corrections}
\label{sec:compcorr}

The ECO catalog is less complete than RESOLVE-B due to two issues: 1)
it does not have as complete of a redshift inventory as RESOLVE-B,
although we have added galaxies where available from several sources,
and 2) it contains massive groups and clusters (including Coma) for
which a substantial number of galaxies may not be included in the
volume due to large peculiar velocities.  To correct for the first
issue, we use empirical completeness corrections based on the galaxy
distributions in the $M_{r,tot}$ vs.\ surface brightness and
M$_{r,tot}$ vs.\ color parameter spaces.  To correct for the second
issue, we employ the group finding algorithm on a larger volume to
estimate the fraction of galaxies in large groups and clusters missed
due to large peculiar velocities.

To calculate the empirical completeness corrections, we follow the
same methodology as presented in M15 and first construct a base sample
for both RESOLVE-B and ECO consisting only of galaxies with SDSS DR7
redshifts, M$_{r,petro}$ $<$ $-17.23$ mag, and having local group
corrected velocities within the respective velocity ranges of each
data set. We call these samples RESOLVE-B-DR7 and ECO-DR7. Second, we
construct a sample for both RESOLVE-B and ECO consisting of all
available local group corrected velocities within the velocity range
for each data set and with M$_{r,tot}$ $<$ $-17.33$ mag.  We call
these samples RESOLVE-B-DR7+ and ECO-DR7+. The RESOLVE-B-DR7+ sample
is complete, while the ECO-DR7+ sample is only partially complete.

To calculate completeness corrections, we adopt the method described
in M15 of adaptively binning each data set above in M$_{r,tot}$ and
$\mu_r$ and M$_{r,tot}$ and \cgi$^m$ parameter space. The adaptive
binning starts out with coarse bins and then refines the bin size
until no more than 10\% of the data set ($\sim$5 galaxies for
RESOLVE-B, $\sim$100 galaxies for ECO) exists in one bin. The
irregularly gridded field is then interpolated onto a smooth density
field.

The RESOLVE-B recovery rate field is simply the RESOLVE-B-DR7+ field
divided by the RESOLVE-B-DR7 field. We cannot apply this RESOLVE-B
recovery rate directly to the ECO catalog data, though, since the ECO
catalog has also been supplemented by other redshift sources, albeit
to a lesser degree.  Thus we create the ECO recovery rate field with
the ECO-DR7+ field divided by the ECO-DR7 field.  The final
completeness correction field that we apply to the ECO catalog is then
the RESOLVE B recovery rate field divided by the ECO recovery rate
field. We perform a box-car smoothing of this final correction field,
replacing values if they are $>$2$\sigma$ above the mean within a
7$\times$7 box and we further do not allow the field to have
correction factors below one. To find the completeness correction
value for a given galaxy, we evaluate the 2D field at the galaxy's
M$_{r,tot}$ and $\mu_r$ or M$_{r,tot}$ and $(g-i)^m$ color. The
completeness correction for each galaxy is saved as a weight
vector.\footnote{We have tested whether dividing our samples into two
  different halo mass bins affects the resulting completeness
  correction fields. We found no difference in the resulting
  completeness corrections, but the small number of RESOLVE-B galaxies
  may limit our ability to detect any group halo mass dependent
  effects.}

\begin{figure}
\plotone{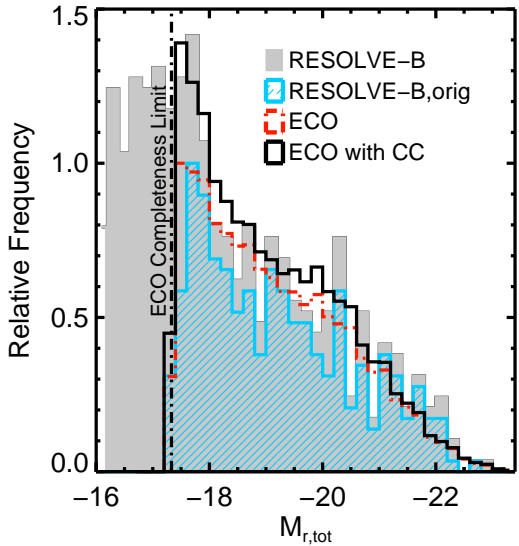}
\epsscale{1.0}
\caption{Relative frequencies of $r$-band absolute magnitude for the
  original and final RESOLVE-B data sets and the raw and
  completeness-corrected ECO data sets. RESOLVE-B is shown in solid
  gray and RESOLVE-B,orig in cross-hatched blue, where both have been
  normalized to the maximum bin height in the RESOLVE-B,orig
  distribution. The raw ECO data set is shown in dashed-dotted red,
  and ECO using the M$_{r,tot}$ vs.\ $\mu_{iso}$ completeness
  corrections is shown in solid black (the result is similar when
  using $(g-i)^m$ color). Both histograms have been normalized to the
  maximum bin height of the raw ECO distribution. We show the
  distributions normalized to the maximum height of the
  \textit{original} data set to emphasize the relative boost factor
  for ECO at each luminosity and how that compares to the difference
  between RESOLVE-B,orig and RESOLVE-B. The ECO luminosity
  completeness limit at M$_{r,tot}$ = $-17.33$ is shown with a black
  dashed-dotted line.}
\label{fg:lumdist}
\end{figure}

The luminosity distribution of the raw ECO data set is shown as a red
dashed-dotted outline histogram in Figure \ref{fg:lumdist} (normalized
to the maximum bin height in the raw ECO data set), and the luminosity
distribution for the completeness-corrected ECO using $\mu_{iso}$ is
shown in black (also normalized to the maximum bin height for the raw
ECO data set). We find that the $\mu_{iso}$ and $(g-i)^m$ completeness
correction fields provide similar corrections for ECO. The
completeness-corrected luminosity distribution for ECO agrees much
better with the RESOLVE-B luminosity distribution.

To correct for the cluster galaxies whose peculiar motions extend
outside the ECO volume, we use the results from M15, who performed
group finding on a larger catalog extending from 1500--12,000
\kms\ using SDSS catalog $r$-band measurements and limiting absolute
$r$-band magnitude to galaxies brighter than \mbox{M$_{r,petro}$ =
  $-18.4$} mag. The groups in the ECO catalog were cross-matched with
groups identified in this larger catalog.  The ratio of the number of
galaxies with \mbox{M$_{r,petro}$ $<$ $-18.4$} mag in the larger
catalog to the number of galaxies with \mbox{M$_{r,petro}$ $<$
  $-18.4$} mag in the ECO catalog is used to find any groups that are
missing significant numbers of galaxies. We apply this ratio as the
correction factor to galaxies in these groups, even for galaxies
fainter than $-18.4$ mag. Because the reprocessed photometry is not
available outside the ECO buffers ($<$2530\kms{} and $>$7470\kms), the
``recovered'' galaxies outside the ECO buffers are not included in the
final ECO group catalog and are used only to calculate the correction
factor.  Only three groups are affected by this issue, including the
Coma cluster.  We correct the masses of these groups by taking into
account the missing galaxies' luminosity.  The group mass estimate for
Coma, though, only changes by 0.06 dex (M15).

\subsubsection{Stellar and Baryonic Mass Completeness}
\label{sec:masscompleteness}

We must determine the stellar and baryonic mass completeness limits of
the RESOLVE-B and ECO data sets to correctly interpret the SMF and
BMF. We have already set the luminosity completeness limits based on
where the M$_{r,tot}$ distribution of the data falls off (Figures
\ref{fg:stripe82skydist}b and \ref{fg:lumdist}). For ECO, the
luminosity completeness limit was taken to be $-17.33$ mag, the SDSS
main redshift survey apparent magnitude limit as converted to our
absolute magnitude system at our largest redshift, and we accounted
for incompleteness in SDSS above its stated completeness limit by
applying empirical completeness corrections based on RESOLVE-B (\S
\ref{sec:compcorr}). For RESOLVE-B the luminosity completeness limit
was taken to be $-17.0$ mag without empirical completeness corrections
due to extended redshift coverage in Stripe 82. The lack of
corrections implies that over the range from $-17.0$ to $-17.33$,
RESOLVE-B underrepresents galaxy counts as illustrated in Figure
\ref{fg:stripe82skydist}b.

\begin{figure*}
\plotone{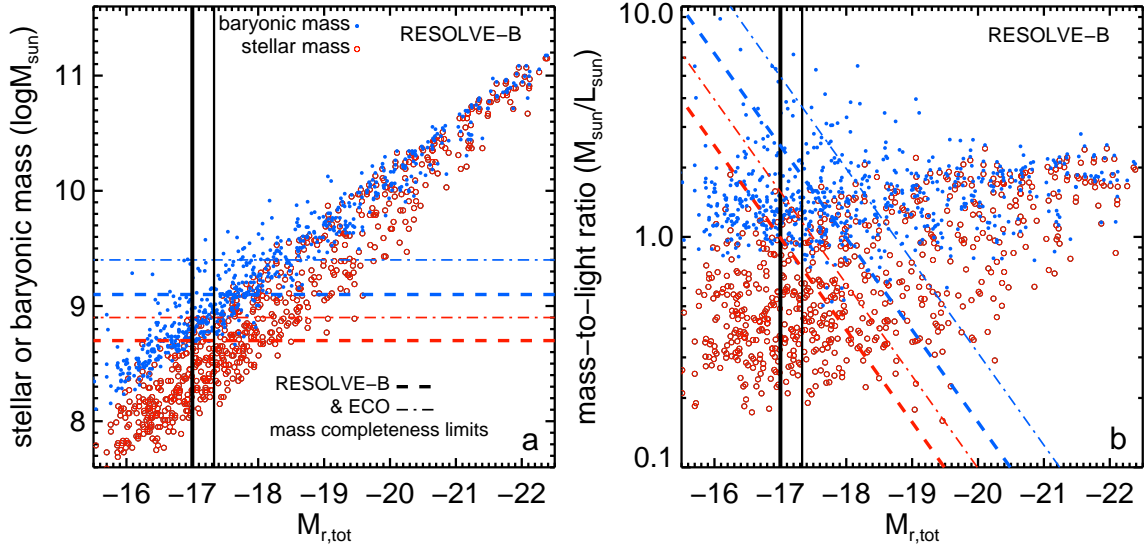}
\epsscale{1.0}
\caption{Determination of stellar and baryonic mass completeness
  limits for ECO and RESOLVE-B. a) Log stellar (open red circles) or
  baryonic (blue points) mass vs.\ absolute $r$-band magnitude for the
  RESOLVE-B data set. The luminosity completeness limits of RESOLVE-B
  (M$_{r,tot}$ = $-17.0$) and ECO (M$_{r,tot}$ = $-17.33$) are shown
  as thick and thin solid black lines respectively. The mass
  completeness limits for both data sets are determined by finding the
  stellar or baryonic mass above which less than 2\% of objects have
  M$_{r,tot}$ fainter than the luminosity completeness limit. For
  RESOLVE-B and ECO, the stellar mass completeness limits are
  log(M$_{star}$)= 8.7 and log(M$_{star}$)= 8.9 respectively (red
  lines), and the baryonic mass completeness limits are
  log(M$_{bary}$) = 9.1 and log(M$_{bary}$) = 9.4 respectively (blue
  lines). b) Stellar and baryonic mass-to-light ratios as a function
  of the absolute $r$-band magnitude M$_{r,tot}$ for RESOLVE-B.  The
  red and blue lines mark the limiting stellar and baryonic
  mass-to-light ratios at a given M$_{r,tot}$ using the stellar and
  baryonic mass completeness limits determined in panel a. RESOLVE-B
  and ECO are treated as complete for galaxies brighter than the
  luminosity completeness limit and having stellar or baryonic
  mass-to-light ratios higher than the respective mass-to-light ratio
  limits.}

\label{fg:masscomp}
\end{figure*}

We can determine stellar and baryonic mass completeness limits by
examining the scatter in stellar and baryonic mass near the luminosity
completeness limits. In Figure \ref{fg:masscomp}a, we plot stellar and
baryonic mass as a function of M$_{r,tot}$ for RESOLVE-B. We estimate
mass completeness limits by finding the percentage of galaxies in
RESOLVE-B with masses above a given mass limit that are fainter than
our luminosity completeness limit for either RESOLVE-B or ECO ($-17.0$
and $-17.33$ respectively). We require this percentage to be
$<$2\%. For ECO, the resulting stellar and baryonic mass limits are
log(M$_{star}$) = 8.9 and log(M$_{bary}$) = 9.4 (marked as thin red
and blue dashed-dotted lines in Figure \ref{fg:masscomp}a). The ECO
mass completeness limits are independent of the completeness
corrections computed for ECO. For RESOLVE-B, the stellar and baryonic
mass limits are \mbox{log(M$_{star}$) = 8.7} and \mbox{log(M$_{bary}$)
  = 9.1} (marked as thick dashed red and blue lines). We note that the
baryonic mass limits for RESOLVE-B and ECO extend to the high-mass
dwarf regime, below the gas-richness threshold scale identified in K13
(\mbox{M$_{bary}$ $\sim$ 10$^{9.9}$}).

Since RESOLVE-B is somewhat incomplete over the range from $-17.0$ to
$-17.33$, we check the robustness of our mass completeness limits by
measuring the percentage of galaxies with masses above the
aforementioned RESOLVE-B mass completeness limits (M$_{star}$ =
10$^{8.7}$~\msun{} and M$_{bary}$ = 10$^{9.1}$~\msun), but with
M$_{r,tot}$ fainter than $-17.33$ rather than $-17.0$. We find that
the percentage of ``missed'' galaxies increases to 4\% for RESOLVE-B
with the higher luminosity limit. Since this increase is modest, we
use the RESOLVE-B mass completeness limits determined at M$_{r,tot}$ =
$-17.0$ in the mass function analysis. However, we have tested our
low-mass slope measurements with a stellar mass limit of
10$^{8.9}$~\msun{} instead of 10$^{8.7}$~\msun{} and the differences
are small and well within the errors.


In Figure \ref{fg:masscomp}b, we show the stellar and baryonic
mass-to-light ratios as a function of M$_{r,tot}$ for RESOLVE-B. The
mass completeness limits are converted to the limiting mass-to-light
ratio at a given M$_{r,tot}$ and shown as lines of constant mass
corresponding to those in panel a. The data sets are most complete for
galaxies brighter than the luminosity completeness limit \textit{and}
having mass-to-light ratios brighter than the limiting ratios for a
given M$_{r,tot}$.



\section{Statistical Analysis of Stellar and Baryonic Mass Functions}
\label{sec:statmassfunctions}

In this section we describe our new ``cross-bin sampling'' method for
measuring the SMF and BMF taking into account the full stellar and
baryonic mass likelihood distributions of galaxies.  We use the full
likelihood distributions of stellar and baryonic mass because the
widths of these distributions are often much larger than the bin size
used to construct the mass function ($\sim$0.1-0.2 dex).  For the LF
this issue is of little concern because photometry errors are
typically smaller than the bin sizes used to construct the LF, so
uncertainties can be assigned to the LF using just the Poisson
counting noise.  In contrast, for the SMF and BMF the uncertainty on
the mass measurement itself may spill over several bins.  Therefore we
have devised the cross-bin sampling method that makes use of the full
mass likelihood distributions to determine uncertainty bands around
the derived mass functions.

We first construct the normalized stellar or baryonic mass likelihood
distributions for the entire RESOLVE-B and ECO data sets from the
outputs of the SED fitting code (\S \ref{sec:mstar}) and the baryonic
mass calculations (\S \ref{sec:computembary}).  The stellar or
baryonic mass likelihood distribution for each galaxy is binned in
0.01 dex intervals, much smaller than the typical mass distribution
1$\sigma$ widths of $\sim$0.25 dex. We then sum the likelihoods from
all the galaxies at each small mass interval and normalize the entire
distribution by the number of galaxies in the data set. As a simple
example, if we started with a sample of 10 galaxies all with the same
stellar mass likelihood distribution, this procedure would yield a
normalized sample likelihood equal to the distribution for any one of
the ten galaxies.

To determine the SMF or BMF and the 68\% and 95\% confidence intervals
around that function for a given data set, we perform repeated Monte
Carlo sampling of the corresponding sample stellar or baryonic mass
likelihood distribution using the inverse transform sampling
method. The inverse transform sampling method allows one to sample
randomly from any probability distribution if its cumulative
distribution function (cdf) is known. A number drawn from a uniform
probability distribution between 0 and 1 (whose cdf is also a uniform
probability distribution between 0--1) can then be used to look up the
non-uniform distribution at the same integrated probability location
within its cdf. To apply this method, we first cumulatively sum the
sample stellar or baryonic mass likelihood distribution to produce the
stellar or baryonic mass cdf.  Second, we simulate a RESOLVE-B or ECO
galaxy population by drawing N values from a uniform probability
distribution (0--1), where N is drawn from a Poisson distribution with
mean value $<$N$>$ equal to the number of galaxies in the data
set. Third, we use the inverse transform sampling method to look up
the stellar or baryonic masses in the stellar or baryonic mass cdfs
that correspond to the N values just selected between 0 and 1,
assigning a stellar or baryonic mass to each galaxy. Lastly, we bin
the mass function into 0.2 dex bins. We perform this procedure 1000
times to create 1000 mass functions. From these 1000 stellar or
baryonic mass functions, we determine the median and the 1 and 2
$\sigma$ upper and lower bounds (16\%--84\% and 2.5\%--97.5\%
percentile ranges) within each bin.

\begin{figure}
\plotone{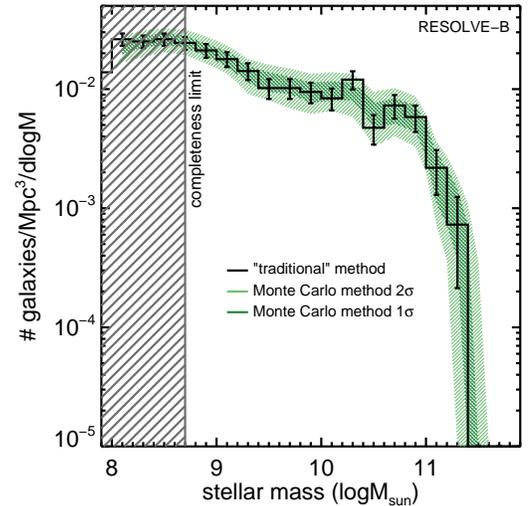}

\epsscale{1.0}
\caption{RESOLVE-B SMF calculated two different ways. The SMF is
  complete down to 10$^{8.7}$ \msun, marked by the dark gray line.
  The solid black histogram shows the number of galaxies per dlogM per
  Mpc$^3$ when using the median stellar mass from each galaxy's
  stellar mass likelihood distribution (see \S \ref{sec:mstar}) with
  Poisson error bars. The dark and light green shaded regions show the
  68\% and 95\% confidence intervals for the SMF sampled from the
  stellar mass likelihood distribution for the entire galaxy data set
  as described in \S \ref{sec:statmassfunctions}. The Poisson error
  bars are similar in width to the 68\% confidence intervals. The
  cross-bin sampling technique, however, yields a smoother SMF that
  takes into account the fact that stellar mass uncertainties can be
  much larger than the typical bin size of $\sim$0.2 dex used to
  construct the SMF.}
\label{fg:comparesmf}
\end{figure}

Figure \ref{fg:comparesmf} shows that our new SMF created through this
cross-bin sampling process (the 1$\sigma$ bounds shown in dark green
and the 2$\sigma$ bounds shown in light green) has a similar shape to
the traditional SMF, which uses the single value stellar mass (median
of each galaxy's stellar mass likelihood distribution) and assumes
Poisson error bars in each bin. There are, however, noticeable
deviations around 10$^{10.5}$ \msun. Using the traditional approach,
we might overinterpret the dip/spike feature occurring
$\sim$10$^{10.5}$ \msun, which has been effectively smoothed over in
the cross-bin sampling method by taking into account the full
likelihood distributions. Throughout the rest of this paper we will
use 1$\sigma$ bounds to show the SMF and BMF uncertainty bands.

For ECO, we slightly modify this general methodology to include
completeness corrections. Before constructing the normalized stellar
or baryonic mass likelihood distribution for the ECO data set, we
weight each galaxy's individual mass likelihood distribution by its
completeness correction factor. The overall ECO mass likelihood
distribution is the sum of these weighted distributions and the
``effective'' total number of galaxies in the data set $<$N$>$ is the
total of all the completeness correction factors rather than the
literal number of galaxies in the observed ECO data set.  We run the
Monte Carlo trials for the ECO data set 1000 times for the
completeness corrections computed using M$_{r,tot}$ and $\mu_r$ and
1000 times for the completeness corrections computed using M$_{r,tot}$
and $(g-i)^m$. We compute the 16th, 50th, and 84th percentiles for the
ECO mass functions based on the two completeness corrections. The ECO
mass function is reported as the average between the two 50th
percentile measurements in each stellar or baryonic mass bin. To
estimate the uncertainty bands for the ECO mass functions, in each bin
we choose the larger of the two 84th percentile values and smaller of
the two 16th percentile values determined for the two
completeness-corrected mass functions.

\section{Stellar and Baryonic Mass Functions}
\label{sec:massfunctions}

In this section, we examine the shape of the SMF and BMF for RESOLVE-B
and ECO using the cross-bin sampling technique described in \S
\ref{sec:statmassfunctions}.  First we check whether the features in
the mass functions are robust to different stellar population
models. We then compare our overall SMF and BMF with each other, with
mass functions from the literature, and with the predicted HMF, paying
special attention to the slope at the low-mass end. Lastly we break
down the SMF and BMF into ``conditional mass functions'' (mass
functions divided into halo mass regimes and further by
central/satellite designation) to analyze how the shape of the mass
function depends on group halo mass.

We provide our raw SMFs and BMFs for RESOLVE-B
and ECO in a machine readable table, the columns of which are given in
Table \ref{tb:allmfs}. For each mass function, we provide the median,
16th, and 84th percentiles. We only include values above the
respective mass completeness limit for each data set.

\input{table2.tex}


\subsection{Choice of Stellar Population Models}
\label{sec:smodelcomp}

To check whether the shape of the SMF is dependent on the assumed grid
of star formation histories, we compare the SMFs resulting from the
cross-bin sampling method for both model sets described in \S
\ref{sec:mstar}. The two SMFs are mostly similar with a $\sim$0.08 dex
zero-point offset such that the model set \textit{a} SMF is shifted
toward lower masses, similar to the $\sim$0.1 dex offset previously
reported in K13. Taking into account this shift, we find that the SMFs
are extremely similar, except for the last two bins above the stellar
mass completeness limit, where the \textit{b} model set is slightly
steeper. At these masses dwarf galaxy stellar mass likelihood
distributions are somewhat sensitive to modeling choices; however, the
two model sets agree within their uncertainty bands after correcting
for the zero-point offset. For simplicity we use model set \textit{a}
for the remainder of this work.  This choice enables direct comparison
of features in our mass functions with the threshold and bimodality
mass scales described in K13. We note that our stellar masses are
consistent with most others in the literature (see \S
\ref{sec:mstar}), apart from those in \citet{2003ApJS..149..289B}.

\subsection{Overall SMF and BMF}
\label{sec:overall}

In this section, we examine the shape of the RESOLVE-B and ECO SMF and
BMF, which are shown in Figures \ref{fg:litcomparesmfbmf} and
\ref{fg:ourcomparesmfbmf}. We examine the results of single and double
Schechter fitting and compare with previous mass functions from the
literature. We also directly compare the SMF and BMF with each
other and with the theoretical HMF in Figure
\ref{fg:dircomparesmfbmf}.

\subsubsection{Mass Function Fit Parameters}
\label{sec:mffits}

We use the Markov chain Monte Carlo Ensemble Sampler \textit{emcee}
\citep{2013PASP..125..306F} to fit both Schechter and double Schechter
functions to the RESOLVE-B and ECO mass functions in a Bayesian
framework. The single Schechter function follows the traditional form
as a function of galaxy mass or $m$:

\begin{equation}
\phi(m)dlog(m)=\phi_{*}(\frac{m}{M_{*}})^{(\alpha+1)}\exp({\frac{-m}{M_{*}}})dlog(m)
\label{eq:ss}
\end{equation}

Here $M_{*}$ is the characteristic mass scale, $\alpha$ is the power
law rise at the low-mass end, and $\phi_{*}$ is the overall normalization
at $M_{*}$. 

We also fit a double Schechter function that allows for two
low-mass power law slopes and two normalization parameters following
the form described in \citet{2008MNRAS.388..945B}:

\begin{equation}
\begin{split}
\phi(m)dlog(m)= &\exp({\frac{-m}{M_{*}}})dlog(m)\times \\
    & [\phi_{*1}(\frac{m}{M_{*}})^{(\alpha_1+1)}+\phi_{*2}(\frac{m}{M_{*}})^{(\alpha_2+1)}]
\end{split}
\label{eq:ds}
\end{equation}

For our Bayesian parameter estimation, we assume uniform priors on all
single Schechter function parameters over ranges encompassing previous
estimates of all the values. We also assume uniform priors on all
double Schechter function parameters with the additional requirement
that $\alpha_2$ be less than $\alpha_1$ (i.e., $\alpha_2$ must have a
steeper power-law slope than $\alpha_1$), similar to
\citet{2008MNRAS.388..945B}. 

The \textit{emcee} code uses a Markov chain Monte Carlo Ensemble
Sampler to fill out the parameter space. For the single Schechter
function we use 100 walkers over 100 steps after a burn-in of 100
steps. For the double Schechter function, we use 400 walkers over 100
steps after a burn-in of 400 steps. To assess the convergence of the
chains, we measured the autocorrelation time for each parameter's
chain, and set the burn-in number of steps to be a few times the
autocorrelation time (per the guidelines discussed in
\citealp{2013PASP..125..306F}). We also visually inspected the chains
to ensure that they properly sampled the parameter space, and we
calculated the mean acceptance fraction of the chains to be $\sim$0.5
and $\sim$0.35 for the single and double Schechter fits respectively
(within the acceptable range 0.2-0.5 as discussed in
\citealp{2013PASP..125..306F}). We report the median of the
marginalized posterior probability distributions for each parameter in
Table \ref{tb:schechfuncparams_emcee} with error bars showing the 16th
and 84th percentiles.

In our parameter fitting, we do not consider any error term due to
cosmic variance, i.e., related to the halo mass mix or overall density
within our data sets, as has been done in some previous work (e.g.,
\citealp{2011ApJ...738...45L,2012MNRAS.426..531S}). Using mock
catalogs, these works have shown that individual bins within the SMF
or LF are correlated with each other for any given data set,
reflecting its overall environmental density, mix of environments, and
the fact that a few high-mass halos contribute most of the galaxies at
the bright end.  \citet{2012MNRAS.426..531S} finds that inclusion of
the full covariance matrix yields Schechter parameter fits differing
by up to 2$\sigma$ compared to fits using only Poisson errors. The
parameters most affected are $\alpha$, which becomes steeper, and
$L_{*}$, which becomes fainter, after taking into account these
covariances. However, since we will show in \S \ref{sec:condmfs} that
the mass function is not universal but depends on halo mass, our
approach does not treat environmental variance (as defined by the
group halo mass distribution) as an ``error'' but rather as a physical
manifestation of the fact that the mass function varies with the group
halo mass distribution in predictable ways. As a result, the variation
in Schechter fit parameters between RESOLVE-B and ECO may be
(unsurprisingly) larger than our quoted errors due to their different
group halo mass distributions.

\input{table3.tex}

\subsubsection{The SMF}
\label{sec:compmstar}

The RESOLVE-B and ECO SMFs drop off steeply for masses $\ga$
10$^{10.8}$\msun, which is near the ``knee'' of the
Schechter function that joins the steep exponential fall-off toward
higher mass galaxies, and the power law rise toward lower mass
galaxies. Based on our single and double Schechter function fits, we
find that the knee occurs at $\sim$10$^{10.8-11.1}$ \msun{} for both
data sets. The knee of the SMF has been measured to be
$\sim$10$^{10.7}$ \msun{} in previous works including
\citep{2007MNRAS.378.1550P,2008MNRAS.388..945B,2010ApJ...721..193P,2012MNRAS.421..621B}.

We note, however, that at high masses, neither the single nor double
Schechter functions fit our SMFs well, indicating that the exponential
fall-off at high masses is not a good model for our two data sets. For
RESOLVE-B, the fall-off is steeper than the fits, while for ECO the
fall-off is shallower. The shallowness of the ECO fall off may be in
line with results from \citet{2013MNRAS.436..697B}, which recovered
more light from bright galaxies using \textit{PyMorph} and found a
shallower fall-off in the LF. RESOLVE-B, on the other hand, has
relatively few extremely bright galaxies, since it has no large
clusters with mass $>$10$^{13.5}$ \msun, potentially leading to the
steeper fall-off. We also note that the GAMA derived SMF from
\citet{2012MNRAS.421..621B} has a fall-off between those of RESOLVE-B
and ECO. While GAMA covers a relatively small volume, they still have
more massive clusters than found in RESOLVE-B
\citep{2011MNRAS.416.2640R}.

\begin{figure*}
\epsscale{1.1}
\plotone{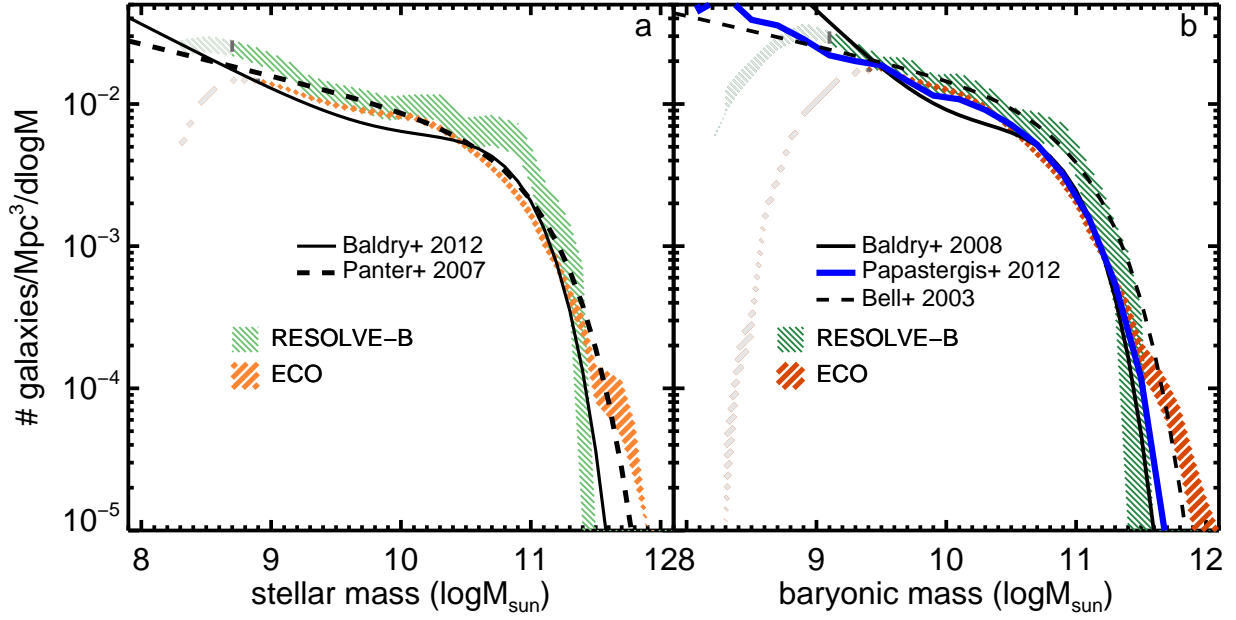}
\caption{RESOLVE-B and ECO SMFs and BMFs using the cross-bin sampling
  technique with comparison to previous work. The uncertainties due to
  cosmic variance are not included in the error budgets of the mass
  functions. a) SMFs for RESOLVE-B (light green) and ECO (light
  orange), with shaded regions showing the 16-84th\% percentile
  confidence intervals. Incomplete regions of the mass functions are
  shaded lighter.  The black solid line shows the double Schechter
  function fit from \citet{2012MNRAS.421..621B}, and the dashed line
  shows the single Schechter function fit from
  \citet{2007MNRAS.378.1550P} (\S \ref{sec:compmstar}). b) BMFs for
  RESOLVE-B (dark green) and ECO (dark orange). The solid blue line
  comes from the measured BMF from \citet{2012ApJ...759..138P} using a
  combination of SDSS and ALFALFA. The solid black line shows the
  inferred BMF from \citet{2008MNRAS.388..945B} and the dashed line is
  from \citet{2003ApJ...585L.117B} (\S \ref{sec:compmbary}). }
\label{fg:litcomparesmfbmf}
\end{figure*}

\begin{figure*}
\epsscale{1.1}
\plotone{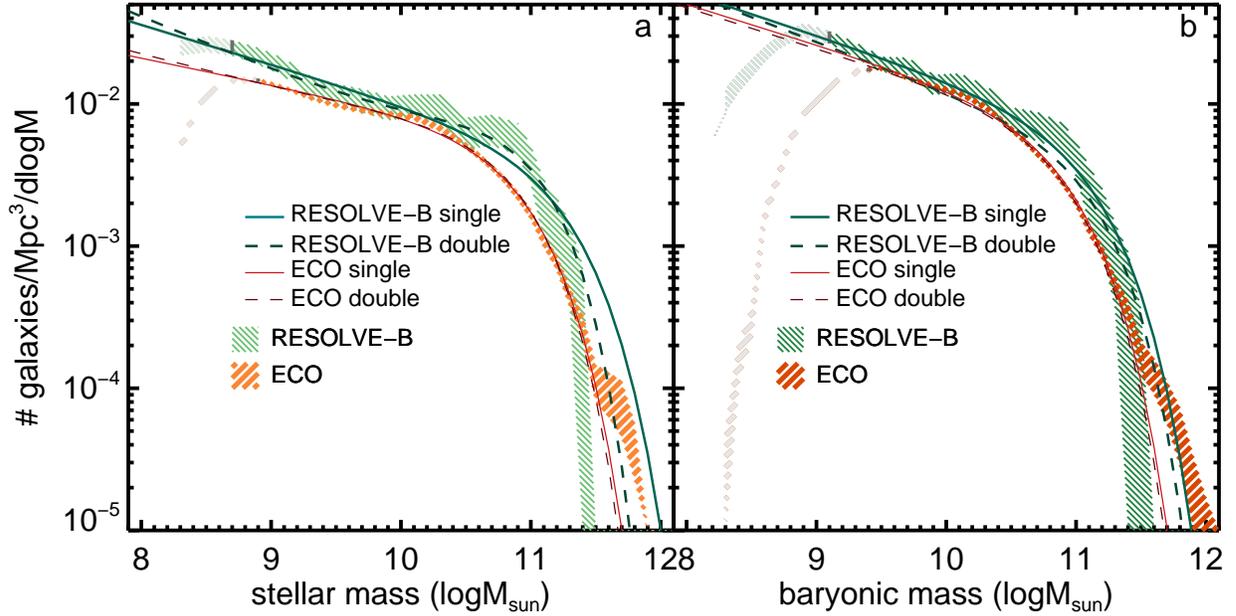}
\caption{RESOLVE-B and ECO SMFs and BMFs using the cross-bin sampling
  technique with single and double Schechter function fits. The
  uncertainties due to cosmic variance are not included in the error
  budgets of the mass functions. a) SMFs for RESOLVE-B (light green)
  and ECO (light orange), with shaded regions showing the 16-84th\%
  percentile confidence intervals. Incomplete regions of the mass
  functions are shaded lighter.  The solid green and orange lines show
  single Schechter function fits to RESOLVE-B and ECO respectively and
  appear to be inadequate fits to the data. The dashed green and
  orange lines show the double Schechter function fits to RESOLVE-B
  and ECO respectively. These fits, however, are not much improved
  over the single Schechter function fits. b) BMFs for RESOLVE-B (dark
  green) and ECO (dark orange).  Again the solid and dashed green and
  orange lines show the single and double Schechter function fits for
  RESOLVE-B and ECO respectively. The additional parameters of the
  double Schechter function are unnecessary for fitting the shape of
  the BMF.}
\label{fg:ourcomparesmfbmf}
\end{figure*}

Below $\sim$10$^{10.8}$\msun, the RESOLVE-B and ECO SMFs rise toward
lower masses. However, the slope plateaus over a mass range of
10$^{9.5-10.2}$ \msun{}. The plateau feature in the SMF has been
observed in several previous studies
\citep{2008MNRAS.388..945B,2009ApJ...707.1595D,2009MNRAS.398.2177L,2010ApJ...721..193P},
motivating the use of the double Schechter function to fit the
low-mass end of the SMF. Single Schechter functions, such as the
dashed black line from \citealt{2007MNRAS.378.1550P} in Figure
\ref{fg:litcomparesmfbmf}a and the dark green and orange solid lines
based on our fits in Figure \ref{fg:ourcomparesmfbmf}a, cannot
reproduce the shape of the SMFs. To model the plateau, others have
implemented double
\citep{2008MNRAS.388..945B,2009ApJ...707.1595D,2010ApJ...721..193P,2012MNRAS.421..621B}
and even triple \citep{2009MNRAS.398.2177L} Schechter functions. We
observe this plateau feature in the SMF (seen more easily in Figure
\ref{fg:dircomparesmfbmf} with reference to the BMF and without
overlapping fit lines), but we find that the double Schechter function
does not yield a much better fit than the single Schechter function
for either the RESOLVE-B or ECO SMFs. Therefore to measure the plateau
slope and the low-mass slope, we fit lines to the SMF over a range
from log(M$_{star}$) = 9.5-10.1 for the plateau, and log(M$_{star}$)
$\leq$ 9.5 for the low-mass end. These values have been determined by
examining where the plateau and steep low-mass upturn features occur
in Figure \ref{fg:ourcomparesmfbmf}. With this technique we measure a
relatively flat slope just below the mass function knee
(\mbox{$\alpha_{plateau}$ = $-1.14$ $\pm$0.18} and \mbox{$-1.14$
  $\pm$0.05} for RESOLVE-B and ECO) and a more steeply rising slope
below the gas-richness threshold mass ($\alpha_{low-mass}$ = $-1.44$
$\pm$0.11 and $-1.30$ $\pm$0.04 for RESOLVE-B and ECO), albeit at
$\sim$1--2$\sigma$ significance. For comparison,
\citet{2012MNRAS.421..621B} find shallow and steep slopes of
$\alpha_1$ = $-0.35$ and $\alpha_2$ = $-1.47$ in their
double-Schechter function fit.


Comparing the normalizations of RESOLVE-B and ECO in the double
Schechter fits, we see that for high masses, $\phi_{*,1}$ is much
lower for ECO than RESOLVE-B (0.0034 vs.\ 0.0090), while $\phi_{*,2}$
is much more similar. The lower normalization of ECO is not unexpected
due to the overall smaller number density in the ECO catalog as
described in \S \ref{sec:cvofds}. Variations in number density with
sample size (i.e., cosmic variance) will affect the overall
normalization, as seen by the vertical displacement of several
previous works in Figure \ref{fg:litcomparesmfbmf}.




\subsubsection{The BMF}
\label{sec:compmbary}

The RESOLVE-B and ECO BMFs also exhibit a steep
drop-off for masses $>$10$^{10.8}$ \msun, which is consistent with
previous work
\citep{2003ApJ...585L.117B,2008MNRAS.388..945B,2012ApJ...759..138P},
although the drop-off occurs at higher masses in
\citet{2003ApJ...585L.117B}, due to different mass scales used.

At masses below 10$^{10.8}$ \msun, however, the BMFs rises as a
straight power law toward lower masses.  The BMF shape is actually
better described by the single Schechter function shape than the
SMF. The double Schechter function fits given in Table
\ref{tb:schechfuncparams_emcee} do not significantly improve on the
single Schechter function fits (see Figure
\ref{fg:ourcomparesmfbmf}b). The low-mass slope given by the single
Schechter function fits for RESOLVE-B and ECO is $\alpha$ $\sim$
$-1.3$, steeper than the low-mass slope of the SMF just below the knee
($\alpha_{plateau}$ $\sim$ $-1.14$). We find a low-mass slope that is
slightly steeper than the slope determined in
\citet{2003ApJ...585L.117B}, which used stellar masses inconsistent
with our own and a different HI mass estimation technique.


Comparing with the observed BMF from \citet{2012ApJ...759..138P},
which is shown as a blue solid line in Figure
\ref{fg:litcomparesmfbmf}b and uses ALFALFA HI measurements to
construct the BMF, we find that it is mostly consistent with the ECO
BMF, except at the largest masses, where the number density of
galaxies is small. Neither the RESOLVE-B nor ECO BMFs are fit well by
the inferred BMF from \citet{2008MNRAS.388..945B}, which is based on
the SMF and infers baryonic mass via the stellar mass-metallicity
relation and other scaling relations between metallicity and stellar
mass fraction (solid black line). This inferred BMF is constructed to
include all baryonic components (including stars, cold, and warm gas),
although we note that it follows the shape of the SMF, from which it
is derived, until rising steeply below M$_{bary}$ $<$ 10$^{9.7}$
\msun.

\subsubsection{Divergence of SMF \& BMF}
\label{sec:compmstarmbary}

\begin{figure}
\epsscale{1.2}
\plotone{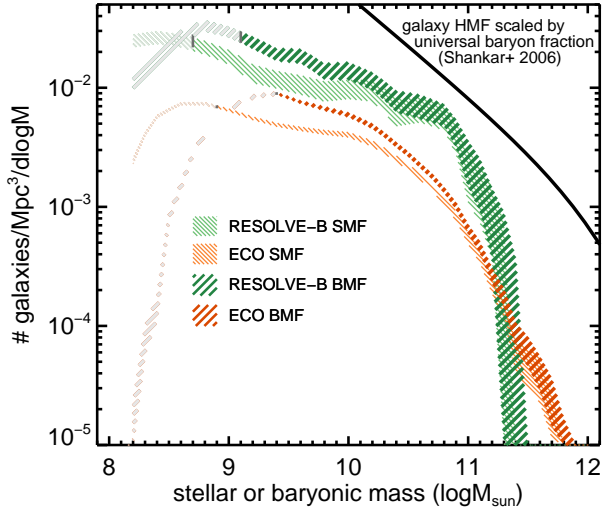}
\caption{Direct comparison of the RESOLVE-B and ECO SMF and BMF using
  the cross-bin sampling technique.  The SMF and BMF are plotted over
  each other for both data sets using the same color scheme as in
  Figure \ref{fg:litcomparesmfbmf}.  The ECO mass functions have been
  scaled down by a factor of 2 for clarity and the incomplete regions
  are shaded gray. For both data sets, the BMF diverges from the SMF,
  beginning near the bimodality mass of M$_{bary}$ $\sim$ 10$^{10.5}$
  \msun{}, and becoming significant below the gas-richness threshold
  mass of M$_{bary}$ $\sim$ 10$^{9.9}$\msun. The galaxy HMF (including
  the contribution from subhalos) from \citet{2006ApJ...643...14S} is
  shown scaled by the universal baryon fraction, 0.15, for
  comparison.}
\label{fg:dircomparesmfbmf}
\end{figure}

To directly compare the RESOLVE-B and ECO SMFs and BMFs, we plot all
of the mass functions in Figure \ref{fg:dircomparesmfbmf}. The ECO
mass functions have been scaled down for clarity. We note that the
relative difference between the SMF and BMF are similar for both
RESOLVE and ECO, with the BMF $\sim$0.17 dex higher than the SMF at
\mbox{10$^{10}$ \msun}, rising to $\sim$0.25 dex higher than the SMF
at \mbox{10$^{9.5}$ \msun}.

It is apparent that the SMF and BMF are effectively the same at large
masses, thus dropping off at a similar knee of M$_{*}$ $\sim$
10$^{10.8}$ \msun.  This result is not surprising since the drop-off
mass scale is above the bimodality mass scale of $\sim$10$^{10.5}$
\msun{} identified in \citet{2003MNRAS.341...54K}, above which
galaxies tend to be bulge-dominated, red, and quenched of star
formation with little to no cold gas (K13).

Going down across the bimodality mass ($\sim$10$^{10.5}$ \msun), we
find that the BMF and SMF start to diverge by $>$0.1 dex (albeit
within the error bars for RESOLVE-B).  The SMF plateaus while the BMF
rises as a straight power law. For RESOLVE-B the divergence between
the two mass functions becomes significant below a stellar or baryonic
mass of 10$^{9.7-9.9}$ \msun, the gas-richness threshold scale
identified in K13. Below the gas-richness threshold mass emerges a
significant population of galaxies that may have as much or more gas
than stellar mass. These gas rich galaxies fill in the plateau region
and push the BMF to rise as a straight power law.

At the lowest galaxy masses (M$_{star}$ $<$ 10$^{9.5}$ \msun), we see that
the SMF starts to rise up more steeply. We cannot
determine whether the BMF follows suit, as our
RESOLVE-B and ECO data sets are limited at M$_{bary}$ $\sim$
10$^{9.1}$ \msun{} and 10$^{9.4}$ \msun{} respectively.

Although the SMF and BMF at high masses are similar when comparing
within each data set, there are large differences between the mass
functions between the RESOLVE-B and ECO data sets. First, as can be
noted in Figures \ref{fg:litcomparesmfbmf} and
\ref{fg:ourcomparesmfbmf}, the RESOLVE-B mass functions are elevated
over the ECO mass functions. This relates to the overall higher
density of RESOLVE-B due to cosmic variance (as described in \S
\ref{sec:cvofds}). Second, while the ECO SMF and BMF decline
gradually, the RESOLVE-B stellar and BMFs drop-off much more abruptly.
The abruptness of the turnover in RESOLVE-B gives the appearance of a
``bump'' at the high mass end of the SMF and BMF. We explore these
differences and tie them to the different group halo mass
distributions sampled within ECO and RESOLVE-B in \S
\ref{sec:condmfs}.

\subsubsection{Relationship of Observed Galaxy to Theoretical HMFs}
\label{sec:comptohalo} 

Finally, we compare the RESOLVE-B and ECO SMFs and BMFs to the galaxy
HMF derived in \citet{2006ApJ...643...14S}, which includes the
contribution from subhalos and removes the subhalo contribution to
large group halos. The galaxy HMF, which has been scaled by the
universal baryon fraction of $\sim$0.15 \citep{2014A&A...571A..16P} to
enable direct comparison to galaxy masses, has a steep slope with
$\alpha_{HMF}$ = $-1.84$ and is shown as a thick solid line
in Figure \ref{fg:dircomparesmfbmf}. Although the low-mass slope of
the BMF is steeper than that of the SMF, it is not nearly as steep as
the galaxy HMF slope. We discuss this result in more detail in \S
\ref{sec:discussion}.

\subsection{Conditional SMF and BMF}
\label{sec:condmfs}

While there are many similarities between the RESOLVE-B and ECO SMF
and BMF, Figures \ref{fg:litcomparesmfbmf} and
\ref{fg:ourcomparesmfbmf} reveal a significant difference in their
shapes at high stellar/baryonic mass.  One possible explanation is the
different group halo mass distributions, since the volumes of
both surveys are too small to escape cosmic variance. For example
RESOLVE-B, unlike ECO, has no high mass clusters $>$10$^{13.5}$
\msun. It does, however, have an overabundance of clusters of mass
$\sim$10$^{12.5}$ \msun{} and 10$^{13.5}$ \msun{} compared to ECO (see
Figure \ref{fg:halodist}).

In this section we investigate how the shape of the mass function
depends on group halo mass. We first define physically motivated halo
mass regimes, then break down the mass functions within each halo mass
regime. We then further break down the mass functions in each group
halo mass regime into the central and satellite components. Finally,
we analyze whether the high-mass discrepancy between RESOLVE-B and ECO
can be explained by distinct group halo mass distributions.

To ensure that group finding and in particular that the choice of
linking lengths described in \S \ref{sec:llchoice} does not drive our
results, we have performed the following analysis with the linking
lengths of \citet{2006ApJS..167....1B} (see \S \ref{sec:llchoice}). We
do not find any significant differences with the results presented in
this section using these alternate linking lengths.

\subsubsection{Definition of Group Halo Mass Regimes}
\label{sec:condmfsebins}

To examine the SMF and BMF in different group halo mass regimes, we
use the group identifications and masses described in \S
\ref{sec:halomass} to divide the RESOLVE-B and ECO data sets into four
group halo mass regimes.  These group halo mass regimes are:
1)~M$_{halo}$~$<$~10$^{11.4}$ \msun{} ``low-mass group,'' 2)
M$_{halo}$ between 10$^{11.4}$ \msun{} and 10$^{12.0}$ \msun{}
``intermediate-mass group,'' 3) M$_{halo}$ between 10$^{12.0}$\msun{}
and 10$^{13.5}$ \msun{} ``large group,'' and 4) M$_{halo}$ $>$
10$^{13.5}$ \msun{} ``cluster,'' which applies only to ECO.

The low-mass group regime includes all group halos below
M$_{halo}$~$=$~10$^{11.4}$~\msun, which is the group halo mass that
roughly corresponds to the gas-richness threshold mass identified in
K13 as 10$^{9.7}$ \msun{} in stellar mass and \mbox{10$^{9.9}$
  \msun{}} in baryonic mass (see Figure \ref{fg:halodist}b).  Galaxies
in halos with masses $<$10$^{11.4}$ \msun{} are generally low-mass
central galaxies of comparable mass to the Large Magellanic Cloud
\citep{1998ApJ...503..674K} and have significant amounts of gas,
resulting in a large increase in their baryonic masses compared to
their stellar masses. It should be noted that we have not included a
lower halo mass floor in defining this regime and that the lowest
extrapolated halo included mass is $\sim$10$^{10.5}$ \msun. Such
low-mass halos mostly indicate low-mass galaxies living in halos by
themselves, at least down to our sample limits.

The intermediate-mass regime ranges from M$_{halo}$ =
10$^{11.4}$--10$^{12.0}$ \msun, which roughly corresponds to the
central galaxy bimodality mass $\sim$~10$^{10.5}$/10$^{10.6}$~\msun{}
in stellar and baryonic mass (\citealp{2003MNRAS.341...54K} and also
K13). In this regime, we find nascent groups with only a few members
(see Figure \ref{fg:halodist} and Figure 8 from M15). Above group halo
mass of \mbox{$\sim$10$^{12.0}$ \msun{}} marks a transition in the
central galaxy mass to halo mass relationship, where for large groups
growth of the integrated galaxy mass in the halo becomes more
dependent on the satellite inventory than central mass growth
\citep{2009ApJ...696..620C,2010ApJ...717..379B,2011ApJ...738...45L,2012ApJ...744..159L,2013ApJ...770...57B}. As
shown in Figure \ref{fg:halodist}b, for nascent groups with mass
\mbox{$<$10$^{12.0}$ \msun}, central galaxies still have appreciable
amounts of gas with a minimal satellite population.  For large group
halos with mass \mbox{$>$10$^{12.0}$ \msun}, the halos start filling
up with satellites and the cold gas becomes less and less important to
the overall baryonic mass of the central.

Our last group halo mass division is placed at
M$_{halo}$~$=$~10$^{13.5}$~\msun{} between the large group and cluster
regimes. While groups less massive than 10$^{13.5}$ \msun{} tend to
live in a range of large scale structure overdensities, clusters above
this halo mass division reside in the most dense structures
\citep{2013ApJ...776...71C}.  Additionally, the group LF
characteristic mass and faint-end slope values converge for groups
more massive than \mbox{10$^{13.5}$ \msun{}}
\citep{2006ApJ...652.1077R}.  In RESOLVE-B, there are no halos more
massive than \mbox{10$^{13.5}$ \msun}, so this cluster regime only
applies to ECO, which includes the Coma cluster.

\subsubsection{SMF and BMF by Group Halo Mass (``Conditional Mass Functions'')}
\label{sec:condmfsineachbin}

\begin{figure*}
\epsscale{1.}
\plotone{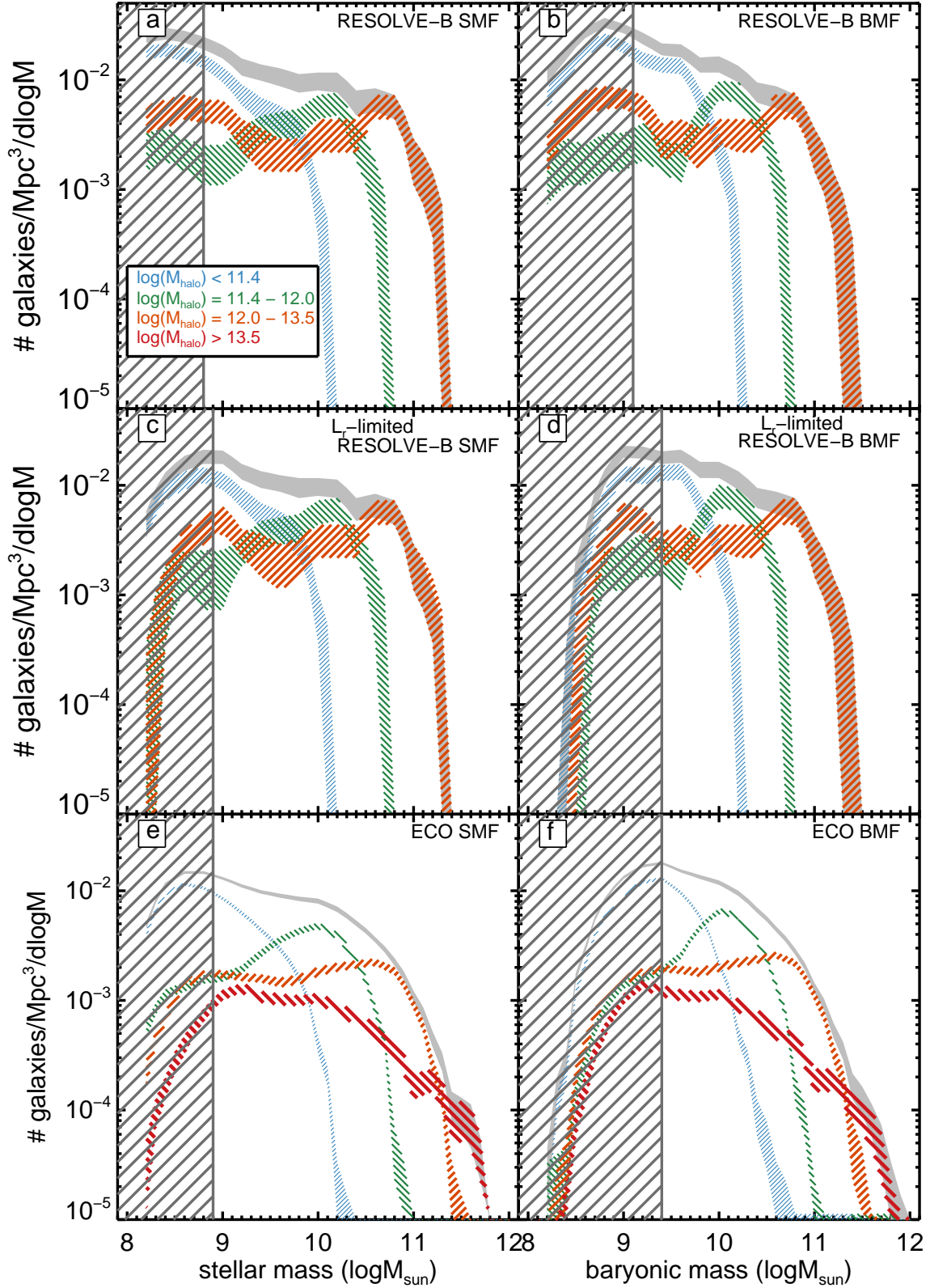}
\caption{Breakdown of RESOLVE-B (top row), L$_{r,tot}$-limited
  RESOLVE-B (middle row), and ECO (bottom row) SMF and BMF into
  different group halo mass regimes. The four group halo mass regimes
  are M$_{halo}$ $<$ 10$^{11.4}$ \msun{} (low-mass groups, often solo
  centrals, blue), M$_{halo}$ between 10$^{11.4}$-10$^{12}$ \msun{}
  (intermediate-mass groups, green), M$_{halo}$ between
  10$^{12}$-10$^{13.5}$ \msun{} (large group, orange), and
  \mbox{M$_{halo}$ $>$ 10$^{13.5}$ \msun{}} (cluster, red, ECO
  only). The dark gray marks the overall SMF or BMF for each data set,
  and the gray hash marked region denotes the incomplete regions for
  each mass function. The conditional mass functions are much more
  complex than the overall mass function, with pronounced bumps and
  dips.}
\label{fg:smfbmfcond}
\end{figure*}

Breaking down the RESOLVE-B and ECO mass functions into these four
group halo mass regimes, shown in Figure \ref{fg:smfbmfcond}, reveals
complex structure within the overall galaxy mass functions. For
instance in increasingly higher halo mass regimes, we observe high
mass drop-offs that occurs at higher galaxy mass. These drop-offs mark
the natural boundary in the largest central galaxy mass for a given
halo mass as seen in Figure \ref{fg:halodist}b.

In both the RESOLVE-B and ECO intermediate (green) and large (orange)
group halo mass regimes, we observe a peak in the SMF and BMF.  For
intermediate group halos, this peak occurs at $\sim$10$^{10.2}$
\msun, right between the gas-richness threshold and bimodality mass
scales.  For large group halos, the peak occurs at \mbox{$\sim$10$^{10.8}$
\msun}, above the bimodality mass scale and also near the knee of the
overall mass functions. The peak in the large group halo regime
appears to be causing the pronounced ``bump'' seen in the overall
RESOLVE-B SMF and BMF.  While the ECO data set also has this
characteristic peak in its large group halo regime mass functions, the
more gradual decline in the overall SMF and BMF for large galaxy
masses seems to be due to the cluster galaxy population in ECO. The
cluster population makes up 17\% of galaxies with \mbox{M$_{star}$ $>$
  10$^{10.5}$ \msun{}} and 31\% of galaxies with \mbox{M$_{star}$ $>$
  10$^{11}$ \msun{}} for ECO. The RESOLVE-B ``bump,'' meanwhile, may
be emphasized by the lack of a cluster galaxy population and the
overabundance of halos in this large group halo regime as shown in
Figure \ref{fg:halodist}a.


It is clear that the low-mass slopes of the SMF and BMF in these
different halo mass regimes are very different. In the cluster halo
mass regime, the rise is quite smooth, although the slope appears to
flatten out at lower masses, at least in part due to incompleteness in
galaxy counts around the clusters caused by high fiber collision rates
and by missing ultra-diffuse galaxies like those recently found in the
Coma and Virgo clusters
\citep{2015ApJ...798L..45V,2015ApJ...809L..21M}. We note that
\citet{2012AJ....144...40Y} found a fairly flat slope for the Coma
cluster LF down to galaxy magnitudes of M$_{R}$ = $-14.0$, with a
steep upturn for galaxies below our luminosity limit. Also, not all
previous studies of cluster LFs have found steep slopes, and many
cluster LF studies have relied on using statistical counts to remove
background galaxies (e.g., \citealp{1978ApJ...223..765D} and
\citealp{2002PASJ...54..515G}).

In the intermediate and large group halo mass regimes, both RESOLVE-B
and ECO have an intriguing fall-off in galaxy number density for
galaxy masses below the peak. While the intermediate group halo mass
regime does not show evidence for a low-mass upturn, the RESOLVE-B
large group halo mass regime does have a steeply rising low-mass
slope. This steeply rising slope is not as apparent in ECO, but it is
shallower. To examine this discrepancy further, we show a version of
RESOLVE-B limited to M$_{r,tot}$ $<$ $-17.33$ to be consistent with
ECO in panels c and d of Figure \ref{fg:smfbmfcond}, and we find that
the low-mass slope in the large group halo mass regime appears less
steep but is still elevated above the intermediate halo mass regime
mass functions. We posit that the difference is due to the
overabundance of such large group mass halos in ECO and investigate
further in \S \ref{sec:reconstruct}.  Finally the low group halo
regime mass functions show a steeply rising slope toward low galaxy
masses. This breakdown shows that the low-mass slope of the BMF across
different group halo mass regimes is not invariant. This result argues
against the conjecture from \citet{2003ApJ...585L.117B} that the
cluster and field BMFs might have similar low-mass slopes.


Finally, Figure \ref{fg:smfbmfcond} reflects the fact that low-mass
galaxies in RESOLVE-B and ECO live in low-mass, mainly isolated group
halos more often than in larger group mass halos. The crossover mass
scales (M$_{star}$ $<$ 10$^{9.5}$ \msun{} and M$_{bary}$ $<$
10$^{9.8}$ \msun) are roughly corresponding with the gas-richness
threshold mass from K13. Thus in the ``high mass dwarf'' regime we
probe, most dwarfs below the threshold mass live in low mass halos
rather than as members of clusters and large groups.



\subsubsection{Central Galaxy Mass Functions in Each Group Halo Mass Bin}
\label{sec:cencondmfsineachbin}

\begin{figure*}
\epsscale{1.2}
\plotone{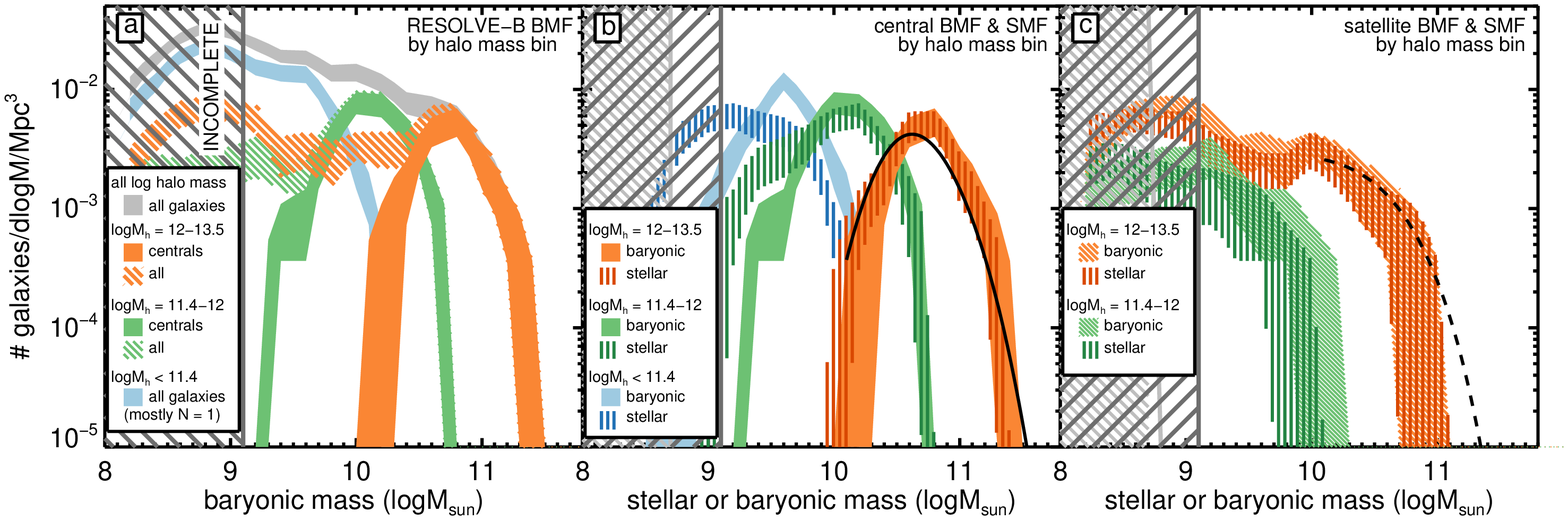}
\caption{RESOLVE-B SMF and BMF broken down by group halo mass regime and central
  vs.\ satellite designation. (a) RESOLVE-B BMF with conditional mass
  functions in cross-hatch and central mass functions shown in
  solid. (b) RESOLVE-B central galaxy SMF (darker cross-hatch) and
  central galaxy BMF (solid). For the central galaxy mass functions, we
  have applied a halo mass floor in the low-mass halo regime at
  M$_{halo}$ = 10$^{11.1}$ \msun. (c) RESOLVE-B satellite galaxy SMF
  (darker cross-hatch) and satellite galaxy BMF (lighter
  cross-hatch). The low-mass slope in the intermediate group halo mass
  regime and the dip in the large group halo mass regime are seen in
  the satellite population. Incomplete regions are shaded in dark gray
  for baryonic mass and light gray for stellar mass.  In the large
  group regime we show the central (solid black line) and satellite
  (dashed black line) conditional SMFs for mock catalogs from
  \citet{2013ApJ...771...30R}, which are in rough agreement with our
  observed mass functions. }
 \label{fg:resbsmfbmfcentsat}
\end{figure*}

\begin{figure*}
\epsscale{1.2}
\plotone{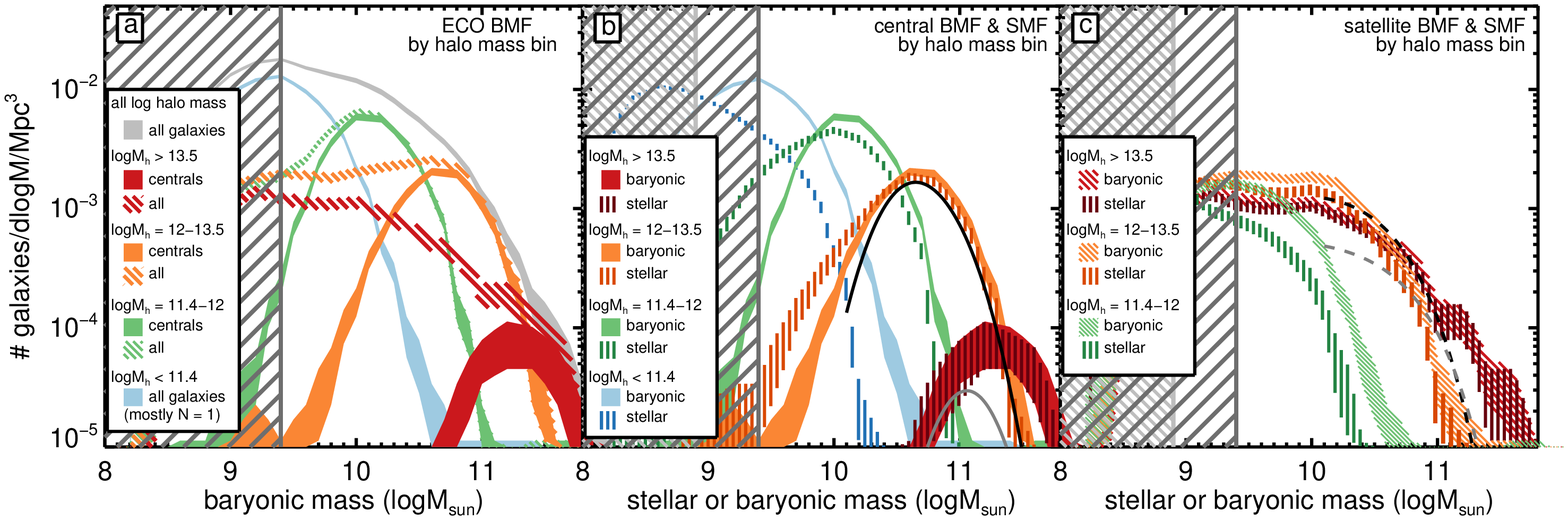}
\caption{Same as Figure \ref{fg:resbsmfbmfcentsat} but for ECO. For
  the central galaxy mass functions, we have applied a halo mass floor
  in the low-mass halo regime at M$_{halo}$ = 10$^{11.1}$ \msun. In
  the large group/cluster regime we show the central (solid black/gray
  lines) and satellite (dashed black/gray lines) SMFs for mock
  catalogs from \citet{2013ApJ...771...30R}. While the
  \citet{2013ApJ...771...30R} large group regime central and satellite
  SMFs are similar to our data, the cluster central galaxy SMF is
  significantly narrower and the satellite galaxy SMF is offset
  significantly lower than the observed ECO cluster SMF.}
 \label{fg:ecosmfbmfcentsat}
\end{figure*}

Next we examine how the central galaxy population in each group halo
mass regime affects the shape of the SMF and BMF by breaking up the
conditional mass functions according to central and satellite
designation for RESOLVE-B and ECO in Figures
\ref{fg:resbsmfbmfcentsat} and \ref{fg:ecosmfbmfcentsat} respectively.

The central galaxy SMFs and BMFs (center panels of Figures
\ref{fg:resbsmfbmfcentsat} and \ref{fg:ecosmfbmfcentsat}) appear in
discrete, narrow ``humps'' whose peak mass value increases with
increasing group halo mass. The centers of these humps correspond with
the peaks seen in the conditional mass functions. This trend is not
surprising in the context of the central galaxy mass to halo mass
relationship shown in Figure \ref{fg:halodist}b, which follows a
monotonic trend. The large drop-off in numbers of centrals in cluster
environments (red, ECO only) underscores the rarity of such large
objects.

Next we examine the difference between the central galaxy SMF and
BMF. It is apparent that in the low group halo mass regime, there is a
significant shift ($\sim$0.5 dex) in the location of the peak of the
central SMFs and BMFs. We have deliberately set the low group halo
mass regime to select groups with central galaxies that are below the
gas-richness threshold mass, which typically have as much or more
neutral gas mass as their stellar mass (K13). The large shift in the
peak of the central galaxy SMF and BMF in the low halo mass regime
simply underscores the importance of including cold gas mass when
considering the masses of low-mass central galaxies.

For larger halo mass regimes, the shift in the location of the peak
for the central galaxy SMF and BMF is much smaller. In the
intermediate halo mass regime, the central stellar and baryonic mass
peaks are $\sim$10$^{10.2}$ \msun{}, although we note that the central
BMF hump becomes noticeably narrower, and thus the lower mass centrals
within this regime still have a significant amount of gas. In the
large group halo mass regime, the central stellar and baryonic mass
peaks are located at $\sim$10$^{10.7}$ \msun{}, and in the cluster
halo mass regime the central mass peaks are located at
$\sim$10$^{11.3}$ \msun{} (the shutdown galaxy mass scale of K13). The
lack of a shift in the peak values of the SMF and BMF reflects that
central galaxies in these larger groups and clusters do not have any
significant cold gas mass, which is expected since they are above the
bimodality mass and they are in dense environments (see e.g.,
\citealp{1984ARA&A..22..445H,1973MNRAS.165..231D}; M15).



\subsubsection{Satellite Galaxy Mass Functions in Each Group Halo Mass Bin}
\label{sec:satcondmfsineachbin}

While the central mass functions follow a pattern of discrete humps,
the satellite mass functions exhibit much more complex structure. In
the right panels of Figures \ref{fg:resbsmfbmfcentsat} and
\ref{fg:ecosmfbmfcentsat}, we show these complex satellite galaxy mass
functions that define the low-mass slopes of the conditional mass
functions. For the low-mass halo regime, we do not show the satellite
galaxy mass functions since there are so few satellites in this regime
with masses greater than the mass completeness limits. We note that
since group finding is not perfect (as discussed in \S
\ref{sec:llchoice}), the satellite mass functions are subject to
issues of purity and completeness of the group halo catalog. As
described previously, however, performing this analysis with alternate
linking lengths yields similar results and we do not think these
issues significantly affect our results.

In the intermediate group halo mass regime (green), we see that the
observed flat low-mass slope described in \S \ref{sec:condmfs} is due
to the satellite population. We also observe that the satellites still
have a gas component, as the BMF is shifted toward higher mass than
the SMF. The shift, however, is not as extreme as for the central
galaxy mass function in low group mass halos (blue), which have
similar stellar masses as the satellites of the intermediate group
halo mass regime but are more gas-rich.

In the large group halo mass regime (orange), the satellite galaxy
mass function has a dip (or possibly flat segment) just below
10$^{10}$ \msun, even in baryonic mass. For RESOLVE-B, we then find
that the satellite galaxy mass function starts to rise again below,
10$^{9.7}$ \msun{}, although this rise is not evident for ECO. We also
note that for more massive satellite galaxies in large group halos,
there is relatively little cold gas as the SMF and BMF are very
similar. For galaxies with mass $<$10$^{10}$ \msun, the baryonic mass
function is shifted toward slightly higher masses, indicating that
some lower mass galaxies do retain a cold gas reservoir.

In contrast, the ECO cluster satellite mass function rises more
smoothly, although it flattens below galaxy masses of $\sim$10$^{10}$
\msun. The SMF and BMF are essentially the same at all masses,
indicating that the satellites of such large clusters have very little
to no cold gas. Although the satellite galaxy mass function in the
cluster appears to have a smoother shape than in the lower halo mass
regimes, the cluster should be made up of what were originally smaller
groups that have fallen into the larger potential well over time. Thus
all the more intricate shapes of the smaller group halos have combined
to form the cluster regime's smoother shape.




The substructure seen in these satellite galaxy mass functions,
particularly for intermediate and large group halo masses, may arise
from the formation of groups. While there is a $\sim$0.3--0.4 shift
between the SMF and BMF of galaxies in the low group halo mass regime
indicating large amounts of cold gas in such galaxies, we find a much
smaller shift between the satellite SMF and BMF at similar galaxy
stellar masses in the intermediate and large group halo mass
regimes. This result suggests that gas-removal processes such as
ram-pressure/viscous stripping
\citep{1972ApJ...176....1G,1982MNRAS.198.1007N} or
starvation/strangulation \citep{1980ApJ...237..692L} may already begin
even in the intermediate group halo mass regime. This lends support
for pre-processing, the idea that galaxies begin to be quenched in
smaller groups before falling into the cluster
\citep{1998ApJ...498L...5Z}. Perhaps even more intriguing are the dips
and varying low-mass slopes seen in the satellite galaxy mass
functions. Conditions within early group formation, such as low
velocity dispersion among group members, may promote merging at
preferred mass scales \citep{2014ApJ...797..127P}, completely removing
satellite galaxies from the overall galaxy population. As these groups
fall into larger clusters, where the velocity dispersion increases,
the satellite galaxies are less likely to merge, resulting in the
smoother appearance of the cluster satellite galaxy mass function.

\subsubsection{Comparison to Previous Work}
\label{sec:comparisonprev}

Previous studies of conditional mass functions in both data and models
have found similar results, fitting the central contribution of the
conditional mass function as a log-normal distribution and the
satellite mass function as a Schechter or truncated Schechter function
\citep{2005ApJ...633..791Z,2009ApJ...695..900Y,2010ApJ...710..903M,2013ApJ...771...30R}. These
works, however, cover only the higher halo mass regimes $>$10$^{12}$
\msun{}. In the comparisons below, we have corrected all other
conditional mass functions to be in units of H$_0$=70 \kms.

We directly compare with the central SMFs of
\citet{2013ApJ...771...30R}, who used the halo abundance matching
technique to assign stellar masses to their simulated halos. To
compare with the observed large group and cluster mass regimes, we
multiply each of their conditional SMFs, which are subdivided more
finely than our own, by the appropriate number of halos in either the
RESOLVE-B or ECO data set, coadding the resulting functions in each of
our halo mass regimes, and then dividing by the volume of the
appropriate data set.  The results of this process are shown for the
large group halo mass (black) and cluster (gray) regimes in
Figures \ref{fg:resbsmfbmfcentsat} and \ref{fg:ecosmfbmfcentsat}.  The
overplotted central SMFs are in good agreement with our observed
central SMFs, although for the cluster regime, the observed SMF has a
wider spread and higher peak mass value than that from
\citet{2013ApJ...771...30R}.  Within \citet{2013ApJ...771...30R} there
is a comparison with the results of \citet{2009ApJ...695..900Y},
finding that while qualitatively similar, the central SMFs of
\citet{2009ApJ...695..900Y} are offset toward higher masses, most
likely due to the difference in stellar mass estimation.  While
\citet{2013ApJ...771...30R} use $kcorrect$ from
\citet{2007AJ....133..734B}, \citet{2009ApJ...695..900Y} use the
stellar mass prescription from \citet{2003ApJS..149..289B}. Given that
we are in good agreement with \citet{2013ApJ...771...30R} and also the
stellar mass discussion in \S \ref{sec:mstar}, we expect a similar
offset with \citet{2009ApJ...695..900Y}.

We also compare our mass functions with the theoretical central SMFs
of \citet{2010ApJ...710..903M}, who used a stellar mass to galaxy halo
mass relationship and stellar mass dependent clustering of galaxies to
assign stellar masses to halos in simulations, constraining the
parameters of these functions by comparing their assigned SMF with the
SMF from \citet{2007MNRAS.378.1550P}. The central galaxy SMFs in
\citet{2010ApJ...710..903M} are binned more finely than in this work,
so we examine the range of peak masses for each halo mass regime.  The
\citet{2010ApJ...710..903M} intermediate halo mass regime has a
stellar mass peak ranging between $\sim$10$^{9.6}$ to 10$^{10.6}$
\msun, encompassing our central mass peak of $\sim$10$^{10.2}$
\msun. In the large group and cluster mass regimes,
\citet{2010ApJ...710..903M} measure stellar mass peak ranges that are
offset toward slightly larger stellar masses than we observe: a range
from \mbox{10$^{10.6}$ \msun{}} to \mbox{10$^{11.3}$ \msun{}} for the
large groups and a peak of 10$^{11.5}$ \msun{} for clusters, while we
observe a peak at 10$^{10.7}$ \msun{} for large groups and 10$^{11.3}$
\msun{} for clusters.




We can also compare the RESOLVE-B and ECO central BMFs with the
theoretical central BMFs predicted in \citet{2005ApJ...633..791Z},
which used an SPH simulation together with the GALFORM semi-analytic
model of galaxy formation \citep{2000MNRAS.319..168C} to measure the
conditional BMF (here baryonic mass includes both the cold atomic and
molecular gas components in addition to the stars).  Comparing the
central mass peaks from the \citet{2005ApJ...633..791Z} BMFs, we find
a range from 10$^{10.5}$ to 10$^{11.3}$~\msun{} for large groups and a
range from 10$^{11.4}$ to 10$^{11.6}$ \msun{} for clusters.  While the
baryonic central mass peak that we measure for large group mass halos
$\sim$10$^{10.7}$ \msun{} is within the range from
\citet{2005ApJ...633..791Z}, in the cluster regime, their peak masses
are slightly ($\sim$0.2 dex) higher than ours ($\sim$10$^{11.3}$
\msun) for clusters.  Similar discrepancies have been noted in the
analysis of \citet{2010ApJ...712..734L}, which compares the galaxy
stellar mass to halo mass relationship predicted by SAMs with
observations, but as we have few massive clusters, we cannot rule out
cosmic variance (see \S \ref{sec:mffits}).

For the satellite mass functions, we can directly compare with those
from \citet{2013ApJ...771...30R}. Our satellite mass functions
typically drop off in numbers around the center of the peak value of
the central galaxy mass function, except in the cluster regime where
the most massive satellites start to outnumber the few centrals
associated with these rare halos. In Figures \ref{fg:resbsmfbmfcentsat} and \ref{fg:ecosmfbmfcentsat}
we show the \citet{2013ApJ...771...30R} satellite galaxy mass
functions in the large group halo mass regime as a black dashed line
and in the cluster regime as a gray dashed line. The large group halo
mass satellite mass functions generally agree with the results in
RESOLVE-B and ECO, although this work could not probe the interesting
substructure that occurs below galaxy mass of $\sim$10$^{10}$
\msun. In the cluster regime, the satellite mass functions do not
agree with our results, which may be related to the discrepancy in the
central mass function.

Overall we find qualitative agreement with previous work in examining
the central and satellite mass functions in different
bins. Differences in stellar mass estimation and prescriptions for
populating mock catalogs may yield offsets in the peak mass values,
but the general pattern for the central conditional mass functions is
evident. With RESOLVE-B we are able to extend our analysis to lower
halo masses than most previous work, finding intriguing patterns in the
intermediate and low group halo mass regimes.


\subsubsection{Mass Function Reconstruction}
\label{sec:reconstruct}

Returning to the problem of the high-mass discrepancy between the
RESOLVE-B and ECO mass functions, we can now examine whether the
overall mass function of a data set simply reflects the survey's group
halo mass distribution. We have seen that it is possible to scale the
mock catalog conditional mass functions from
\citet{2013ApJ...771...30R} by the appropriate number of halos to
match the observed conditional mass functions we observe (\S
\ref{sec:comparisonprev}). If the particular halo mass regime sets the
shape of the mass function, then given a basis set of ``per group
halo'' mass functions and a group halo mass distribution for a survey,
we should be able to reconstruct the observed galaxy mass function of
the survey.


\begin{figure*}
\epsscale{1.}
\plotone{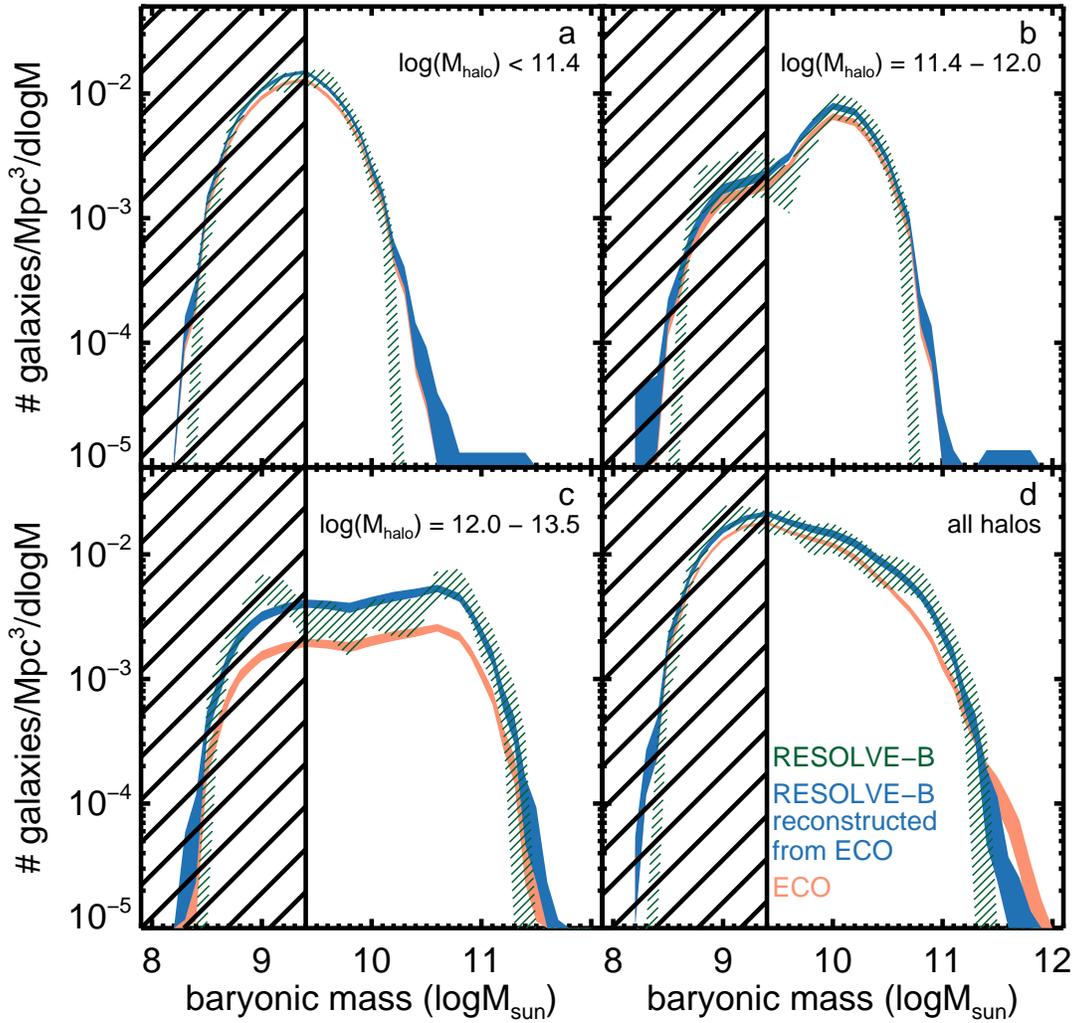}
\caption{Reconstruction of the RESOLVE-B BMF using the ECO conditional
  mass functions.  Panels a-c show the conditional BMFs in
  progressively higher mass halo regimes.  The RESOLVE-B BMFs are
  constructed from an L$_{r,tot}$-limited data set (M$_{r,tot}$ $<$
  $-17.33$) to be consistent with ECO and are shown in cross-hatched
  green. The original ECO BMFs are shown in solid light-pink. The
  reconstructed RESOLVE-B BMFs based on scaling the ECO BMFs catalog
  are shown in cross-hatched blue. The difference in abundance of
  large group halos in RESOLVE-B and ECO is very apparent in panel c,
  but the overall shape of the large group mass functions is
  similar. In panel d, we add the three conditional BMFs based on ECO
  (blue) to compare with the overall RESOLVE-B BMF (green). The
  reconstructed BMF matches that of RESOLVE-B very well. The observed
  ECO BMF is offset lower near the large group bump in RESOLVE-B and
  extends to higher masses since it includes the cluster regime not
  represented in RESOLVE-B.}
\label{fg:reconstbmf}
\end{figure*}


To determine whether such a reconstruction would reconcile RESOLVE-B
and ECO, we first derive a basis set of BMFs from ECO. We
start with the completeness-corrected BMFs from the
four halo mass regimes defined in the previous section, then normalize
these four conditional mass functions by the number of group halos in
each halo mass regime. Since the ECO mass functions have been
completeness corrected, we use the weighted total of centrals in each
halo mass regime as the total number of halos in each halo mass
regime. We next determine the number of group halos in each group halo
mass regime for the data set that we want to reconstruct: RESOLVE-B in
this case. For this analysis, we limit RESOLVE-B to galaxies brighter
than $-17.33$ in order to match the selection for ECO. Since RESOLVE-B
has no halos with mass $>$10$^{13.5}$ \msun, we consider only the
low, intermediate, and large group halo mass regimes from ECO, then
multiply the ECO basis functions by the number of RESOLVE-B halos
in each group halo mass regime.

The resulting RESOLVE-B and scaled ECO basis BMFs for each halo mass
regime are shown in panels a-c of Figure \ref{fg:reconstbmf}, where
the RESOLVE-B conditional mass functions are shown in green and the
scaled ECO basis mass functions are shown in blue. The raw ECO mass
functions are shown in pink for comparison. To create the total BMF,
we simply add the mass functions in each halo mass regime to compare
with the actual data.  We find good agreement between the observed and
reconstructed RESOLVE-B BMFs, including the bump near $\sim$10$^{11}$
\msun{} and the overall normalization.

The ability to reconstruct the RESOLVE-B BMF from the ECO basis BMFs
suggests that the shape of the mass function for a given survey is
dependent on the particular group halo mass distribution sampled. This
conclusion is not to say that large scale structure does not play a
role, since the group halo mass distribution depends on the large
scale structure. We infer that the group halo mass distribution
contributes to differences between the mass functions observed in
different studies, particularly for small surveys.  For larger data
sets where the halo distribution closely resembles the true halo mass
distribution, the mass functions should all tend toward the same
shape.  For smaller data sets, like RESOLVE-B, we expect to find
interesting differences in the mass function, such as the prominent
high mass bump in Figure \ref{fg:reconstbmf}c. This result also
supports the claim in \citet{2010MNRAS.408.1113F} that the high-mass
bump seen in other galaxy mass functions is primarily due to the
central galaxy populations, and in particular centrals from the large
group halo mass regime. The apparent contribution of a bump from red
galaxies \citep{2010A&A...524A..76B}, then, is due to the fact that
centrals in high mass halos are more likely to be red.

\section{Discussion}
\label{sec:discussion}

In this section, we discuss implications of our results for the
physics of group formation. We also revisit the discrepancy in
low-mass slope between the observed galaxy and theoretical halo mass
function in light of our analysis.

\subsection{How Group Halo Mass Environment Shapes the Galaxy Mass Function}
\label{sec:shape}

In \S \ref{sec:condmfs}, we showed that underneath the relatively
simple shape of the overall galaxy mass function, the conditional
galaxy mass functions have much more complex structure. These include
regular humps from central galaxies and both dips and varying low-mass
slopes from satellite galaxies (e.g., a flat slope in
intermediate-mass halos and a steeply rising slope in large group
halos). Furthermore, we showed that differences in the overall mass
function of different surveys, such as the prominent bump around
$\sim$10$^{11}$ \msun{} in RESOLVE-B, can be explained by scaling a
basis set of conditional mass functions by the appropriate group halo
mass distribution (see \S \ref{sec:reconstruct} and Figure
\ref{fg:reconstbmf}). Together, these observations suggest that groups
have a profound effect on shaping the galaxy population. In
particular, group formation processes, such as merging and stripping,
appear to occur from the onset of nascent group formation in
intermediate-mass halos and create the complex structure seen in the
conditional mass functions.

In the intermediate and large group halo mass regimes groups still
have relatively few members. In the low group halo mass regime, with
M$_{halo}$ $<$ 10$^{11.4}$ \msun, $>$90\% of our groups are N=1 galaxy
systems consisting only of a central galaxy without any satellites
brighter than the luminosity completeness limit used for group
finding. Transitioning into the intermediate and large group halo mass
regimes (M$_{halo}$ between 10$^{11.4}$ \msun{} and 10$^{13.5}$
\msun), we find a population of small groups consisting of 2 to 7
members. In the cluster regime (M$_{halo}$ $>$ 10$^{13.5}$ \msun),
groups consist of tens to hundreds of members.

Intermediate-mass groups are nascent groups, the site of galaxies'
first coming together before entering larger groups and clusters. The
flat low-mass slope seen in this first group halo mass regime suggests
that within nascent groups, satellite stripping and destruction is
already occurring and shaping the galaxy mass functions.  Mergers and
stripping remove galaxies from the galaxy population or shift them to
lower masses, depressing the mass function in the low-mass range. Only
relatively rare major mergers substantially increase galaxy masses,
while minor mergers repeatedly eliminate counts from the low-mass end
of the mass function.  Nascent groups may be preferred environments
for merging, as the relative speeds of the galaxies in these groups
are much smaller than in larger groups and clusters (as found in
\citealp{2014ApJ...797..127P}). These first groups (and more solitary
field galaxies) eventually combine into larger groups, so that
clusters are essentially made up of a range of smaller group
masses. Thus the cluster mass function represents a linear combination
of all of the smaller group mass functions, potentially smoothing the
dips and varying low-mass slopes into one broadly rising composite
that hides the complexity of its past. This scenario is supported by
prior work suggesting the existence of pre-processing, in which
galaxies begin quenching in smaller groups before falling into larger
clusters (e.g.,
\citealp{1998ApJ...498L...5Z,2009MNRAS.400..937M,2013MNRAS.432..336W,2015ApJ...806..101H}).
In follow-up work, we will show that the group integrated baryonic to
halo mass ratio is relatively flat over this intermediate halo mass
range, as these groups are efficiently converting gas to stars (Eckert
et al.\ in prep.).



\subsection{Comparison Between Galaxy and Halo Mass Functions}
\label{sec:discrepancy}

We now revisit the discrepancy between the low-mass slopes of the
overall galaxy SMF and BMF and the theoretical HMF, focusing on the
combined subhalo and group halo mass function, which should map to
satellites and centrals.  Based on the studies of the baryonic
Tully-Fisher relation, which reveal a tight correlation between galaxy
cold baryonic mass and rotation velocity even for low-mass galaxies
(e.g., \citealp{2000ApJ...533L..99M}), we might expect that the BMF
should trace the theoretical HMF scaled by the universal baryon
fraction. Examining Figure \ref{fg:dircomparesmfbmf}, we do in fact
find that the low-mass slope of the BMF is steeper than that of the
SMF over the mass range below the knee of the mass functions
($\alpha_{BMF}$ $\sim$ $-1.3$ vs.\ $\alpha_{SMF-plateau}$ $\sim$
$-1.2$). However, the BMF is still significantly shallower than the
universal baryon fraction scaled galaxy HMF including subhalos (see \S
\ref{sec:comptohalo} and Figure \ref{fg:dircomparesmfbmf}), which has
a low-mass slope of $\alpha_{HMF}$ $\sim$ $-1.84$
\citep{2006ApJ...643...14S}.


To alleviate this tension, we consider first that the baryonic rotation
velocity V$_{disk}$ may not be directly related to halo circular
velocity V$_{halo}$, since the halo extends much farther than the
optical galaxy.  Using a combination of lensing to determine halo
velocities and Tully-Fisher derived disk velocities,
\citet{2012MNRAS.425.2610R} find that for galaxies ranging in mass
from 10$^{9}$~\msun\ to 10$^{11}$ \msun\ the ratio of V$_{disk}$ to
V$_{halo}$ ranges from 1.27-1.39.  An abundance matching analysis,
however, finds the ratio of the V$_{disk}$ to V$_{halo}$ to be
$\sim$1.5, but only over a narrow velocity range from 110 \kms\ to 170
\kms, significantly underestimating V$_{halo}$ at lower masses
\citep{2011ApJ...739...38P}. Thus lensing and abundance matching
disagree, again returning us to the discrepancy in counts of low-mass
galaxies.

Assuming that the V$_{disk}$-V$_{halo}$ correspondence does hold, the
next consideration might be that the theoretical HMF coming from
N-body simulations is incorrect. This second possibility runs counter
to the body of evidence supporting the $\Lambda$CDM cosmological
framework including measurement of the fluctuations in the CMB
\citep{2003ApJS..148..175S,2014A&A...571A..16P} and successful
production of large scale structure by $\Lambda$CDM dark matter
simulations \citep{2005Natur.435..629S}.  Despite these successes,
however, the low-mass slope mismatch and other discrepancies between
observations and simulations (e.g., the core vs.\ cusp problem
reviewed in \citealt{2010AdAst2010E...5D}) have prompted widespread
investigation of different forms of dark matter, e.g., warm dark
matter, scalar field dark matter, etc.\
\citep{2014MNRAS.439..300L,2012JPhCS.378a2012M,2014PhRvD..90l3526V}.

An alternative to considering other forms of dark matter may be that
the HMF from N-body simulations is incorrect because the simulations
lack baryonic physics. Including baryonic physics of feedback may
re-distribute dark matter within halos, flattening the inner cuspy
profiles of dark matter halos. Very high resolution simulations of
low-mass galaxies reveal that even the smallest dwarf galaxies can
transform their dark matter cusps to cores as long as star formation
can proceed \citep{2015arXiv150804143R}. The authors argue that these
dark matter halo transformations may allow for halos to be tidally
destroyed more easily. Interpreting this idea in the context of
galaxies being destroyed in nascent groups (\S 6.1), we might consider
that current N-body simulations without baryonic physics are not
destroying enough subhalos, leading to an artificially steep low-mass
slope and also throwing off abundance matching. However, this answer
may be only partial as \citet{2015A&A...574A.113P} argue that there
is a missing dwarfs problem in the field, outside the group
environment.

A third consideration thus arises that cold atomic gas mass may
significantly underestimate the collapsed gas mass in dwarfs. Low-mass
galaxies often show notably smaller baryon fractions than expected
even when including atomic gas
\citep{2010ApJ...708L..14M,2012ApJ...759..138P}.  These extremely low
baryon fractions imply deviations from the baryonic Tully-Fisher
relation and indeed some studies find that multiplying the neutral gas
content by a factor of 3-11 can tighten the baryonic Tully-Fisher
relation further
\citep{2005A&A...431..511P,2008MNRAS.386..138B,2009A&A...501..171R}. These
results point toward the potential existence of large undetected gas
reservoirs of either ionized or ultra-cold molecular gas in low-mass
galaxies. If our observed baryonic masses significantly underestimate
true baryonic masses, the true BMF might be steeper more like the
theoretical HMF. We plan to examine this possibility in more detail in
a follow-up paper.

For completeness, we note the orthogonal possibility that gas loss
driven by feedback explains the slope discrepancy as lower mass
galaxies are less efficient at forming stars.  This scenario relies on
supernova feedback expelling gas from low-mass halos
\citep{1986ApJ...303...39D}, but based on realistic feedback energies,
even dwarf galaxies with masses as low as 10$^{7}$ \msun{} are able to
retain their gas \citep{1999ApJ...513..142M}.  A more likely candidate
to suppress star formation and reduce cold gas in low mass galaxies is
the strong ionizing UV background from the epoch of reionization,
leaving only early-forming halos able to form stars
\citep{2000ApJ...539..517B}. In this context
\citet{2003ApJ...585L.117B} conjecture a later re-cooling that could
produce an environment independent BMF, but our conditional mass
function results argue against this picture.

One final consideration is the issue of fly-by interactions or ejected
satellite galaxies \citep{2012ApJ...751...17S}, which have interacted
with a group or cluster but are no longer within that group or
cluster's halo. In \citet{2014MNRAS.439.2687W}, the authors find that
the stellar-to-halo mass relationship for fly-bys/ejected satellites
is much higher than usual, as the halo is stripped during its
encounter with the cluster halo. This halo stripping could increase
the numbers of low-mass dark matter halos (in simulations as well as
the real universe) relative to low-mass galaxies, further increasing
the mismatch in low-mass slope.



It is possible that there is an element of all of these considerations
play in explaining the differences between observed galaxy and
theoretical halo mass functions. In this work we have stressed the
role of nascent groups in shaping the galaxy population through
satellite stripping and/or merging.  In future work we will more
closely examine the possibility that unobservable gas can help to
reconcile the galaxy and integrated group BMFs with theoretical HMFs.


\section{Conclusions}
\label{sec:conclusions}

We have constructed two volume-limited data sets, RESOLVE-B and ECO,
for studying the galaxy SMF and BMF. The RESOLVE-B data set has
unprecedented completeness (see Figures \ref{fg:stripe82skydist}b and
\ref{fg:sbcomp}), which allows us to study the entire galaxy
population and produce empirical completeness corrections for the ECO
catalog. Using volume-limited surveys allows us to study mass
functions without the statistical completeness corrections necessary
for magnitude-limited surveys and to define groups of galaxies and
assign group halo masses via Friends-of-Friends group finding and halo
abundance matching. In \S \ref{sec:statmassfunctions} we present a
novel cross-bin sampling method for constructing galaxy mass functions
using the full stellar and gas mass likelihood distributions. Our
findings are:

\textbullet\ The SMF and BMF start to diverge for masses
$<$10$^{10.5}$ \msun{} and become significantly different below
$\sim$10$^{9.9}$ \msun{} (near the gas-richness threshold scale of
K13). The BMF rises as a straight power law, following the traditional
Schechter function form, while the SMF plateaus before appearing to
rise more steeply below $\sim$10$^{9.5}$ \msun{} (see \S
\ref{sec:compmstarmbary} and Figure \ref{fg:dircomparesmfbmf}).


\textbullet\ While steeper than the SMF's low-mass slope, the BMF's
low-mass slope is still much shallower (\mbox{$\alpha_{BMF}$ $\sim$
  $-1.3$}) than the predicted slope from theoretical HMFs, alleviating
some tension but not fully explaining the discrepancy (Figure
\ref{fg:dircomparesmfbmf}).


\textbullet\ The conditional SMF and BMF broken down into four halo
mass regimes have more complex structure than the overall galaxy mass
functions and reveal that the majority of low-mass galaxies are
centrals in low-mass halos without satellites above our survey limits
(see \S \ref{sec:condmfs} and Figure \ref{fg:smfbmfcond}).

\textbullet\ The conditional mass functions for central galaxies are
divided into narrow humps at discrete mass intervals as expected from the
monotonic relationship between galaxy mass and halo mass (see \S
\ref{sec:condmfs} and Figures \ref{fg:resbsmfbmfcentsat} and
\ref{fg:ecosmfbmfcentsat}).

\textbullet\ The low-mass slopes of the conditional satellite mass
functions vary significantly in different halo mass regimes. In larger
group halos the low-mass slope in RESOLVE-B rises quite steeply below
an initial dip, but the same feature is not clear at ECO's shallower
depth. In the intermediate group halo masses of nascent
multiple-galaxy groups, we find a flat low-mass slope (Figures
\ref{fg:smfbmfcond} and \ref{fg:resbsmfbmfcentsat}).

\textbullet\ These features seen in galaxy mass functions of
intermediate and large group halos suggest the possibility that even
in nascent groups, satellite merging and/or stripping are already
shaping the galaxy population. Recent work suggesting that the
theoretical HMF may be much shallower in hydrodynamic simulations that
include baryonic physics driving dark matter core formation and
facilitating tidal stripping \citep{2015arXiv150804143R} raises the
possibility that group formation and satellite destruction may help to
explain the discrepancy between the observed galaxy BMF and
theoretical HMF (see \S \ref{sec:discussion}).




\textbullet\ As evidence of the primacy of group-scale physics in
determining the galaxy mass function, we show that scaling a basis set
of conditional mass functions from the ECO data set by the group halo
mass distribution in RESOLVE-B recovers the shape of the RESOLVE-B
data set, including a bump due to centrals in large groups (Figure
\ref{fg:reconstbmf}).

In future work, we plan to compare the RESOLVE-B and ECO BMFs with
semi-analytic models and hydrodynamic simulations to explore the role
of group formation in shaping galaxy mass functions. We will also
study the integrated group SMF and BMF to examine the role of
unobserved gas in reconciling observed and theoretical mass functions.

\acknowledgements

We acknowledge the anonymous referee, whose comments have improved
this work. We would like to thank Tom Loredo and Michael Blanton for
productive conversations on statistical analysis of mass functions. We
are grateful for helpful insights and suggestions from Ashley Baker,
Jillian Bellovary, Chris Clemens, Gerald Cecil, Art Champagne,
Adrienne Erickcek, and Andrew Baker. We also thank Victor Calderon for
providing custom mock catalogs for the computation of cosmic variance
and group finding errors, and we thank Jessi Cisewski for helpful
discussions regarding cosmic variance. KE, SK, DS, and MN acknowledge
support from NSF CAREER grant AST-0955368. KE and DS were also
supported by GAANN Fellowships, KE, DS, and AM were supported by NC
Space Grant Fellowships. AM acknowledges support from the NASA Harriet
Jenkins fellowship. AM, DS, and KE were supported by the University of
North Carolina Royster Society of Fellows.

This work is based on observations from the SDSS. Funding for SDSS-III
has been provided by the Alfred P. Sloan Foundation, the Participating
Institutions, the National Science Foundation, and the U.S. Department
of Energy Office of Science. The SDSS-III web site is
http://www.sdss3.org/. SDSS-III is managed by the Astrophysical
Research Consortium for the Participating Institutions of the SDSS-III
Collaboration including the University of Arizona, the Brazilian
Participation Group, Brookhaven National Laboratory, Carnegie Mellon
University, University of Florida, the French Participation Group, the
German Participation Group, Harvard University, the Instituto de
Astrofisica de Canarias, the Michigan State/Notre Dame/JINA
Participation Group, Johns Hopkins University, Lawrence Berkeley
National Laboratory, Max Planck Institute for Astrophysics, Max Planck
Institute for Extraterrestrial Physics, New Mexico State University,
New York University, Ohio State University, Pennsylvania State
University, University of Portsmouth, Princeton University, the
Spanish Participation Group, University of Tokyo, University of Utah,
Vanderbilt University, University of Virginia, University of
Washington, and Yale University. This work is based on observations
made with the NASA Galaxy Evolution Explorer. GALEX is operated for
NASA by the California Institute of Technology under NASA contract
NAS5-98034.  This publication makes use of data products from the Two
Micron All Sky Survey, which is a joint project of the University of
Massachusetts and the Infrared Processing and Analysis
Center/California Institute of Technology, funded by the National
Aeronautics and Space Administration and the National Science
Foundation. This work is based in part on data obtained as part of the
UKIRT Infrared Deep Sky Survey.  This work uses data from the Arecibo
observatory. The Arecibo Observatory is operated by SRI International
under a cooperative agreement with the National Science Foundation
(AST-1100968), and in alliance with Ana G. Méndez-Universidad
Metropolitana, and the Universities Space Research Association. This
work is based on observations using the Green Bank Telescope. The
National Radio Astronomy Observatory is a facility of the National
Science Foundation operated under cooperative agreement by Associated
Universities, Inc.

\bibliographystyle{apj}
\bibliography{/srv/one/keckert/papers/bib/paperbib}

\end{document}

%% file: table1.tex
\begin{deluxetable*}{cccccc}
\tablecaption{Central Stellar and Baryonic Mass to Group Halo Mass Fit Parameters}
\startdata
\tablehead{\colhead{data} & \colhead{log$\phi_{0}$} & \colhead{log$M_{0}$} & \colhead{$\alpha$} & \colhead{$\beta$} & \colhead{$x_{0}$}\\ \colhead{} & \colhead{log(M$_{\odot}$)} & \colhead{log(M$_{\odot}$)} & \colhead{} & \colhead{} & \colhead{}} 
RESOLVE-B stellar & 10.73 $\pm$ 0.72 & 11.85 $\pm$ 2.04 & 3.49 $\pm$ 0.21 & 3.14 $\pm$ 0.19 & 0.32 $\pm$ 1.52 \\
RESOLVE-B baryonic & 10.69 $\pm$ 0.87 & 11.77 $\pm$ 2.40 & 2.67 $\pm$ 0.22 & 2.31 $\pm$ 0.19 & 0.39 $\pm$ 2.18 \\
ECO stellar & 10.50 $\pm$ 0.22 & 11.55 $\pm$ 0.56 & 6.56 $\pm$ 0.24 & 6.16 $\pm$ 0.23 & 0.24 $\pm$ 0.31 \\
ECO baryonic & 10.76 $\pm$ 0.23 & 11.83 $\pm$ 0.68 & 2.34 $\pm$ 0.05 & 2.00 $\pm$ 0.05 & 0.44 $\pm$ 0.68 \\
\enddata
\label{tb:cent}
\end{deluxetable*}

%% file: table2.tex
\begin{deluxetable}{cl}
\tablecaption{All RESOLVE-B and ECO stellar and baryonic mass functions}
\tablehead{\colhead{column} & \colhead{Mass Function Description}\\ \colhead{number} & \colhead{[$\phi_{16}$,$\phi_{med}$,$\phi_{84}$]}}
\startdata
1 &  stellar or baryonic mass \\
2-4 &  RESOLVE-B SMF \\
5-7 &  RESOLVE-B SMF log(M$_{halo}$) $<$ 11.4 \\
8-10 &  RESOLVE-B SMF 11.4 $<$ log(M$_{halo}$) $<$ 12.0\\
11-13 &  RESOLVE-B SMF 12.0 $<$ log(M$_{halo}$) $<$ 13.5 \\
14-16 &  RESOLVE-B SMF central log(M$_{halo}$) $<$ 11.4 \\
17-19 &  RESOLVE-B SMF central 11.4 $<$ log(M$_{halo}$) $<$ 12.0\\
20-22 &  RESOLVE-B SMF central 12.0 $<$ log(M$_{halo}$) $<$ 13.5 \\
23-25 &  RESOLVE-B SMF satellite 11.4 $<$ log(M$_{halo}$) $<$ 12.0\\
26-28 &  RESOLVE-B SMF satellite 12.0 $<$ log(M$_{halo}$) $<$ 13.5\\
29-31 &  RESOLVE-B BMF \\
32-34 &  RESOLVE-B BMF log(M$_{halo}$) $<$ 11.4 \\
35-37 &  RESOLVE-B BMF 11.4 $<$ log(M$_{halo}$) $<$ 12.0\\
38-40 &  RESOLVE-B BMF 12.0 $<$ log(M$_{halo}$) $<$ 13.5 \\
41-43 &  RESOLVE-B BMF central log(M$_{halo}$) $<$ 11.4 \\
44-46 &  RESOLVE-B BMF central 11.4 $<$ log(M$_{halo}$) $<$ 12.0\\
47-49 &  RESOLVE-B BMF central 12.0 $<$ log(M$_{halo}$) $<$ 13.5 \\
50-52 &  RESOLVE-B BMF satellite 11.4 $<$ log(M$_{halo}$) $<$ 12.0\\
53-55 &  RESOLVE-B BMF satellite 12.0 $<$ log(M$_{halo}$) $<$ 13.5\\
56-58 &  ECO SMF \\
59-61 &  ECO SMF log(M$_{halo}$) $<$ 11.4 \\
62-64 &  ECO SMF 11.4 $<$ log(M$_{halo}$) $<$ 12.0\\
65-67 &  ECO SMF 12.0 $<$ log(M$_{halo}$) $<$ 13.5 \\
68-70 &  ECO SMF log(M$_{halo}$) $>$ 13.5 \\
71-73 &  ECO SMF central log(M$_{halo}$) $<$ 11.4 \\
74-76 &  ECO SMF central 11.4 $<$ log(M$_{halo}$) $<$ 12.0\\
77-79 &  ECO SMF central 12.0 $<$ log(M$_{halo}$) $<$ 13.5 \\
80-82 &  ECO SMF central log(M$_{halo}$) $>$ 13.5 \\
83-85 &  ECO SMF satellite 11.4 $<$ log(M$_{halo}$) $<$ 12.0\\
86-88 &  ECO SMF satellite 12.0 $<$ log(M$_{halo}$) $<$ 13.5\\
89-91 &  ECO SMF satellite log(M$_{halo}$) $>$ 13.5\\
92-94 &  ECO BMF \\
95-97 &  ECO BMF log(M$_{halo}$) $<$ 11.4 \\
98-100 &  ECO BMF 11.4 $<$ log(M$_{halo}$) $<$ 12.0\\
101-103 &  ECO BMF 12.0 $<$ log(M$_{halo}$) $<$ 13.5 \\
104-106 &  ECO BMF log(M$_{halo}$) $>$ 13.5 \\
107-109 &  ECO BMF central log(M$_{halo}$) $<$ 11.4 \\
110-112 &  ECO BMF central 11.4 $<$ log(M$_{halo}$) $<$ 12.0\\
113-115 &  ECO BMF central 12.0 $<$ log(M$_{halo}$) $<$ 13.5 \\
116-118 &  ECO BMF central log(M$_{halo}$) $>$ 13.5 \\
119-121 &  ECO BMF satellite 11.4 $<$ log(M$_{halo}$) $<$ 12.0\\
122-124 &  ECO BMF satellite 12.0 $<$ log(M$_{halo}$) $<$ 13.5\\
125-127 &  ECO BMF satellite log(M$_{halo}$) $>$ 13.5\\
\tablecomments{The first column has units of logM$_{sun}$ and all other columns have units of dlogM$^{-1}$Mpc$^{-3}$. The three columns for each mass function represent the 16th, 50th, and 84th percentiles.}
\enddata
\label{tb:allmfs}
\end{deluxetable}

%% file: table3.tex
\begin{deluxetable*}{cccccc}
\tablecaption{Single and Double Schechter Function Parameters for RESOLVE-B and ECO stellar and baryonic mass functions}
\tablehead{\colhead{mass function} & \colhead{log(M$_{*}$)} & \colhead{$\phi_{*1}$} & \colhead{$\alpha_{1}$} & \colhead{$\phi_{*2}$} & \colhead{$\alpha_{2}$}\\ \colhead{} & \colhead{log(M$_{\odot}$)} & \colhead{10$^{3}$$\times$(Mpc$^3$dlogM)$^{-1}$} & \colhead{} & \colhead{10$^{3}$$\times$(Mpc$^3$dlogM)$^{-1}$} & \colhead{}}
\startdata
RESOLVE-B SMF & 11.25$^{+0.25}_{-0.19}$ & 4.47$^{+1.82}_{-1.53}$ & -1.28$^{+0.06}_{-0.05}$ & ... & ... \\
RESOLVE-B SMF & 10.87$^{+0.33}_{-0.27}$ & 9.00$^{+6.36}_{-8.47}$ & -0.52$^{+0.87}_{-0.49}$ & 3.25$^{+3.00}_{-2.81}$ & -1.38$^{+0.13}_{-0.35}$ \\
ECO SMF & 10.92$^{+0.03}_{-0.03}$ & 5.95$^{+0.41}_{-0.42}$ & -1.19$^{+0.02}_{-0.02}$ & ... & ... \\
ECO SMF & 10.87$^{+0.05}_{-0.06}$ & 3.44$^{+2.25}_{-1.93}$ & -0.91$^{+0.23}_{-0.15}$ & 3.62$^{+1.49}_{-1.78}$ & -1.26$^{+0.06}_{-0.11}$ \\
RESOLVE-B BMF & 11.11$^{+0.19}_{-0.16}$ & 6.93$^{+2.56}_{-2.33}$ & -1.30$^{+0.06}_{-0.07}$ & ... & ... \\
RESOLVE-B BMF & 10.98$^{+0.22}_{-0.25}$ & 2.74$^{+10.0}_{-2.66}$ & -0.48$^{+1.80}_{-0.64}$ & 5.54$^{+3.86}_{-4.36}$ & -1.35$^{+0.10}_{-0.28}$ \\
ECO BMF & 10.92$^{+0.03}_{-0.04}$ & 7.48$^{+0.85}_{-0.79}$ & -1.28$^{+0.03}_{-0.03}$ & ... & ... \\
ECO BMF & 10.89$^{+0.05}_{-0.07}$ & 1.55$^{+2.70}_{-1.17}$ & -1.02$^{+1.69}_{-0.21}$ & 6.27$^{+2.40}_{-2.54}$ & -1.30$^{+0.05}_{-0.06}$ \\
\enddata
\label{tb:schechfuncparams_emcee}
\end{deluxetable*}

%% file: ms.bbl
\begin{thebibliography}{128}
\expandafter\ifx\csname natexlab\endcsname\relax\def\natexlab#1{#1}\fi

\bibitem[{{Aihara} {et~al.}(2011){Aihara}, {Allende Prieto}, {An}, {Anderson},
  {Aubourg}, {Balbinot}, {Beers}, {Berlind}, {Bickerton}, {Bizyaev}, {Blanton},
  {Bochanski}, {Bolton}, {Bovy}, {Brandt}, {Brinkmann}, {Brown}, {Brownstein},
  {Busca}, {Campbell}, {Carr}, {Chen}, {Chiappini}, {Comparat}, {Connolly},
  {Cortes}, {Croft}, {Cuesta}, {da Costa}, {Davenport}, {Dawson}, {Dhital},
  {Ealet}, {Ebelke}, {Edmondson}, {Eisenstein}, {Escoffier}, {Esposito},
  {Evans}, {Fan}, {Femen{\'{\i}}a Castell{\'a}}, {Font-Ribera}, {Frinchaboy},
  {Ge}, {Gillespie}, {Gilmore}, {Gonz{\'a}lez Hern{\'a}ndez}, {Gott}, {Gould},
  {Grebel}, {Gunn}, {Hamilton}, {Harding}, {Harris}, {Hawley}, {Hearty}, {Ho},
  {Hogg}, {Holtzman}, {Honscheid}, {Inada}, {Ivans}, {Jiang}, {Johnson},
  {Jordan}, {Jordan}, {Kazin}, {Kirkby}, {Klaene}, {Knapp}, {Kneib},
  {Kochanek}, {Koesterke}, {Kollmeier}, {Kron}, {Lampeitl}, {Lang}, {Le Goff},
  {Lee}, {Lin}, {Long}, {Loomis}, {Lucatello}, {Lundgren}, {Lupton}, {Ma},
  {MacDonald}, {Mahadevan}, {Maia}, {Makler}, {Malanushenko}, {Malanushenko},
  {Mandelbaum}, {Maraston}, {Margala}, {Masters}, {McBride}, {McGehee},
  {McGreer}, {M{\'e}nard}, {Miralda-Escud{\'e}}, {Morrison}, {Mullally},
  {Muna}, {Munn}, {Murayama}, {Myers}, {Naugle}, {Neto}, {Nguyen}, {Nichol},
  {O'Connell}, {Ogando}, {Olmstead}, {Oravetz}, {Padmanabhan},
  {Palanque-Delabrouille}, {Pan}, {Pandey}, {P{\^a}ris}, {Percival},
  {Petitjean}, {Pfaffenberger}, {Pforr}, {Phleps}, {Pichon}, {Pieri}, {Prada},
  {Price-Whelan}, {Raddick}, {Ramos}, {Reyl{\'e}}, {Rich}, {Richards}, {Rix},
  {Robin}, {Rocha-Pinto}, {Rockosi}, {Roe}, {Rollinde}, {Ross}, {Ross},
  {Rossetto}, {S{\'a}nchez}, {Sayres}, {Schlegel}, {Schlesinger}, {Schmidt},
  {Schneider}, {Sheldon}, {Shu}, {Simmerer}, {Simmons}, {Sivarani}, {Snedden},
  {Sobeck}, {Steinmetz}, {Strauss}, {Szalay}, {Tanaka}, {Thakar}, {Thomas},
  {Tinker}, {Tofflemire}, {Tojeiro}, {Tremonti}, {Vandenberg}, {Vargas
  Maga{\~n}a}, {Verde}, {Vogt}, {Wake}, {Wang}, {Weaver}, {Weinberg}, {White},
  {White}, {Yanny}, {Yasuda}, {Yeche}, \& {Zehavi}}]{2011ApJS..193...29A}
{Aihara}, H., {Allende Prieto}, C., {An}, D., {et~al.} 2011, \apjs, 193, 29

\bibitem[{{Baldry} {et~al.}(2006){Baldry}, {Balogh}, {Bower}, {Glazebrook},
  {Nichol}, {Bamford}, \& {Budavari}}]{2006MNRAS.373..469B}
{Baldry}, I.~K., {Balogh}, M.~L., {Bower}, R.~G., {et~al.} 2006, \mnras, 373,
  469

\bibitem[{{Baldry} {et~al.}(2008){Baldry}, {Glazebrook}, \&
  {Driver}}]{2008MNRAS.388..945B}
{Baldry}, I.~K., {Glazebrook}, K., \& {Driver}, S.~P. 2008, \mnras, 388, 945

\bibitem[{{Baldry} {et~al.}(2012){Baldry}, {Driver}, {Loveday}, {Taylor},
  {Kelvin}, {Liske}, {Norberg}, {Robotham}, {Brough}, {Hopkins}, {Bamford},
  {Peacock}, {Bland-Hawthorn}, {Conselice}, {Croom}, {Jones}, {Parkinson},
  {Popescu}, {Prescott}, {Sharp}, \& {Tuffs}}]{2012MNRAS.421..621B}
{Baldry}, I.~K., {Driver}, S.~P., {Loveday}, J., {et~al.} 2012, \mnras, 421,
  621

\bibitem[{{Becker} {et~al.}(2001){Becker}, {Fan}, {White}, {Strauss},
  {Narayanan}, {Lupton}, {Gunn}, {Annis}, {Bahcall}, {Brinkmann}, {Connolly},
  {Csabai}, {Czarapata}, {Doi}, {Heckman}, {Hennessy}, {Ivezi{\'c}}, {Knapp},
  {Lamb}, {McKay}, {Munn}, {Nash}, {Nichol}, {Pier}, {Richards}, {Schneider},
  {Stoughton}, {Szalay}, {Thakar}, \& {York}}]{2001AJ....122.2850B}
{Becker}, R.~H., {Fan}, X., {White}, R.~L., {et~al.} 2001, \aj, 122, 2850

\bibitem[{{Begum} {et~al.}(2008){Begum}, {Chengalur}, {Karachentsev}, \&
  {Sharina}}]{2008MNRAS.386..138B}
{Begum}, A., {Chengalur}, J.~N., {Karachentsev}, I.~D., \& {Sharina}, M.~E.
  2008, \mnras, 386, 138

\bibitem[{{Behroozi} {et~al.}(2010){Behroozi}, {Conroy}, \&
  {Wechsler}}]{2010ApJ...717..379B}
{Behroozi}, P.~S., {Conroy}, C., \& {Wechsler}, R.~H. 2010, \apj, 717, 379

\bibitem[{{Behroozi} {et~al.}(2013){Behroozi}, {Wechsler}, \&
  {Conroy}}]{2013ApJ...770...57B}
{Behroozi}, P.~S., {Wechsler}, R.~H., \& {Conroy}, C. 2013, \apj, 770, 57

\bibitem[{{Bell} \& {de Jong}(2001)}]{2001ApJ...550..212B}
{Bell}, E.~F., \& {de Jong}, R.~S. 2001, \apj, 550, 212

\bibitem[{{Bell} {et~al.}(2003{\natexlab{a}}){Bell}, {McIntosh}, {Katz}, \&
  {Weinberg}}]{2003ApJ...585L.117B}
{Bell}, E.~F., {McIntosh}, D.~H., {Katz}, N., \& {Weinberg}, M.~D.
  2003{\natexlab{a}}, \apjl, 585, L117

\bibitem[{{Bell} {et~al.}(2003{\natexlab{b}}){Bell}, {McIntosh}, {Katz}, \&
  {Weinberg}}]{2003ApJS..149..289B}
---. 2003{\natexlab{b}}, \apjs, 149, 289

\bibitem[{{Berlind} {et~al.}(2006){Berlind}, {Frieman}, {Weinberg}, {Blanton},
  {Warren}, {Abazajian}, {Scranton}, {Hogg}, {Scoccimarro}, {Bahcall},
  {Brinkmann}, {Gott}, {Kleinman}, {Krzesinski}, {Lee}, {Miller}, {Nitta},
  {Schneider}, {Tucker}, {Zehavi}, \& {SDSS
  Collaboration}}]{2006ApJS..167....1B}
{Berlind}, A.~A., {Frieman}, J., {Weinberg}, D.~H., {et~al.} 2006, \apjs, 167,
  1

\bibitem[{{Bernardi} {et~al.}(2013){Bernardi}, {Meert}, {Sheth}, {Vikram},
  {Huertas-Company}, {Mei}, \& {Shankar}}]{2013MNRAS.436..697B}
{Bernardi}, M., {Meert}, A., {Sheth}, R.~K., {et~al.} 2013, \mnras, 436, 697

\bibitem[{{Blanton} {et~al.}(2011){Blanton}, {Kazin}, {Muna}, {Weaver}, \&
  {Price-Whelan}}]{2011AJ....142...31B}
{Blanton}, M.~R., {Kazin}, E., {Muna}, D., {Weaver}, B.~A., \& {Price-Whelan},
  A. 2011, \aj, 142, 31

\bibitem[{{Blanton} {et~al.}(2003{\natexlab{a}}){Blanton}, {Lin}, {Lupton},
  {Maley}, {Young}, {Zehavi}, \& {Loveday}}]{2003AJ....125.2276B}
{Blanton}, M.~R., {Lin}, H., {Lupton}, R.~H., {et~al.} 2003{\natexlab{a}}, \aj,
  125, 2276

\bibitem[{{Blanton} {et~al.}(2005){Blanton}, {Lupton}, {Schlegel}, {Strauss},
  {Brinkmann}, {Fukugita}, \& {Loveday}}]{2005ApJ...631..208B}
{Blanton}, M.~R., {Lupton}, R.~H., {Schlegel}, D.~J., {et~al.} 2005, \apj, 631,
  208

\bibitem[{{Blanton} \& {Roweis}(2007)}]{2007AJ....133..734B}
{Blanton}, M.~R., \& {Roweis}, S. 2007, \aj, 133, 734

\bibitem[{{Blanton} {et~al.}(2003{\natexlab{b}}){Blanton}, {Hogg}, {Bahcall},
  {Brinkmann}, {Britton}, {Connolly}, {Csabai}, {Fukugita}, {Loveday},
  {Meiksin}, {Munn}, {Nichol}, {Okamura}, {Quinn}, {Schneider}, {Shimasaku},
  {Strauss}, {Tegmark}, {Vogeley}, \& {Weinberg}}]{2003ApJ...592..819B}
{Blanton}, M.~R., {Hogg}, D.~W., {Bahcall}, N.~A., {et~al.} 2003{\natexlab{b}},
  \apj, 592, 819

\bibitem[{{Bolzonella} {et~al.}(2010){Bolzonella}, {Kova{\v c}}, {Pozzetti},
  {Zucca}, {Cucciati}, {Lilly}, {Peng}, {Iovino}, {Zamorani}, {Vergani},
  {Tasca}, {Lamareille}, {Oesch}, {Caputi}, {Kampczyk}, {Bardelli}, {Maier},
  {Abbas}, {Knobel}, {Scodeggio}, {Carollo}, {Contini}, {Kneib}, {Le
  F{\`e}vre}, {Mainieri}, {Renzini}, {Bongiorno}, {Coppa}, {de la Torre}, {de
  Ravel}, {Franzetti}, {Garilli}, {Le Borgne}, {Le Brun}, {Mignoli},
  {Pell{\'o}}, {Perez-Montero}, {Ricciardelli}, {Silverman}, {Tanaka},
  {Tresse}, {Bottini}, {Cappi}, {Cassata}, {Cimatti}, {Guzzo}, {Koekemoer},
  {Leauthad}, {Maccagni}, {Marinoni}, {McCracken}, {Memeo}, {Meneux},
  {Porciani}, {Scaramella}, {Aussel}, {Capak}, {Halliday}, {Ilbert},
  {Kartaltepe}, {Salvato}, {Sanders}, {Scarlata}, {Scoville}, {Taniguchi}, \&
  {Thompson}}]{2010A&A...524A..76B}
{Bolzonella}, M., {Kova{\v c}}, K., {Pozzetti}, L., {et~al.} 2010, \aap, 524,
  A76

\bibitem[{{Boselli} {et~al.}(2014){Boselli}, {Cortese}, {Boquien}, {Boissier},
  {Catinella}, {Lagos}, \& {Saintonge}}]{2014A&A...564A..66B}
{Boselli}, A., {Cortese}, L., {Boquien}, M., {et~al.} 2014, \aap, 564, A66

\bibitem[{{Bruzual} \& {Charlot}(2003)}]{2003MNRAS.344.1000B}
{Bruzual}, G., \& {Charlot}, S. 2003, \mnras, 344, 1000

\bibitem[{{Bullock} {et~al.}(2000){Bullock}, {Kravtsov}, \&
  {Weinberg}}]{2000ApJ...539..517B}
{Bullock}, J.~S., {Kravtsov}, A.~V., \& {Weinberg}, D.~H. 2000, \apj, 539, 517

\bibitem[{{Calzetti}(2001)}]{2001PASP..113.1449C}
{Calzetti}, D. 2001, \pasp, 113, 1449

\bibitem[{{Carollo} {et~al.}(2013){Carollo}, {Cibinel}, {Lilly}, {Miniati},
  {Norberg}, {Silverman}, {van Gorkom}, {Cameron}, {Finoguenov}, {Peng},
  {Pipino}, \& {Rudick}}]{2013ApJ...776...71C}
{Carollo}, C.~M., {Cibinel}, A., {Lilly}, S.~J., {et~al.} 2013, \apj, 776, 71

\bibitem[{{Casoli} {et~al.}(1998){Casoli}, {Sauty}, {Gerin}, {Boselli},
  {Fouque}, {Braine}, {Gavazzi}, {Lequeux}, \& {Dickey}}]{1998A&A...331..451C}
{Casoli}, F., {Sauty}, S., {Gerin}, M., {et~al.} 1998, \aap, 331, 451

\bibitem[{{Catinella} {et~al.}(2013){Catinella}, {Schiminovich}, {Cortese},
  {Fabello}, {Hummels}, {Moran}, {Lemonias}, {Cooper}, {Wu}, {Heckman}, \&
  {Wang}}]{2013MNRAS.436...34C}
{Catinella}, B., {Schiminovich}, D., {Cortese}, L., {et~al.} 2013, \mnras, 436,
  34

\bibitem[{{Chabrier}(2003)}]{2003PASP..115..763C}
{Chabrier}, G. 2003, \pasp, 115, 763

\bibitem[{{Cole} {et~al.}(2000){Cole}, {Lacey}, {Baugh}, \&
  {Frenk}}]{2000MNRAS.319..168C}
{Cole}, S., {Lacey}, C.~G., {Baugh}, C.~M., \& {Frenk}, C.~S. 2000, \mnras,
  319, 168

\bibitem[{{Colless} {et~al.}(2001){Colless}, {Dalton}, {Maddox}, {Sutherland},
  {Norberg}, {Cole}, {Bland-Hawthorn}, {Bridges}, {Cannon}, {Collins}, {Couch},
  {Cross}, {Deeley}, {De Propris}, {Driver}, {Efstathiou}, {Ellis}, {Frenk},
  {Glazebrook}, {Jackson}, {Lahav}, {Lewis}, {Lumsden}, {Madgwick}, {Peacock},
  {Peterson}, {Price}, {Seaborne}, \& {Taylor}}]{2001MNRAS.328.1039C}
{Colless}, M., {Dalton}, G., {Maddox}, S., {et~al.} 2001, \mnras, 328, 1039

\bibitem[{{Conroy} \& {Wechsler}(2009)}]{2009ApJ...696..620C}
{Conroy}, C., \& {Wechsler}, R.~H. 2009, \apj, 696, 620

\bibitem[{{Cooray}(2006)}]{2006MNRAS.365..842C}
{Cooray}, A. 2006, \mnras, 365, 842

\bibitem[{{Davies} \& {Lewis}(1973)}]{1973MNRAS.165..231D}
{Davies}, R.~D., \& {Lewis}, B.~M. 1973, \mnras, 165, 231

\bibitem[{{de Blok}(2010)}]{2010AdAst2010E...5D}
{de Blok}, W.~J.~G. 2010, Advances in Astronomy, 2010, 5

\bibitem[{{Dekel} \& {Silk}(1986)}]{1986ApJ...303...39D}
{Dekel}, A., \& {Silk}, J. 1986, \apj, 303, 39

\bibitem[{{Dressler}(1978)}]{1978ApJ...223..765D}
{Dressler}, A. 1978, \apj, 223, 765

\bibitem[{{Driver} \& {Robotham}(2010)}]{2010MNRAS.407.2131D}
{Driver}, S.~P., \& {Robotham}, A.~S.~G. 2010, \mnras, 407, 2131

\bibitem[{{Driver} {et~al.}(2011){Driver}, {Hill}, {Kelvin}, {Robotham},
  {Liske}, {Norberg}, {Baldry}, {Bamford}, {Hopkins}, {Loveday}, {Peacock},
  {Andrae}, {Bland-Hawthorn}, {Brough}, {Brown}, {Cameron}, {Ching}, {Colless},
  {Conselice}, {Croom}, {Cross}, {de Propris}, {Dye}, {Drinkwater}, {Ellis},
  {Graham}, {Grootes}, {Gunawardhana}, {Jones}, {van Kampen}, {Maraston},
  {Nichol}, {Parkinson}, {Phillipps}, {Pimbblet}, {Popescu}, {Prescott},
  {Roseboom}, {Sadler}, {Sansom}, {Sharp}, {Smith}, {Taylor}, {Thomas},
  {Tuffs}, {Wijesinghe}, {Dunne}, {Frenk}, {Jarvis}, {Madore}, {Meyer},
  {Seibert}, {Staveley-Smith}, {Sutherland}, \& {Warren}}]{2011MNRAS.413..971D}
{Driver}, S.~P., {Hill}, D.~T., {Kelvin}, L.~S., {et~al.} 2011, \mnras, 413,
  971

\bibitem[{{Drory} {et~al.}(2009){Drory}, {Bundy}, {Leauthaud}, {Scoville},
  {Capak}, {Ilbert}, {Kartaltepe}, {Kneib}, {McCracken}, {Salvato}, {Sanders},
  {Thompson}, \& {Willott}}]{2009ApJ...707.1595D}
{Drory}, N., {Bundy}, K., {Leauthaud}, A., {et~al.} 2009, \apj, 707, 1595

\bibitem[{{Duarte} \& {Mamon}(2014)}]{2014MNRAS.440.1763D}
{Duarte}, M., \& {Mamon}, G.~A. 2014, \mnras, 440, 1763

\bibitem[{{Dunkley} {et~al.}(2009){Dunkley}, {Spergel}, {Komatsu}, {Hinshaw},
  {Larson}, {Nolta}, {Odegard}, {Page}, {Bennett}, {Gold}, {Hill}, {Jarosik},
  {Weiland}, {Halpern}, {Kogut}, {Limon}, {Meyer}, {Tucker}, {Wollack}, \&
  {Wright}}]{2009ApJ...701.1804D}
{Dunkley}, J., {Spergel}, D.~N., {Komatsu}, E., {et~al.} 2009, \apj, 701, 1804

\bibitem[{{Eckert} {et~al.}(2015){Eckert}, {Kannappan}, {Stark}, {Moffett},
  {Norris}, {Snyder}, \& {Hoversten}}]{2015ApJ...810..166E}
{Eckert}, K.~D., {Kannappan}, S.~J., {Stark}, D.~V., {et~al.} 2015, \apj, 810,
  166

\bibitem[{{Falco} {et~al.}(1999){Falco}, {Kurtz}, {Geller}, {Huchra}, {Peters},
  {Berlind}, {Mink}, {Tokarz}, \& {Elwell}}]{1999PASP..111..438F}
{Falco}, E.~E., {Kurtz}, M.~J., {Geller}, M.~J., {et~al.} 1999, \pasp, 111, 438

\bibitem[{{Faltenbacher}(2010)}]{2010MNRAS.408.1113F}
{Faltenbacher}, A. 2010, \mnras, 408, 1113

\bibitem[{{Foreman-Mackey} {et~al.}(2013){Foreman-Mackey}, {Hogg}, {Lang}, \&
  {Goodman}}]{2013PASP..125..306F}
{Foreman-Mackey}, D., {Hogg}, D.~W., {Lang}, D., \& {Goodman}, J. 2013, \pasp,
  125, 306

\bibitem[{{Gnedin} \& {Ostriker}(1997)}]{1997ApJ...486..581G}
{Gnedin}, N.~Y., \& {Ostriker}, J.~P. 1997, \apj, 486, 581

\bibitem[{{Goto} {et~al.}(2002){Goto}, {Okamura}, {McKay}, {Bahcall}, {Annis},
  {Bernard}, {Brinkmann}, {G{\'o}mez}, {Hansen}, {Kim}, {Sekiguchi}, \&
  {Sheth}}]{2002PASJ...54..515G}
{Goto}, T., {Okamura}, S., {McKay}, T.~A., {et~al.} 2002, \pasj, 54, 515

\bibitem[{{Gunn} \& {Gott}(1972)}]{1972ApJ...176....1G}
{Gunn}, J.~E., \& {Gott}, III, J.~R. 1972, \apj, 176, 1

\bibitem[{{Haines} {et~al.}(2015){Haines}, {Pereira}, {Smith}, {Egami},
  {Babul}, {Finoguenov}, {Ziparo}, {McGee}, {Rawle}, {Okabe}, \&
  {Moran}}]{2015ApJ...806..101H}
{Haines}, C.~P., {Pereira}, M.~J., {Smith}, G.~P., {et~al.} 2015, \apj, 806,
  101

\bibitem[{{Hambly} {et~al.}(2008){Hambly}, {Collins}, {Cross}, {Mann}, {Read},
  {Sutorius}, {Bond}, {Bryant}, {Emerson}, {Lawrence}, {Rimoldini}, {Stewart},
  {Williams}, {Adamson}, {Hirst}, {Dye}, \& {Warren}}]{2008MNRAS.384..637H}
{Hambly}, N.~C., {Collins}, R.~S., {Cross}, N.~J.~G., {et~al.} 2008, \mnras,
  384, 637

\bibitem[{{Haynes} {et~al.}(1984){Haynes}, {Giovanelli}, \&
  {Chincarini}}]{1984ARA&A..22..445H}
{Haynes}, M.~P., {Giovanelli}, R., \& {Chincarini}, G.~L. 1984, \araa, 22, 445

\bibitem[{{Haynes} {et~al.}(2011){Haynes}, {Giovanelli}, {Martin}, {Hess},
  {Saintonge}, {Adams}, {Hallenbeck}, {Hoffman}, {Huang}, {Kent}, {Koopmann},
  {Papastergis}, {Stierwalt}, {Balonek}, {Craig}, {Higdon}, {Kornreich},
  {Miller}, {O'Donoghue}, {Olowin}, {Rosenberg}, {Spekkens}, {Troischt}, \&
  {Wilcots}}]{2011AJ....142..170H}
{Haynes}, M.~P., {Giovanelli}, R., {Martin}, A.~M., {et~al.} 2011, \aj, 142,
  170

\bibitem[{{Huang} {et~al.}(2012){Huang}, {Haynes}, {Giovanelli}, \&
  {Brinchmann}}]{2012ApJ...756..113H}
{Huang}, S., {Haynes}, M.~P., {Giovanelli}, R., \& {Brinchmann}, J. 2012, \apj,
  756, 113

\bibitem[{{Jones} {et~al.}(2009){Jones}, {Read}, {Saunders}, {Colless},
  {Jarrett}, {Parker}, {Fairall}, {Mauch}, {Sadler}, {Watson}, {Burton},
  {Campbell}, {Cass}, {Croom}, {Dawe}, {Fiegert}, {Frankcombe}, {Hartley},
  {Huchra}, {James}, {Kirby}, {Lahav}, {Lucey}, {Mamon}, {Moore}, {Peterson},
  {Prior}, {Proust}, {Russell}, {Safouris}, {Wakamatsu}, {Westra}, \&
  {Williams}}]{2009MNRAS.399..683J}
{Jones}, D.~H., {Read}, M.~A., {Saunders}, W., {et~al.} 2009, \mnras, 399, 683

\bibitem[{{Kannappan}(2004)}]{2004ApJ...611L..89K}
{Kannappan}, S.~J. 2004, \apjl, 611, L89

\bibitem[{{Kannappan} \& {Gawiser}(2007)}]{2007ApJ...657L...5K}
{Kannappan}, S.~J., \& {Gawiser}, E. 2007, \apjl, 657, L5

\bibitem[{{Kannappan} {et~al.}(2009){Kannappan}, {Guie}, \&
  {Baker}}]{2009AJ....138..579K}
{Kannappan}, S.~J., {Guie}, J.~M., \& {Baker}, A.~J. 2009, \aj, 138, 579

\bibitem[{{Kannappan} \& {Wei}(2008)}]{2008AIPC.1035..163K}
{Kannappan}, S.~J., \& {Wei}, L.~H. 2008, in American Institute of Physics
  Conference Series, Vol. 1035, The Evolution of Galaxies Through the Neutral
  Hydrogen Window, ed. {R.~Minchin \& E.~Momjian}, 163--168

\bibitem[{{Kannappan} {et~al.}(2013){Kannappan}, {Stark}, {Eckert}, {Moffett},
  {Wei}, {Pisano}, {Baker}, {Vogel}, {Fabricant}, {Laine}, {Norris}, {Jogee},
  {Lepore}, {Hough}, \& {Weinberg-Wolf}}]{2013ApJ...777...42K}
{Kannappan}, S.~J., {Stark}, D.~V., {Eckert}, K.~D., {et~al.} 2013, \apj, 777,
  42

\bibitem[{{Kauffmann} {et~al.}(2003{\natexlab{a}}){Kauffmann}, {Heckman},
  {White}, {Charlot}, {Tremonti}, {Brinchmann}, {Bruzual}, {Peng}, {Seibert},
  {Bernardi}, {Blanton}, {Brinkmann}, {Castander}, {Cs{\'a}bai}, {Fukugita},
  {Ivezic}, {Munn}, {Nichol}, {Padmanabhan}, {Thakar}, {Weinberg}, \&
  {York}}]{2003MNRAS.341...33K}
{Kauffmann}, G., {Heckman}, T.~M., {White}, S.~D.~M., {et~al.}
  2003{\natexlab{a}}, \mnras, 341, 33

\bibitem[{{Kauffmann} {et~al.}(2003{\natexlab{b}}){Kauffmann}, {Heckman},
  {White}, {Charlot}, {Tremonti}, {Peng}, {Seibert}, {Brinkmann}, {Nichol},
  {SubbaRao}, \& {York}}]{2003MNRAS.341...54K}
---. 2003{\natexlab{b}}, \mnras, 341, 54

\bibitem[{{Kim} {et~al.}(1998){Kim}, {Staveley-Smith}, {Dopita}, {Freeman},
  {Sault}, {Kesteven}, \& {McConnell}}]{1998ApJ...503..674K}
{Kim}, S., {Staveley-Smith}, L., {Dopita}, M.~A., {et~al.} 1998, \apj, 503, 674

\bibitem[{{Kravtsov} {et~al.}(2014){Kravtsov}, {Vikhlinin}, \&
  {Meshscheryakov}}]{2014arXiv1401.7329K}
{Kravtsov}, A., {Vikhlinin}, A., \& {Meshscheryakov}, A. 2014, ArXiv e-prints

\bibitem[{{Larson} {et~al.}(1980){Larson}, {Tinsley}, \&
  {Caldwell}}]{1980ApJ...237..692L}
{Larson}, R.~B., {Tinsley}, B.~M., \& {Caldwell}, C.~N. 1980, \apj, 237, 692

\bibitem[{{Leauthaud} {et~al.}(2011){Leauthaud}, {Tinker}, {Behroozi}, {Busha},
  \& {Wechsler}}]{2011ApJ...738...45L}
{Leauthaud}, A., {Tinker}, J., {Behroozi}, P.~S., {Busha}, M.~T., \&
  {Wechsler}, R.~H. 2011, \apj, 738, 45

\bibitem[{{Leauthaud} {et~al.}(2012){Leauthaud}, {Tinker}, {Bundy}, {Behroozi},
  {Massey}, {Rhodes}, {George}, {Kneib}, {Benson}, {Wechsler}, {Busha},
  {Capak}, {Cort{\^e}s}, {Ilbert}, {Koekemoer}, {Le F{\`e}vre}, {Lilly},
  {McCracken}, {Salvato}, {Schrabback}, {Scoville}, {Smith}, \&
  {Taylor}}]{2012ApJ...744..159L}
{Leauthaud}, A., {Tinker}, J., {Bundy}, K., {et~al.} 2012, \apj, 744, 159

\bibitem[{{Li} \& {White}(2009)}]{2009MNRAS.398.2177L}
{Li}, C., \& {White}, S.~D.~M. 2009, \mnras, 398, 2177

\bibitem[{{Liu} {et~al.}(2010){Liu}, {Yang}, {Mo}, {van den Bosch}, \&
  {Springel}}]{2010ApJ...712..734L}
{Liu}, L., {Yang}, X., {Mo}, H.~J., {van den Bosch}, F.~C., \& {Springel}, V.
  2010, \apj, 712, 734

\bibitem[{{Lovell} {et~al.}(2014){Lovell}, {Frenk}, {Eke}, {Jenkins}, {Gao}, \&
  {Theuns}}]{2014MNRAS.439..300L}
{Lovell}, M.~R., {Frenk}, C.~S., {Eke}, V.~R., {et~al.} 2014, \mnras, 439, 300

\bibitem[{{Mac Low} \& {Ferrara}(1999)}]{1999ApJ...513..142M}
{Mac Low}, M.-M., \& {Ferrara}, A. 1999, \apj, 513, 142

\bibitem[{{Maga{\~n}a} \& {Matos}(2012)}]{2012JPhCS.378a2012M}
{Maga{\~n}a}, J., \& {Matos}, T. 2012, Journal of Physics Conference Series,
  378, 012012

\bibitem[{{Markwardt}(2009)}]{2009ASPC..411..251M}
{Markwardt}, C.~B. 2009, in Astronomical Society of the Pacific Conference
  Series, Vol. 411, Astronomical Data Analysis Software and Systems XVIII, ed.
  D.~A. {Bohlender}, D.~{Durand}, \& P.~{Dowler}, 251

\bibitem[{{McGaugh} {et~al.}(2000){McGaugh}, {Schombert}, {Bothun}, \& {de
  Blok}}]{2000ApJ...533L..99M}
{McGaugh}, S.~S., {Schombert}, J.~M., {Bothun}, G.~D., \& {de Blok}, W.~J.~G.
  2000, \apjl, 533, L99

\bibitem[{{McGaugh} {et~al.}(2010){McGaugh}, {Schombert}, {de Blok}, \&
  {Zagursky}}]{2010ApJ...708L..14M}
{McGaugh}, S.~S., {Schombert}, J.~M., {de Blok}, W.~J.~G., \& {Zagursky}, M.~J.
  2010, \apjl, 708, L14

\bibitem[{{McGee} {et~al.}(2009){McGee}, {Balogh}, {Bower}, {Font}, \&
  {McCarthy}}]{2009MNRAS.400..937M}
{McGee}, S.~L., {Balogh}, M.~L., {Bower}, R.~G., {Font}, A.~S., \& {McCarthy},
  I.~G. 2009, \mnras, 400, 937

\bibitem[{{Mihos} {et~al.}(2015){Mihos}, {Durrell}, {Ferrarese}, {Feldmeier},
  {C{\^o}t{\'e}}, {Peng}, {Harding}, {Liu}, {Gwyn}, \&
  {Cuillandre}}]{2015ApJ...809L..21M}
{Mihos}, J.~C., {Durrell}, P.~R., {Ferrarese}, L., {et~al.} 2015, \apjl, 809,
  L21

\bibitem[{{Moffett} {et~al.}(2015){Moffett}, {Kannappan}, {Berlind}, {Eckert},
  {Stark}, {Hendel}, {Norris}, \& {Grogin}}]{2015ApJ...812...89M}
{Moffett}, A.~J., {Kannappan}, S.~J., {Berlind}, A.~A., {et~al.} 2015, \apj,
  812, 89

\bibitem[{Mor\'{e}(1978)}]{more}
Mor\'{e}, J. 1978, in Lecture Notes in Mathematics, Vol. 630, Numerical
  Analysis, ed. G.~Watson (Springer Berlin Heidelberg), 105--116

\bibitem[{{More} {et~al.}(2009){More}, {van den Bosch}, {Cacciato}, {Mo},
  {Yang}, \& {Li}}]{2009MNRAS.392..801M}
{More}, S., {van den Bosch}, F.~C., {Cacciato}, M., {et~al.} 2009, \mnras, 392,
  801

\bibitem[{{Morrissey} {et~al.}(2007){Morrissey}, {Conrow}, {Barlow}, {Small},
  {Seibert}, {Wyder}, {Budav{\'a}ri}, {Arnouts}, {Friedman}, {Forster},
  {Martin}, {Neff}, {Schiminovich}, {Bianchi}, {Donas}, {Heckman}, {Lee},
  {Madore}, {Milliard}, {Rich}, {Szalay}, {Welsh}, \&
  {Yi}}]{2007ApJS..173..682M}
{Morrissey}, P., {Conrow}, T., {Barlow}, T.~A., {et~al.} 2007, \apjs, 173, 682

\bibitem[{{Moster} {et~al.}(2010){Moster}, {Somerville}, {Maulbetsch}, {van den
  Bosch}, {Macci{\`o}}, {Naab}, \& {Oser}}]{2010ApJ...710..903M}
{Moster}, B.~P., {Somerville}, R.~S., {Maulbetsch}, C., {et~al.} 2010, \apj,
  710, 903

\bibitem[{{Moster} {et~al.}(2011){Moster}, {Somerville}, {Newman}, \&
  {Rix}}]{2011ApJ...731..113M}
{Moster}, B.~P., {Somerville}, R.~S., {Newman}, J.~A., \& {Rix}, H.-W. 2011,
  \apj, 731, 113

\bibitem[{{Nulsen}(1982)}]{1982MNRAS.198.1007N}
{Nulsen}, P.~E.~J. 1982, \mnras, 198, 1007

\bibitem[{{Panter} {et~al.}(2007){Panter}, {Jimenez}, {Heavens}, \&
  {Charlot}}]{2007MNRAS.378.1550P}
{Panter}, B., {Jimenez}, R., {Heavens}, A.~F., \& {Charlot}, S. 2007, \mnras,
  378, 1550

\bibitem[{{Papastergis} {et~al.}(2012){Papastergis}, {Cattaneo}, {Huang},
  {Giovanelli}, \& {Haynes}}]{2012ApJ...759..138P}
{Papastergis}, E., {Cattaneo}, A., {Huang}, S., {Giovanelli}, R., \& {Haynes},
  M.~P. 2012, \apj, 759, 138

\bibitem[{{Papastergis} {et~al.}(2015){Papastergis}, {Giovanelli}, {Haynes}, \&
  {Shankar}}]{2015A&A...574A.113P}
{Papastergis}, E., {Giovanelli}, R., {Haynes}, M.~P., \& {Shankar}, F. 2015,
  \aap, 574, A113

\bibitem[{{Papastergis} {et~al.}(2011){Papastergis}, {Martin}, {Giovanelli}, \&
  {Haynes}}]{2011ApJ...739...38P}
{Papastergis}, E., {Martin}, A.~M., {Giovanelli}, R., \& {Haynes}, M.~P. 2011,
  \apj, 739, 38

\bibitem[{{Paturel} {et~al.}(2003){Paturel}, {Petit}, {Prugniel}, {Theureau},
  {Rousseau}, {Brouty}, {Dubois}, \& {Cambr{\'e}sy}}]{2003A&A...412...45P}
{Paturel}, G., {Petit}, C., {Prugniel}, P., {et~al.} 2003, \aap, 412, 45

\bibitem[{{Peng} {et~al.}(2010){Peng}, {Lilly}, {Kova{\v c}}, {Bolzonella},
  {Pozzetti}, {Renzini}, {Zamorani}, {Ilbert}, {Knobel}, {Iovino}, {Maier},
  {Cucciati}, {Tasca}, {Carollo}, {Silverman}, {Kampczyk}, {de Ravel},
  {Sanders}, {Scoville}, {Contini}, {Mainieri}, {Scodeggio}, {Kneib}, {Le
  F{\`e}vre}, {Bardelli}, {Bongiorno}, {Caputi}, {Coppa}, {de la Torre},
  {Franzetti}, {Garilli}, {Lamareille}, {Le Borgne}, {Le Brun}, {Mignoli},
  {Perez Montero}, {Pello}, {Ricciardelli}, {Tanaka}, {Tresse}, {Vergani},
  {Welikala}, {Zucca}, {Oesch}, {Abbas}, {Barnes}, {Bordoloi}, {Bottini},
  {Cappi}, {Cassata}, {Cimatti}, {Fumana}, {Hasinger}, {Koekemoer},
  {Leauthaud}, {Maccagni}, {Marinoni}, {McCracken}, {Memeo}, {Meneux}, {Nair},
  {Porciani}, {Presotto}, \& {Scaramella}}]{2010ApJ...721..193P}
{Peng}, Y.-j., {Lilly}, S.~J., {Kova{\v c}}, K., {et~al.} 2010, \apj, 721, 193

\bibitem[{{Pfenniger} \& {Revaz}(2005)}]{2005A&A...431..511P}
{Pfenniger}, D., \& {Revaz}, Y. 2005, \aap, 431, 511

\bibitem[{{Pipino} {et~al.}(2014){Pipino}, {Cibinel}, {Tacchella}, {Carollo},
  {Lilly}, {Miniati}, {Silverman}, {van Gorkom}, \&
  {Finoguenov}}]{2014ApJ...797..127P}
{Pipino}, A., {Cibinel}, A., {Tacchella}, S., {et~al.} 2014, \apj, 797, 127

\bibitem[{{Planck Collaboration} {et~al.}(2014){Planck Collaboration}, {Ade},
  {Aghanim}, {Armitage-Caplan}, {Arnaud}, {Ashdown}, {Atrio-Barandela},
  {Aumont}, {Baccigalupi}, {Banday}, \& et~al.}]{2014A&A...571A..16P}
{Planck Collaboration}, {Ade}, P.~A.~R., {Aghanim}, N., {et~al.} 2014, \aap,
  571, A16

\bibitem[{{Popesso} {et~al.}(2006){Popesso}, {Biviano}, {B{\"o}hringer}, \&
  {Romaniello}}]{2006A&A...445...29P}
{Popesso}, P., {Biviano}, A., {B{\"o}hringer}, H., \& {Romaniello}, M. 2006,
  \aap, 445, 29

\bibitem[{{Press} \& {Schechter}(1974)}]{1974ApJ...187..425P}
{Press}, W.~H., \& {Schechter}, P. 1974, \apj, 187, 425

\bibitem[{{Pritchard} {et~al.}(2010){Pritchard}, {Loeb}, \&
  {Wyithe}}]{2010MNRAS.408...57P}
{Pritchard}, J.~R., {Loeb}, A., \& {Wyithe}, J.~S.~B. 2010, \mnras, 408, 57

\bibitem[{{Read} {et~al.}(2015){Read}, {Agertz}, \&
  {Collins}}]{2015arXiv150804143R}
{Read}, J.~I., {Agertz}, O., \& {Collins}, M.~L.~M. 2015, ArXiv e-prints

\bibitem[{{Read} \& {Trentham}(2005)}]{2005RSPTA.363.2693R}
{Read}, J.~I., \& {Trentham}, N. 2005, Royal Society of London Philosophical
  Transactions Series A, 363, 2693

\bibitem[{{Reddick} {et~al.}(2013){Reddick}, {Wechsler}, {Tinker}, \&
  {Behroozi}}]{2013ApJ...771...30R}
{Reddick}, R.~M., {Wechsler}, R.~H., {Tinker}, J.~L., \& {Behroozi}, P.~S.
  2013, \apj, 771, 30

\bibitem[{{Revaz} {et~al.}(2009){Revaz}, {Pfenniger}, {Combes}, \&
  {Bournaud}}]{2009A&A...501..171R}
{Revaz}, Y., {Pfenniger}, D., {Combes}, F., \& {Bournaud}, F. 2009, \aap, 501,
  171

\bibitem[{{Reyes} {et~al.}(2012){Reyes}, {Mandelbaum}, {Gunn}, {Nakajima},
  {Seljak}, \& {Hirata}}]{2012MNRAS.425.2610R}
{Reyes}, R., {Mandelbaum}, R., {Gunn}, J.~E., {et~al.} 2012, \mnras, 425, 2610

\bibitem[{{Robotham} {et~al.}(2006){Robotham}, {Wallace}, {Phillipps}, \& {De
  Propris}}]{2006ApJ...652.1077R}
{Robotham}, A., {Wallace}, C., {Phillipps}, S., \& {De Propris}, R. 2006, \apj,
  652, 1077

\bibitem[{{Robotham} {et~al.}(2011){Robotham}, {Norberg}, {Driver}, {Baldry},
  {Bamford}, {Hopkins}, {Liske}, {Loveday}, {Merson}, {Peacock}, {Brough},
  {Cameron}, {Conselice}, {Croom}, {Frenk}, {Gunawardhana}, {Hill}, {Jones},
  {Kelvin}, {Kuijken}, {Nichol}, {Parkinson}, {Pimbblet}, {Phillipps},
  {Popescu}, {Prescott}, {Sharp}, {Sutherland}, {Taylor}, {Thomas}, {Tuffs},
  {van Kampen}, \& {Wijesinghe}}]{2011MNRAS.416.2640R}
{Robotham}, A.~S.~G., {Norberg}, P., {Driver}, S.~P., {et~al.} 2011, \mnras,
  416, 2640

\bibitem[{{Roediger} \& {Courteau}(2015)}]{2015MNRAS.452.3209R}
{Roediger}, J.~C., \& {Courteau}, S. 2015, \mnras, 452, 3209

\bibitem[{{Salim} {et~al.}(2005){Salim}, {Charlot}, {Rich}, {Kauffmann},
  {Heckman}, {Barlow}, {Bianchi}, {Byun}, {Donas}, {Forster}, {Friedman},
  {Jelinsky}, {Lee}, {Madore}, {Malina}, {Martin}, {Milliard}, {Morrissey},
  {Neff}, {Schiminovich}, {Seibert}, {Siegmund}, {Small}, {Szalay}, {Welsh}, \&
  {Wyder}}]{2005ApJ...619L..39S}
{Salim}, S., {Charlot}, S., {Rich}, R.~M., {et~al.} 2005, \apjl, 619, L39

\bibitem[{{Salim} {et~al.}(2007){Salim}, {Rich}, {Charlot}, {Brinchmann},
  {Johnson}, {Schiminovich}, {Seibert}, {Mallery}, {Heckman}, {Forster},
  {Friedman}, {Martin}, {Morrissey}, {Neff}, {Small}, {Wyder}, {Bianchi},
  {Donas}, {Lee}, {Madore}, {Milliard}, {Szalay}, {Welsh}, \&
  {Yi}}]{2007ApJS..173..267S}
{Salim}, S., {Rich}, R.~M., {Charlot}, S., {et~al.} 2007, \apjs, 173, 267

\bibitem[{{Shankar} {et~al.}(2006){Shankar}, {Lapi}, {Salucci}, {De Zotti}, \&
  {Danese}}]{2006ApJ...643...14S}
{Shankar}, F., {Lapi}, A., {Salucci}, P., {De Zotti}, G., \& {Danese}, L. 2006,
  \apj, 643, 14

\bibitem[{{Sinha} \& {Holley-Bockelmann}(2012)}]{2012ApJ...751...17S}
{Sinha}, M., \& {Holley-Bockelmann}, K. 2012, \apj, 751, 17

\bibitem[{{Skrutskie} {et~al.}(2006){Skrutskie}, {Cutri}, {Stiening},
  {Weinberg}, {Schneider}, {Carpenter}, {Beichman}, {Capps}, {Chester},
  {Elias}, {Huchra}, {Liebert}, {Lonsdale}, {Monet}, {Price}, {Seitzer},
  {Jarrett}, {Kirkpatrick}, {Gizis}, {Howard}, {Evans}, {Fowler}, {Fullmer},
  {Hurt}, {Light}, {Kopan}, {Marsh}, {McCallon}, {Tam}, {Van Dyk}, \&
  {Wheelock}}]{2006AJ....131.1163S}
{Skrutskie}, M.~F., {Cutri}, R.~M., {Stiening}, R., {et~al.} 2006, \aj, 131,
  1163

\bibitem[{{Smith}(2012)}]{2012MNRAS.426..531S}
{Smith}, R.~E. 2012, \mnras, 426, 531

\bibitem[{{Somerville}(2002)}]{2002ApJ...572L..23S}
{Somerville}, R.~S. 2002, \apjl, 572, L23

\bibitem[{{Spergel} {et~al.}(2003){Spergel}, {Verde}, {Peiris}, {Komatsu},
  {Nolta}, {Bennett}, {Halpern}, {Hinshaw}, {Jarosik}, {Kogut}, {Limon},
  {Meyer}, {Page}, {Tucker}, {Weiland}, {Wollack}, \&
  {Wright}}]{2003ApJS..148..175S}
{Spergel}, D.~N., {Verde}, L., {Peiris}, H.~V., {et~al.} 2003, \apjs, 148, 175

\bibitem[{{Springel} {et~al.}(2005){Springel}, {White}, {Jenkins}, {Frenk},
  {Yoshida}, {Gao}, {Navarro}, {Thacker}, {Croton}, {Helly}, {Peacock}, {Cole},
  {Thomas}, {Couchman}, {Evrard}, {Colberg}, \& {Pearce}}]{2005Natur.435..629S}
{Springel}, V., {White}, S.~D.~M., {Jenkins}, A., {et~al.} 2005, \nat, 435, 629

\bibitem[{{Stoughton} {et~al.}(2002){Stoughton}, {Lupton}, {Bernardi},
  {Blanton}, {Burles}, {Castander}, {Connolly}, {Eisenstein}, {Frieman},
  {Hennessy}, {Hindsley}, {Ivezi{\'c}}, {Kent}, {Kunszt}, {Lee}, {Meiksin},
  {Munn}, {Newberg}, {Nichol}, {Nicinski}, {Pier}, {Richards}, {Richmond},
  {Schlegel}, {Smith}, {Strauss}, {SubbaRao}, {Szalay}, {Thakar}, {Tucker},
  {Vanden Berk}, {Yanny}, {Adelman}, {Anderson}, {Anderson}, {Annis},
  {Bahcall}, {Bakken}, {Bartelmann}, {Bastian}, {Bauer}, {Berman},
  {B{\"o}hringer}, {Boroski}, {Bracker}, {Briegel}, {Briggs}, {Brinkmann},
  {Brunner}, {Carey}, {Carr}, {Chen}, {Christian}, {Colestock}, {Crocker},
  {Csabai}, {Czarapata}, {Dalcanton}, {Davidsen}, {Davis}, {Dehnen},
  {Dodelson}, {Doi}, {Dombeck}, {Donahue}, {Ellman}, {Elms}, {Evans}, {Eyer},
  {Fan}, {Federwitz}, {Friedman}, {Fukugita}, {Gal}, {Gillespie}, {Glazebrook},
  {Gray}, {Grebel}, {Greenawalt}, {Greene}, {Gunn}, {de Haas}, {Haiman},
  {Haldeman}, {Hall}, {Hamabe}, {Hansen}, {Harris}, {Harris}, {Harvanek},
  {Hawley}, {Hayes}, {Heckman}, {Helmi}, {Henden}, {Hogan}, {Hogg}, {Holmgren},
  {Holtzman}, {Huang}, {Hull}, {Ichikawa}, {Ichikawa}, {Johnston}, {Kauffmann},
  {Kim}, {Kimball}, {Kinney}, {Klaene}, {Kleinman}, {Klypin}, {Knapp},
  {Korienek}, {Krolik}, {Kron}, {Krzesi{\'n}ski}, {Lamb}, {Leger},
  {Limmongkol}, {Lindenmeyer}, {Long}, {Loomis}, {Loveday}, {MacKinnon},
  {Mannery}, {Mantsch}, {Margon}, {McGehee}, {McKay}, {McLean}, {Menou},
  {Merelli}, {Mo}, {Monet}, {Nakamura}, {Narayanan}, {Nash}, {Neilsen},
  {Newman}, {Nitta}, {Odenkirchen}, {Okada}, {Okamura}, {Ostriker}, {Owen},
  {Pauls}, {Peoples}, {Peterson}, {Petravick}, {Pope}, {Pordes}, {Postman},
  {Prosapio}, {Quinn}, {Rechenmacher}, {Rivetta}, {Rix}, {Rockosi}, {Rosner},
  {Ruthmansdorfer}, {Sandford}, {Schneider}, {Scranton}, {Sekiguchi}, {Sergey},
  {Sheth}, {Shimasaku}, {Smee}, {Snedden}, {Stebbins}, {Stubbs}, {Szapudi},
  {Szkody}, {Szokoly}, {Tabachnik}, {Tsvetanov}, {Uomoto}, {Vogeley}, {Voges},
  {Waddell}, {Walterbos}, {Wang}, {Watanabe}, {Weinberg}, {White}, {White},
  {Wilhite}, {Wolfe}, {Yasuda}, {York}, {Zehavi}, \&
  {Zheng}}]{2002AJ....123..485S}
{Stoughton}, C., {Lupton}, R.~H., {Bernardi}, M., {et~al.} 2002, \aj, 123, 485

\bibitem[{{Strauss} {et~al.}(2002){Strauss}, {Weinberg}, {Lupton}, {Narayanan},
  {Annis}, {Bernardi}, {Blanton}, {Burles}, {Connolly}, {Dalcanton}, {Doi},
  {Eisenstein}, {Frieman}, {Fukugita}, {Gunn}, {Ivezi{\'c}}, {Kent}, {Kim},
  {Knapp}, {Kron}, {Munn}, {Newberg}, {Nichol}, {Okamura}, {Quinn}, {Richmond},
  {Schlegel}, {Shimasaku}, {SubbaRao}, {Szalay}, {Vanden Berk}, {Vogeley},
  {Yanny}, {Yasuda}, {York}, \& {Zehavi}}]{2002AJ....124.1810S}
{Strauss}, M.~A., {Weinberg}, D.~H., {Lupton}, R.~H., {et~al.} 2002, \aj, 124,
  1810

\bibitem[{{Thoul} \& {Weinberg}(1996)}]{1996ApJ...465..608T}
{Thoul}, A.~A., \& {Weinberg}, D.~H. 1996, \apj, 465, 608

\bibitem[{{Trenti} \& {Stiavelli}(2008)}]{2008ApJ...676..767T}
{Trenti}, M., \& {Stiavelli}, M. 2008, \apj, 676, 767

\bibitem[{{Tully} {et~al.}(2002){Tully}, {Somerville}, {Trentham}, \&
  {Verheijen}}]{2002ApJ...569..573T}
{Tully}, R.~B., {Somerville}, R.~S., {Trentham}, N., \& {Verheijen}, M.~A.~W.
  2002, \apj, 569, 573

\bibitem[{{Valotto} {et~al.}(1997){Valotto}, {Nicotra}, {Muriel}, \&
  {Lambas}}]{1997ApJ...479...90V}
{Valotto}, C.~A., {Nicotra}, M.~A., {Muriel}, H., \& {Lambas}, D.~G. 1997,
  \apj, 479, 90

\bibitem[{{van Dokkum} {et~al.}(2015){van Dokkum}, {Abraham}, {Merritt},
  {Zhang}, {Geha}, \& {Conroy}}]{2015ApJ...798L..45V}
{van Dokkum}, P.~G., {Abraham}, R., {Merritt}, A., {et~al.} 2015, \apjl, 798,
  L45

\bibitem[{{Velten} {et~al.}(2014){Velten}, {Caram{\^e}s}, {Fabris}, {Casarini},
  \& {Batista}}]{2014PhRvD..90l3526V}
{Velten}, H., {Caram{\^e}s}, T.~R.~P., {Fabris}, J.~C., {Casarini}, L., \&
  {Batista}, R.~C. 2014, \prd, 90, 123526

\bibitem[{{Warren} {et~al.}(2006){Warren}, {Abazajian}, {Holz}, \&
  {Teodoro}}]{2006ApJ...646..881W}
{Warren}, M.~S., {Abazajian}, K., {Holz}, D.~E., \& {Teodoro}, L. 2006, \apj,
  646, 881

\bibitem[{{Wetzel} {et~al.}(2013){Wetzel}, {Tinker}, {Conroy}, \& {van den
  Bosch}}]{2013MNRAS.432..336W}
{Wetzel}, A.~R., {Tinker}, J.~L., {Conroy}, C., \& {van den Bosch}, F.~C. 2013,
  \mnras, 432, 336

\bibitem[{{Wetzel} {et~al.}(2014){Wetzel}, {Tinker}, {Conroy}, \& {van den
  Bosch}}]{2014MNRAS.439.2687W}
---. 2014, \mnras, 439, 2687

\bibitem[{{Yamanoi} {et~al.}(2012){Yamanoi}, {Komiyama}, {Yagi}, {Okamura},
  {Iye}, {Kashikawa}, {Takata}, {Furusawa}, \& {Yoshida}}]{2012AJ....144...40Y}
{Yamanoi}, H., {Komiyama}, Y., {Yagi}, M., {et~al.} 2012, \aj, 144, 40

\bibitem[{{Yang} {et~al.}(2008){Yang}, {Mo}, \& {van den
  Bosch}}]{2008ApJ...676..248Y}
{Yang}, X., {Mo}, H.~J., \& {van den Bosch}, F.~C. 2008, \apj, 676, 248

\bibitem[{{Yang} {et~al.}(2009){Yang}, {Mo}, \& {van den
  Bosch}}]{2009ApJ...695..900Y}
---. 2009, \apj, 695, 900

\bibitem[{{York} {et~al.}(2000){York}, {Adelman}, {Anderson}, {Anderson},
  {Annis}, {Bahcall}, {Bakken}, {Barkhouser}, {Bastian}, {Berman}, {Boroski},
  {Bracker}, {Briegel}, {Briggs}, {Brinkmann}, {Brunner}, {Burles}, {Carey},
  {Carr}, {Castander}, {Chen}, {Colestock}, {Connolly}, {Crocker}, {Csabai},
  {Czarapata}, {Davis}, {Doi}, {Dombeck}, {Eisenstein}, {Ellman}, {Elms},
  {Evans}, {Fan}, {Federwitz}, {Fiscelli}, {Friedman}, {Frieman}, {Fukugita},
  {Gillespie}, {Gunn}, {Gurbani}, {de Haas}, {Haldeman}, {Harris}, {Hayes},
  {Heckman}, {Hennessy}, {Hindsley}, {Holm}, {Holmgren}, {Huang}, {Hull},
  {Husby}, {Ichikawa}, {Ichikawa}, {Ivezi{\'c}}, {Kent}, {Kim}, {Kinney},
  {Klaene}, {Kleinman}, {Kleinman}, {Knapp}, {Korienek}, {Kron}, {Kunszt},
  {Lamb}, {Lee}, {Leger}, {Limmongkol}, {Lindenmeyer}, {Long}, {Loomis},
  {Loveday}, {Lucinio}, {Lupton}, {MacKinnon}, {Mannery}, {Mantsch}, {Margon},
  {McGehee}, {McKay}, {Meiksin}, {Merelli}, {Monet}, {Munn}, {Narayanan},
  {Nash}, {Neilsen}, {Neswold}, {Newberg}, {Nichol}, {Nicinski}, {Nonino},
  {Okada}, {Okamura}, {Ostriker}, {Owen}, {Pauls}, {Peoples}, {Peterson},
  {Petravick}, {Pier}, {Pope}, {Pordes}, {Prosapio}, {Rechenmacher}, {Quinn},
  {Richards}, {Richmond}, {Rivetta}, {Rockosi}, {Ruthmansdorfer}, {Sandford},
  {Schlegel}, {Schneider}, {Sekiguchi}, {Sergey}, {Shimasaku}, {Siegmund},
  {Smee}, {Smith}, {Snedden}, {Stone}, {Stoughton}, {Strauss}, {Stubbs},
  {SubbaRao}, {Szalay}, {Szapudi}, {Szokoly}, {Thakar}, {Tremonti}, {Tucker},
  {Uomoto}, {Vanden Berk}, {Vogeley}, {Waddell}, {Wang}, {Watanabe},
  {Weinberg}, {Yanny}, {Yasuda}, \& {SDSS Collaboration}}]{2000AJ....120.1579Y}
{York}, D.~G., {Adelman}, J., {Anderson}, Jr., J.~E., {et~al.} 2000, \aj, 120,
  1579

\bibitem[{{Zabludoff} \& {Mulchaey}(1998)}]{1998ApJ...498L...5Z}
{Zabludoff}, A.~I., \& {Mulchaey}, J.~S. 1998, \apjl, 498, L5

\bibitem[{{Zheng} {et~al.}(2005){Zheng}, {Berlind}, {Weinberg}, {Benson},
  {Baugh}, {Cole}, {Dav{\'e}}, {Frenk}, {Katz}, \&
  {Lacey}}]{2005ApJ...633..791Z}
{Zheng}, Z., {Berlind}, A.~A., {Weinberg}, D.~H., {et~al.} 2005, \apj, 633, 791

\end{thebibliography}
